\documentclass[11pt]{article}
\usepackage{style}
\usepackage{shortcuts}
\date{}

 \author{Amir Abboud\thanks{Weizmann Institute of Science. Part of this work was done while visiting the Courant Institute of Mathematical Sciences at New York University. This work is part of the project CONJEXITY that has received funding from the European Research Council (ERC) under the European Union's Horizon Europe research and innovation programme (grant agreement No.~101078482).}
 \and Shyan Akmal\orcidlink{0000-0002-1825-0097}\thanks{Max Planck Institute for Informatics, Saarbr\"ucken, Germany. This work received funding from the Klaus Tschira Boost Fund, a joint initiative of GSO Guidance, Skills \& Opportunities e.V. and the Klaus Tschira Stiftung.}
 \and Ron Safier\thanks{Weizmann Institute of Science.}
 \and Nathan Wallheimer\footnotemark[3]
 }

\title{Lower Bounds for Structured OMv, APSP, and Triangle Detection}

\begin{document}
   \pagenumbering{gobble}
   \hypersetup{pageanchor=false}
   \maketitle 
   \begin{abstract}
      We study the complexity of core graph and matrix problems under two fundamental notions of structure: bounded \VC-dimension and exclusion of a fixed subgraph pattern. 
      We obtain the first \VC{} and distance-\VC{} parameterized lower bounds for OMv, APSP, and Triangle Detection.
      These results show that existing $\Theta(1/d)$-type savings in the exponent for inputs with dimension at most $d$ are the ``right answer'' for the  complexity of these problems.

      We prove that restricting OMv to matrices of \VC-dimension at most $d$ can reduce the optimal running time of OMv by a factor of at most $\tilde \Order(n^{1/d + O(1/d^2)})$. 
      Assuming the OMv Conjecture, this implies a nearly-matching lower bound for the $\tOrder(n^{3-1/d})$ algorithm of Anand, van den Brand, and McCarty for solving OMv on instances with dimension at most $d$. 
      We also obtain tight bounds for OMv on matrices of \VC-dimension $2$, and on bounded \emph{\VC-density}. 

      For Triangle Detection and unweighted APSP, we show that for every sufficiently large $d$, 
      restricting the distance \VC-dimension to at most $d$
      can improve upon the optimal running time of unrestricted Triangle Detection by a factor of at most $n^{O(1/d)}$.
      In particular, assuming the Boolean Matrix Multiplication (BMM) Conjecture, combinatorial algorithms for APSP and Triangle Detection in graphs of distance \VC-dimension at most $d$ require $n^{3-O(1/d)}$ time,
      matching existing algorithms up to the constant in the $O(1/d)$ savings.

      Our main conceptual contribution is to introduce \emph{filtering} as a general framework for proving fine-grained lower bounds on structured inputs. 
      Intuitively, we apply a family of filters to an arbitrary input to eliminate all configurations forbidden by a given structural promise, while ensuring that every solution in the original input survives in at least one filtered instance.
      For bounded \VC-dimension, the filters prevent complicated shattering patterns, and for $H$-freeness, the filters remove appearances of the pattern $H$. 
      
      Using this method, we prove that $H$-freeness can reduce the optimal running time of Triangle Detection by a factor of at most $\tilde \Order(n^{3v(H)/e(H)})$. 
      In particular, under the BMM Conjecture, combinatorial algorithms for Triangle Detection
      require $n^{3-3v(H)/e(H) - o(1)}$ time 
      on such instances whenever $e(H)>3v(H)$. 
      We complement this lower bound with a matching $n^{3-\Theta(v(H)/e(H))}$ upper bound for an infinite family of patterns $H$.
      We also present stronger, specialized lower bounds when $H$ is a biclique.

      Finally, we uncover a new algorithmic connection between these two notions of structure: for every bipartite pattern $H$, induced $H$-freeness allows Triangle Detection to be reduced to instances of bounded \VC-dimension. 
      This yields truly subcubic combinatorial algorithms for all bipartite $H$. 
      For comparison, previously such an algorithm was unknown even for $H = P_7$.
   \end{abstract}

   \clearpage
   \tableofcontents
   \clearpage
   \pagenumbering{arabic}
   \setcounter{page}{1}
   \hypersetup{pageanchor=true}

   \section{Introduction}
      \label{sec:intro}
      Fine-grained complexity seeks to characterize the hardness of important problems by identifying a small set of \emph{core} problems,
each explaining the difficulty of obtaining faster algorithms for
broad classes of computational tasks.
Some of the core problems most fundamental to static and dynamic graph algorithms
are Online Matrix-Vector Multiplication (OMv),
Boolean Matrix Multiplication (BMM), and Triangle Detection.

These problems play two complementary roles. On the one hand, their
conjectured hardness forms the basis of numerous conditional lower bounds.
On the other hand, they are natural algorithmic questions in their own
right. Despite sustained community effort, these problems have resisted
major improvements in their running times. Such improvements are therefore
expected to require new ideas whose consequences would extend
far beyond any individual core problem.

Indeed, recent algorithmic breakthroughs illustrate this
effect. 
They include truly subcubic algorithms for weighted All-Pairs
Shortest Paths (APSP) and truly subquadratic algorithms for
3SUM~\cite{alman2026truly}, as well as subexponential algorithms for
$k$-SAT for every fixed $k$~\cite{claude2026algorithm,openai2026eth}.
These problems were central in fine-grained complexity, and the ideas behind their improved runtimes have led to faster algorithms for many other problems like
Clique Detection~\cite{williams2026kclique},
Hyperclique Detection~\cite{fischer2026hypercliqu,alppw2026hyperclique,cimy2026hyperclique}, and 
Dynamic Matrix Inversion~\cite{alman2026subquadratic}.

All the advances above rely on \emph{algebraic methods}: tensor 
identities and polynomial techniques. 
They highlight another longstanding question of 
whether \emph{combinatorial} approaches can also yield comparable speedups. 
Combinatorial algorithms which avoid such algebraic subroutines
are desired because they tend to be more \emph{interpretable}, closer to being \emph{practical}, and \emph{generalize} to additional settings. 
A recent example is provided in \cite{chan20263sum}: they design 
a truly subquadratic algorithm for 3SUM using dynamic programming
and inclusion--exclusion,
not relying on a tensor identity for fast matrix multiplication as the original algebraic approach
of~\cite{alman2026truly} did.
Thanks to their combinatorial approach, \cite{chan20263sum} is able to generalize the 3SUM algorithm to improve the runtime of
$k$-SUM for every odd $k\ge 3$, an extension left open by the earlier paper~\cite[Section 6]{alman2026truly}.

To try and improve the runtimes for core problems in fine-grained complexity, and design algorithms more applicable in practice, 
a growing line of work has studied these core problems on \emph{structured} inputs.
This viewpoint has led to progress for solving APSP~\cite{bonnet2026fast,abboud2025all,bonnet2026time,chan2019apsp,planken2012computing},
3SUM~\cite{chan2015cluster,chan2023fredman,abboud2023stronger,jin2023removing}, Triangle
Detection~\cite{abboud2026triangle,abboud2026equivalent,Gkasieniec2003improved,coudert2019fpt,chan2023triangle,bonnet2026time},
and OMv~\cite{gsieniec2021pushing,anand2026structural,cardinal2026biclique,bonnet2026fast,bonnet2026time,kozma2026fast} on restricted classes of inputs. 

There are two main motivations for studying structured instances. 
The first is
a \emph{beyond worst-case} perspective. Although these problems are conjectured 
to be hard in general, an algorithm can be substantially faster when an 
appropriate structural parameter is far from its worst-case value. Such 
parameters may also help explain why instances arising in practice 
are often observed to be easier than worst-case ones.
Most work motivated by this perspective has focused on algorithms.
Comparatively few results establish lower bounds whose exponents track the
structural parameter (see, e.g.,~\cite{kopelowitz2016higher,bringmann2021current,fischer2023effect,bringman2018multivariate}),
leaving open whether the known dependence on these parameters is optimal, or whether substantially faster algorithms are possible.

The second motivation is to better understand the anatomy of worst-case
instances: which kinds of structure make a core problem easier, and which
do not? An answer to this question may isolate the true obstacles to faster
algorithms. Indeed, a recurring strategy in combinatorial attacks on these problems is to
exploit a structure-versus-pseudorandomness dichotomy ~\cite{abboud2024new,fischer2026universe}. 
Articulating which
notions of structure are algorithmically useful is central to this approach.

\paragraph{\VC-dimension and OMv.}
One fundamental notion of structure is \VC-dimension~\cite{vapnik71dim},
which measures the richness of the patterns realized by a set system.
It is central in learning theory~\cite{shalev2014understanding,blumer1989learnability}
and computational geometry~\cite{har2011geometric,matouvsek2002lectures,chan2025truly},
and has received increasing attention in combinatorics and
algorithms~\cite{fox2019erdHos,nguyen2025induced,ducoffe2022diameter,duraj2023better}.
It arises naturally in matrix and graph problems: the rows of a Boolean matrix
can be viewed as subsets of its columns, while the neighborhoods of a graph
form a set system over its vertices.

A particularly sharp question about the algorithmic power of bounded
\VC-dimension arises for OMv. In this problem, we are given an $n\times n$
Boolean matrix $M$ that may be preprocessed in polynomial time, 
and then receive a sequence of $n$
Boolean vectors $v_1,\ldots,v_n$ arriving one at a time. Upon receiving
$v_i$, the algorithm must output $Mv_i$ before seeing the next vector. The OMv
Conjecture~\cite{henzinger2015unifying} asserts that this problem requires
$n^{3-o(1)}$ total time.
It is one of the main sources of conditional
lower bounds for dynamic graph
problems~\cite{dahlgaard2016,abboud2016popular,henzinger2017conditional,dallant2025conditional}.

Anand, van den Brand, and McCarty~\cite{anand2026structural} recently showed
that OMv in matrices of \VC-dimension at most $d$ can be solved in
$\tOrder(n^{3-1/d})$ total time.\footnote{Throughout, $\tOrder,\tilde \Omega,$ and $\tilde\Theta$ hide $\poly \log{n}$ factors.} 
Thus, bounded \VC-dimension  makes
OMv easier. For $d>1$, however, this bound remains polynomially larger than $n^2$. Could a different technique achieve quadratic time for every
constant $d$? Even if quadratic time is impossible, is the $1/d$ saving in the
exponent best possible, or could it be improved, say, to $1/\sqrt d$? This motivates the main question of our work:

\begin{quote}
\emph{Does OMv in matrices of constant \VC-dimension admit a quadratic-time
algorithm?}
\end{quote}

This question was posed explicitly by Anand et
al.~\cite[Section 4]{anand2026structural}, and reiterated in a presentation by van
den Brand, who suggested the known upper bound might not be optimal~\cite{vdB-talk}.

\paragraph{Distance \VC-dimension and APSP}

A conceptually similar question arises for combinatorial algorithms for Triangle
Detection and APSP. For APSP, the task is to compute the full matrix of exact
distances in an unweighted undirected graph.
A hypothesis known as the BMM Conjecture (defined in \Cref{conj:bmm}) rules out truly subcubic combinatorial algorithms for
either problem in the worst case~\cite{williams2010subcubic}. We ask how this picture changes
on graphs of bounded \VC-dimension.

For distance problems such as APSP, ordinary \VC-dimension, defined as the 
\VC-dimension of the neighborhoods set system, does not fully capture the relevant structure: it controls 
only balls of radius one. The appropriate metric notion is 
\emph{distance \VC-dimension}, defined as the \VC-dimension of the set system
of all closed balls

\[
   \mathcal{B}(G)=
   \set{N_r[v]\mid v\in V(G),\ r \geq 0}
\]
where $N_r[v]$ denotes the set of vertices in the graph at distance at most $r$ from $v$.

Existing techniques obtain an $n^{1/d}$-factor improvement over the cubic
running time for combinatorial APSP in graphs of distance \VC-dimension at
most $d$~\cite[Theorem 3]{bonnet2026time}.\footnote{This improvement applies more generally to graphs of \emph{\VC-density} at most $d$, a related notion we discuss later.} This raises the analogue of the OMv
question: is the $1/d$ saving in the exponent inherent, or can it be improved
substantially---perhaps all the way to a quadratic-time algorithm for every
constant $d$?

This question is particularly relevant to recent work on the Diameter problem
for planar and geometric graphs. 
Planar graphs have distance \VC-dimension
at most $4$. 
More generally,  many well-studied geometric intersection graphs also have distance \VC-dimension at most
$4$~\cite{chan2025truly}, and every $K_h$-minor-free graph has distance
\VC-dimension at most $(h-1)$~\cite{ducoffe2022diameter}.

These bounds suggest a research program towards achieving linear-time Diameter
algorithms on these graph classes. 
One proposed step in this program is to
develop a quadratic-time algorithm for APSP in graphs of distance
\VC-dimension at most $4$, or, more ambitiously, in graphs of any constant
distance \VC-dimension. Such an algorithm could then be combined with
class-specific machinery such as $r$-divisions and Voronoi diagrams.
Thus, the case of small constant distance \VC-dimension, and particularly
$d=4$, is an important target within this broader research project.

This leads to the following question:

\begin{quote}
\emph{Do APSP and Triangle Detection admit
quadratic-time algorithms on graphs of constant distance \VC-dimension?}
\end{quote}

\subsection{Overview of Results}
We give conditional negative answers to the questions raised above. 
We do this by designing a \emph{filtering framework} that converts an
unrestricted input into a controlled collection of structured instances whose
relevant set systems have bounded \VC-dimension. 
The filters eliminate 
complicated shattering patterns while ensuring that every witness to the
original input survives in at least one filtered instance. Consequently,
a sufficiently fast algorithm on instances with bounded \VC-dimension would violate
the conjectured hardness of the corresponding unrestricted problem.

For OMv, we show that the known $n^{1/d}$-type saving is essentially optimal. 
Our lower bound matches the known upper bound up to subpolynomial 
factors for $d=2$, and up to an $n^{O(1/d^2)}$ factor for every $d\geq3$. We
also obtain tight lower bounds in terms of \VC-density, a quantitative notion
based on the full shatter function.

For graph problems, we prove new conditional lower bounds for both
combinatorial APSP and combinatorial Triangle Detection in graphs of bounded
distance \VC-dimension. For every sufficiently large constant $d$, our lower
bounds require time  $n^{3-O(1/d)}
$ under the BMM Conjecture. Together with the known
$n^{3-\Omega(1/d)}$-time algorithms, this determines the correct order of
magnitude of the exponent saving: bounded distance \VC-dimension permits a
$\Theta(1/d)$ saving, but no asymptotically larger one. In particular, it
prohibits a quadratic-time algorithm that works for every constant value of
$d$, and even small parameterized improvements such as replacing the $1/d$ savings by
$1/\sqrt d$.

This rules out the general form of the research program described above.
However, our lower bounds do not apply to $d=4$, and therefore leave open the
case relevant to planar graphs, for example. 
Any positive result in this
case would have to exploit the particularly small value $d=4$, or leverage further
structure beyond just low distance \VC-dimension.

\paragraph{Beyond \VC-dimension: $H$-free graphs.}
The filtering framework has applications to structured inputs beyond
\VC-dimension. We apply it to Triangle Detection in $H$-free graphs, a setting
that has received considerable recent attention~\cite{abboud2026triangle,abboud2026equivalent}. 
For every sufficiently dense fixed pattern $H$, we prove a new conditional lower bound whose exponent varies
with the density of $H$. We complement this result with new deterministic
upper bounds for a broad family of patterns, improving the saving over cubic
time from $n^{\Theta(2^{-v(H)})}$ to $n^{\Omega(1/v(H))}$, where $v(H)$ denotes the number of vertices of $H$. 
For a natural infinite family of dense patterns, the upper and lower bounds have matching
dependence on $v(H)$ up to constant factors in the exponent. 
For $K_{s,t}$-free graphs, we obtain a stronger lower bound, leaving
a gap of $1/s$ between the lower and upper exponents. 

Finally, we identify a direct algorithmic connection between $H$-freeness and
\VC-dimension: for certain induced forbidden patterns, $H$-freeness forces the 
relevant set systems to have bounded \VC-dimension. This observation yields
truly subcubic algorithms for Triangle Detection in induced $H$-free graphs 
for every bipartite $H$, and in particular answers the recently raised open 
question for induced paths~\cite{abboud2026triangle}. 

\subsection{Results in More Detail}

We next state the results more precisely. We organize them by the relevant
structural promise, while emphasizing that they arise from a common framework.

Let $T_{\mathrm{OMv}}$ and $T_{\mathrm{TD}}$ denote the optimal
randomized worst-case running times for $n\times n$ OMv and
$n$-vertex Triangle Detection, respectively, with high probability (\whp). 
We use \whp to mean a success probability of at least $2/3$. 
For OMv, total time includes preprocessing and all $n$ online 
queries, and success requires answering the entire query sequence correctly.

We state our lower bounds in a \emph{relative} form, meaning that they hold 
\emph{unconditionally} in terms of the true complexities $T_{\mathrm{OMv}}$ and $T_{\mathrm{TD}}$. 
These bounds are summarized in  \Cref{tab:lower-bounds}.
The lower bounds for APSP and Triangle Detection also hold for
randomized combinatorial algorithms when $T_{\mathrm{TD}}$ is taken over
that class.

\begin{table}[H]
   \centering
   \small
   \setlength{\tabcolsep}{4pt}
   \begin{tabular}{@{}lccccc@{}}
      \toprule
      \textbf{Problem} & \textbf{\VC-dimension} & \textbf{\VC-density} &
      \textbf{Distance \VC-dimension} &
      \textbf{$H$-freeness} & \textbf{$K_{s,t}$-freeness} \\
      & $d$ & $d$ & $d$ & $\rho=\dfrac{3v(H)}{e(H)}$ & $t\geq(s-1)!+1$ \\
      \midrule
      OMv &
      $\widetilde{\Omega}\left(\dfrac{T_{\mathrm{OMv}}}{n^{1/d+\Order(1/d^2)}}\right)$ &
      $\widetilde{\Omega}\left(\dfrac{T_{\mathrm{OMv}}}{n^{1/d}}\right)$ &
      $\widetilde{\Omega}\left(\dfrac{T_{\mathrm{OMv}}}{n^{1/d+\Order(1/d^2)}}\right)$ &
      \textemdash & \textemdash \\
      & {\footnotesize\Cref{thm:omv-vc-lb}} & {\footnotesize\Cref{thm:shattering}} &
      {\footnotesize\Cref{cor:omv-dvc}} & & \\
      \addlinespace
      \makecell[l]{Triangle\\Detection} & \textemdash & \textemdash &
      $\widetilde{\Omega}\left(\dfrac{T_{\mathrm{TD}}}{n^{\Order(1/d)}}\right)$ &
      $\widetilde{\Omega}\left(\dfrac{T_{\mathrm{TD}}}{n^\rho}\right)$ &
      $\widetilde{\Omega}\left(\dfrac{T_{\mathrm{TD}}}{n^{3/s}}\right)$ \\
      & & & {\footnotesize\Cref{thm:triangle-dist}} &
      {\footnotesize\Cref{thm:triangle-lb}} & {\footnotesize\Cref{thm:triangle-kst}} \\
      \addlinespace
      APSP & \textemdash & \textemdash &
      $\widetilde{\Omega}\left(\dfrac{T_{\mathrm{TD}}}{n^{\Order(1/d)}}\right)$ &
      \textemdash & \textemdash \\
      & & & {\footnotesize\Cref{thm:APSP-lb}} & & \\
      \bottomrule
   \end{tabular}
   \caption{Our lower bounds for graph and matrix problems on structured inputs.
   Technically, the lower bound for Triangle Detection also needs an additive $-n^2$ term.}
   \label{tab:lower-bounds}
\end{table}

\subsubsection{Online Matrix-Vector Multiplication and \VC-Dimension}
Given an $n\times n$ Boolean matrix $M$ and a Boolean vector $v\in
\set{0,1}^n$, their Boolean product is the vector $w=Mv$ defined by
\[
   w[i]=\bigvee_{j=1}^n \bigl(M[i,j]\land v[j]\bigr).
\]
\begin{definition}[Online Matrix-Vector Multiplication~\cite{henzinger2015unifying}]
   In the Online Matrix-Vector Multiplication (OMv) problem, we are given a
   Boolean matrix $M\in\set{0,1}^{n\times n}$ that may be preprocessed in
   polynomial time. We then receive Boolean vectors
   $v_1,\ldots,v_n\in\set{0,1}^n$ one at a time. After receiving $v_i$, we
   must output $Mv_i$ before receiving $v_{i+1}$.
\end{definition}

\begin{conjecture}[OMv Conjecture~\cite{henzinger2015unifying}]
   \label{conj:omv}
   There is no $n^{3-\Omega(1)}$-time randomized algorithm that solves OMv \whp. 
\end{conjecture}

In a structured OMv instance, we promise that the matrix $M$ belongs to a
restricted class, but permit the online query vectors $v_i$ to be arbitrary.

To define the relevant restriction, identify each row of $M$ with the set of
columns in which it has a $1$. A set $T$ of columns is \emph{shattered} if the
rows of $M$, restricted to $T$, realize all $2^{|T|}$ subsets of $T$. The
\VC-dimension of $M$ is the maximum size of such a \emph{shattered set}.

Anand, van den Brand, and McCarty~\cite{anand2026structural} studied OMv in
matrices of bounded \VC-dimension and obtained the following.

\begin{proposition}[{Anand et al.~\cite[Theorem 1.3]{anand2026structural}}]
   \label{thm:omv-ub}
   OMv in $n\times n$ matrices of \VC-dimension at most $d$ can be solved in
   $\tOrder(n^{3-1/d})$ total time \whp
\end{proposition}
This algorithm saves a factor of $n^{1/d}$ over the standard cubic runtime for OMv. 
We show that in general, restricting to inputs with \VC-dimension at most $d$ can speed up algorithms for OMv by a factor of at most $n^{1/d + O(1/d^2)}$. 

   \begin{theorem}
      \label{thm:omv-vc-lb}
      For every integer $d\ge3$, any algorithm that solves OMv in
      $n\times n$ matrices of \VC-dimension at most $d$ \whp requires
      total time $\tilde{\Omega}(T_{\mathrm{OMv}}/n^{\alpha_d})$, where
      \begin{equation}
      \label{eq:ad-exponent}
         \alpha_d=
         \frac{d^3+3d^2+14d+18}{(d+1)^2(d^2-d+6)}
         = \frac{1}{d}+\Order(1/d^2).
      \end{equation}
   \end{theorem}

   For dimension $d=2$, we show that a speedup of $n^{1/2}$ is the best possible. 

   \begin{theorem}
      \label{thm:omv-vc-2-lb}
      Any algorithm that solves OMv in $n\times n$ matrices of
      \VC-dimension at most $2$ \whp requires total time
      $\tilde{\Omega}(T_{\mathrm{OMv}}/n^{1/2})$.
   \end{theorem}
Under the OMv Conjecture, our results imply that known algorithms are essentially optimal for dimension $d=2$, and leave only a lower order $O(1/d^2)$ gap in the exponent for $d \geq 3$. 

   \begin{corollary}
      Assuming \Cref{conj:omv}, for every integer $d\geq2$, 
      any algorithm that solves OMv on $n\times n$ matrices of 
      \VC-dimension at most $d$ \whp requires total time at least 
      \[
         \begin{cases}
            n^{3-\alpha_d-o(1)} & \text{if $d\geq3$,} \\
            n^{5/2-o(1)} & \text{if $d=2$,}
         \end{cases}
      \]
      where $\alpha_d$ is defined in \Cref{eq:ad-exponent}.
   \end{corollary}

It is known that OMv in matrices of \VC-dimension $1$ admits an $\tOrder(n^2)$-time algorithm,
which is optimal up to logarithmic factors given the problem's quadratic input and output
size. 
Thus, we have a tight understanding of the complexity of OMv at each constant \VC-dimension.

\paragraph{OMv in matrices of bounded \VC-density.}
Our next result concerns a more quantitative restriction on the same set
system. Whereas \VC-dimension records the size of the largest shattered set,
the \emph{shatter function} records the maximum number of patterns that can
be realized on $t$ columns.

   \begin{definition}[Shatter Function]
    \label{def:shatter}
      Let $\mathcal{F}$ be a set family over $[n]$.
      The shatter function $\pi_{\mathcal{F}}\colon [n]\to \mathbb{N}$
      assigns to each $t\in[n]$ the maximum number of distinct sets obtained
      by intersecting members of $\mathcal{F}$ with a subset of $[n]$ of size $t$:
      \[
         \pi_{\mathcal{F}}(t)
         :=
         \max_{\substack{S\subseteq[n]\\ |S|=t}}
         \abs[\big]{\set{S\cap U\mid U\in\mathcal{F}}}.
      \]
      Given a Boolean matrix $M\in\set{0,1}^{m\times n}$, we let $\pi_M$ be
      the shatter function of its row set system. Equivalently, $\pi_M(t)$ is
      the maximum number of distinct row patterns obtained by restricting $M$
      to a set of $t$ columns.
   \end{definition}

    \begin{definition}[\VC-Dimension of Set Systems]
        \label{def:vc-dim-set-system}
        Let $\mathcal{F}$ be a set family.
        Its \VC-dimension is defined to be the largest integer $t$ such that $\pi_{\mathcal{F}}(t) = 2^t$.
    \end{definition}
   
By definition, $\pi_M(t)\leq 2^t$, and the \VC-dimension of $M$ is the largest
$t$ for which equality holds. If $M$ has \VC-dimension at most $d\geq1$, the
Sauer--Shelah lemma~\cite{sauer1972density,shelah1972combinatorial} gives
   \begin{equation}
   \label{eq:sauer-shelah-intro}
      \pi_M(t)
      \leq \sum_{i=0}^d \binom{t}{i}
      \leq (d+1)t^d
   \end{equation}
for
every $1\leq t\leq n$.

We say that a family $\mathcal M$ of Boolean matrices has \VC-density at most
$d$ if there exists a  $c>0$ independent of the dimensions of the
matrices, such that every $M\in\mathcal M$ with $n$ columns satisfies
\[
   \pi_M(t)\leq c t^d
   \qquad\text{for every integer }1\leq t\leq n.
\]
By \Cref{eq:sauer-shelah-intro}, a family of matrices of \VC-dimension at most
$d$ has \VC-density at most $d$. Thus, bounded \VC-density is a weaker promise than bounded \VC-dimension in general.

The biclique decompositions of Krapivin et al.~\cite[Theorem 39]{krapivin2026optimal}
imply an $\Order(n^{3-1/(d+1)})$-time algorithm for OMv in matrix families
of \VC-density at most $d$.\footnote{This bound also follows from a direct lookup-table 
   approach for OMv: Partition the columns into blocks of size $s=n^{1/(d+1)}$ 
and list the $\Order(s^d)$ distinct rows in each block. This takes $\Order(n^2)$ time. 
Given a query vector, we can compute the corresponding block-vector product in time $\Order(s^{d+1} + n)$ per block. 
Over all query vectors and all blocks, the total time is $\Order(n \cdot (n/s) \cdot (s^{d+1} + n)) = \Order(n^{3-1/(d+1)})$.}
   The improved decomposition of Cardinal, McCarty, 
and Yuditsky~\cite[Theorem 1]{cardinal2026biclique} yields a 
randomized $\tOrder(n^{3-1/d})$-time algorithm. 

We show that in general,  \VC-density $d$ can guarantee a speed up factor of at most $n^{1/d}$ over the optimal running time for OMv. 
\begin{theorem}
   \label{thm:shattering}
   For every integer $d\geq1$, there exists a family $\mathcal M_d$ of
   square Boolean matrices with \VC-density at most $d$ such that any
   algorithm that solves OMv in $n\times n$ matrices from $\mathcal M_d$
   \whp requires total time $\tilde{\Omega}(T_{\mathrm{OMv}}/n^{1/d})$.
\end{theorem}

Under the OMv Conjecture, this implies existing algorithms for OMv on instances of bounded \VC-density are essentially optimal.

\begin{corollary}
   Assuming \Cref{conj:omv}, for every integer $d\geq1$, any algorithm
   that solves OMv on $n\times n$ matrices from the family $\mathcal M_d$
   in \Cref{thm:shattering} \whp requires $n^{3-1/d-o(1)}$ time.
\end{corollary}

Beyond giving a tight lower bound for a natural class of matrices,
\Cref{thm:shattering} identifies a barrier for future algorithms: any
improvement to \Cref{thm:omv-ub} would have to exploit a property specific to
bounded \VC-dimension that fails for bounded \VC-density.

\subsubsection{Distance \VC-Dimension}
Typically, the \VC-dimension of a graph is defined to the \VC-dimension of the set system of its open
neighborhoods.
Unfortunately, this definition does not appear to capture the distance information of graphs: for example,
subdividing every edge of a graph once doubles all distances between original
vertices, yet the resulting graph always has ordinary \VC-dimension at most
$2$.
A  metric-sensitive alternative is the \emph{distance
\VC-dimension} (used, e.g., in
~\cite{chan2025truly,duraj2023better,ducoffe2022diameter}). 

For a graph $G$,
consider the set system of all closed balls
\begin{equation}
    \label{eq:closed-ball-system}
   \mathcal B(G)=
   \set{N_r[v]\mid v\in V(G),\ r\in\mathbb Z_{\geq 0}}.
\end{equation}
We define
\[
   \VCB(G)=\VC\bigl(\mathcal B(G)\bigr)
\]
to be the distance \VC-dimension of $G$.

Our metric lower bounds are based on the Boolean Matrix Multiplication (BMM) Conjecture.
In Triangle
Detection, the task is to decide whether an $n$-vertex graph contains three
mutually adjacent vertices. By the reduction of Vassilevska W. and
Williams~\cite{williams2010subcubic}, a truly subcubic combinatorial algorithm
for Triangle Detection is equivalent to one for BMM.

\begin{conjecture}[BMM Conjecture]
   \label{conj:bmm}
   There is no truly subcubic randomized combinatorial algorithm for Triangle Detection,
   or equivalently for BMM, that succeeds \whp
\end{conjecture}

We show that restricting APSP to graphs of distance
$\VC$-dimension $d$ can improve upon the optimal running time of unrestricted 
Triangle Detection by a factor of at most $n^{O(1/d)}$.

\begin{restatable}{theorem}{apspDistanceVCLowerBound}
   \label{thm:APSP-lb}
   For all sufficiently large integers $d$, there exists $\delta>0$
   with $\delta=\Order(1/d)$ such that any randomized algorithm that
   solves APSP in $n$-vertex graphs $G$ satisfying $\VCB(G)\leq d$ \whp
   requires time $\tilde{\Omega}(T_{\mathrm{TD}}/n^\delta)$.
\end{restatable}

In particular, the BMM conjecture implies the following lower bound.

\begin{corollary}
   Assuming \Cref{conj:bmm}, for all sufficiently large integers $d$ and
   every constant $\eps>0$, no randomized combinatorial algorithm solves
   APSP in $n$-vertex graphs $G$ satisfying $\VCB(G)\leq d$ in time
   $n^{3-\delta-\eps}$ \whp, for some $\delta = O(1/d)$.
\end{corollary}

Similarly, we obtain the following lower bounds for Triangle Detection.

\begin{restatable}{theorem}{triangleDistanceVCLowerBound}
   \label{thm:triangle-dist}
   For all sufficiently large integers $d$, there exists $\tau>0$ with
   $\tau=\Order(1/d)$ such that any randomized algorithm that solves
   Triangle Detection in $n$-vertex graphs $G$ satisfying $\VCB(G)\leq d$
   \whp has running time at least
   $\tilde\Omega(T_{\mathrm{TD}}/n^\tau)-n^2$.
\end{restatable}
\begin{corollary}
   Assuming \Cref{conj:bmm}, for all sufficiently large integers $d$ and
   every constant $\eps>0$, no randomized combinatorial algorithm solves
   Triangle Detection in $n$-vertex graphs $G$ satisfying $\VCB(G)\leq d$
   in time $n^{3-\tau-\eps}$ \whp, where $\tau$ is as in \Cref{thm:triangle-dist}.
\end{corollary}

Relative to the standard cubic-time combinatorial algorithms for APSP and Triangle Detection, our lower bounds leave only
an $\Order(1/d)$ gap in the exponent. 

We analogously define distance \VC-dimension for square Boolean matrices.
Recall that for Boolean matrices $A$ and $B$, their Boolean product $C$ is
defined by
\[
   C[i,j]=\bigvee_k A[i,k]\land B[k,j]
\]
for all indices $i,j$. For a nonnegative integer $r$, the $r$th power $M^r$
of a Boolean matrix $M$ is the Boolean product of $r$ copies of $M$. By
convention, $M^0=I$ is the identity matrix.

\begin{definition}[Matrix Distance \VC-dimension]
   For a Boolean matrix $M\in\set{0,1}^{n\times n}$ and a binary vector $x$,
   let $\supp(x)=\set{j\mid x_j=1}$. We define the distance \VC-dimension of
   $M$ as the \VC-dimension of the set system
   \[
      \mathcal B(M)=
      \set[\Big]{\supp\bigl((M^r)[i,\cdot]\bigr)
      \Bigm| i\in[n],\ r\in\mathbb Z_{\geq 0}}.
   \]
   Equivalently, this is the \VC-dimension of the multiset of rows of
   $M^0,M^1,M^2,\ldots$, viewed as characteristic vectors.
\end{definition}
We use the same notation $\VCB(M)$ for this matrix parameter.

If $M$ is the adjacency matrix of an undirected graph with a $1$ in every
diagonal entry, then the support of row $i$ of $M^r$ is precisely $N_r[i]$,
so the graph and matrix definitions agree. Moreover, $M=M^1$, and hence
\[
   \VC(M)\leq \VCB(M).
\]
Thus OMv in matrices of \VCB-dimension at most $d$ is a special case of OMv
in matrices of \VC-dimension at most $d$. A converse reduction is less
immediate. We obtain such a reduction while losing only an additive constant
in the parameter, showing that bounded \VC-dimension and bounded
\VCB-dimension are nearly equivalent notions of structure for OMv.

\begin{theorem}
   \label{thm:omv-vc-to-dvc}
   There is a linear-time reduction from OMv in $n\times n$ matrices of 
   \VC-dimension at most $d$ to OMv in $(n+1)\times(n+1)$ matrices of
   \VCB-dimension at most $d+2$.
\end{theorem}
\begin{corollary}
   \label{cor:omv-dvc}
   For every integer $d\ge5$, any algorithm that solves OMv in
   $n\times n$ matrices of \VCB-dimension at most $d$ \whp requires
   total time $\tilde{\Omega}(T_{\mathrm{OMv}}/n^{\alpha_{d-2}})$,
   where $\alpha_{d-2} = 1/(d-2) + \Order(1/d^2) = 1/d + \Order(1/d^2)$ is as in \Cref{eq:ad-exponent}.
\end{corollary}

\subsubsection{Triangle Detection in \texorpdfstring{$H$}{H}-Free Graphs}

We next apply our filtering framework to Triangle Detection when the input
excludes a fixed graph $H$ as a subgraph. This gives a continuum of lower
bounds governed by the density of $H$. 
We let $v(H)$ and $e(H)$ denote the numbers of vertices and edges in $H$ respectively.

   \begin{theorem}
      \label{thm:triangle-lb}
      For every fixed graph $H$ with $e(H)>0$, any randomized algorithm
      that solves Triangle Detection in $n$-vertex $H$-free graphs \whp
      has worst-case running time at least
      \begin{equation}
        \label{eq:additive-n^2}
         \tilde\Omega\left(\frac{T_{\mathrm{TD}}}{n^{3v(H)/e(H)}}\right)-n^2.
      \end{equation}
   \end{theorem}

   \begin{remark}[Additive Quadratic Term]
      \label{remark:quadratic}
      The subtracted $n^2$ term from \Cref{eq:additive-n^2} matters if the leading term from \Cref{eq:additive-n^2} is $\Order(n^2)$. 
      In many cases of interest however, like when the input is given as an adjacency matrix, Triangle Detection admits a trivial unconditional quadratic-time
      lower bound, and this term can be omitted.
   \end{remark}

   \begin{corollary}
      \label{cor:triangle-lb}
      Assuming \Cref{conj:bmm}, for every fixed graph $H$ with $e(H)>3v(H)$,
      any randomized combinatorial algorithm that solves Triangle Detection
      on $n$-vertex $H$-free graphs \whp requires
      $n^{3-\frac{3v(H)}{e(H)}-o(1)}$ time.
   \end{corollary}
   Notice that the condition $e(H)>3v(H)$ in the corollary is precisely the regime where 
   \[3-3v(H)/e(H)>2,\] so the negative quadratic term from \Cref{eq:additive-n^2} vanishes. 

   We next give a sharper lower bound for Triangle Detection in $H$-free graphs when $H$ is an unbalanced biclique 
   $K_{s,t}$ with parts of sizes $s$ and $t$.
   \begin{theorem}
      \label{thm:triangle-kst}
      For all fixed integers $s\geq2$ and $t\geq(s-1)!+1$, any randomized
      algorithm that solves Triangle Detection in $n$-vertex
      $K_{s,t}$-free graphs \whp has running time at least
      $\tilde\Omega(T_{\mathrm{TD}}/n^{3/s})-n^2$.
   \end{theorem}

   Since $T_{\mathrm{TD}}\le \Order(n^3)$,
   the lower bound above is interesting for $s\ge 4$. 

   \begin{corollary}
      \label{cor:triangle-kst}
      Assuming \Cref{conj:bmm}, for all fixed integers $s\geq4$ and
      $t\geq(s-1)!+1$, any randomized combinatorial algorithm that solves
      Triangle Detection on $n$-vertex $K_{s,t}$-free graphs \whp requires
      $n^{3-3/s-o(1)}$ time.
   \end{corollary}

   When $st>3(s+t)$, \Cref{cor:triangle-lb} gives a lower bound
   of $n^{3-3/s-3/t-o(1)}$ for $K_{s,t}$-free graphs, so
   \Cref{cor:triangle-kst} improves the lower-bound exponent by $3/t$.

   For comparison, let us discuss the known upper bounds for Triangle Detection in $H$-free graphs, 
   starting with $H = K_{s,t}$. The classic KST theorem~\cite{kovari1954problem}
   implies that $K_{s,t}$-free graphs have at most
   $m=\Order(n^{2-1/s})$ edges and degeneracy
   $\Order(n^{1-1/s})$. The algorithm of Chiba and Nishizeki~\cite{chiba1985arboricity} 
   for bounded degeneracy graphs therefore solves
   Triangle Detection in $\Order(mn^{1-1/s})$ time, giving
   an upper bound of $\Order(n^{3-2/s})$. This leaves a gap of $1/s$ between the exponents of the lower and upper bounds. 

   Assuming \Cref{conj:bmm}, the dependence on the pattern size in
   \Cref{cor:triangle-lb} is tight up to constant factors in the exponent
   for an infinite family of dense patterns.
   To explain this, we use the notion of a degenerate coloring, introduced
   in~\cite{abboud2026triangle}.

   \begin{restatable}[Degenerate Coloring]{definition}{degeneratecoloring}
      \label{def:degenerate}
      A proper $3$-coloring of a graph $H$ 
      is a \emph{degenerate coloring} if every nonempty induced subgraph $J$ of
      $H$ which is not isomorphic to a triangle has a vertex whose neighborhood in $J$ is
      monochromatic. We consider an empty neighborhood to be monochromatic.
   \end{restatable}
   Such a graph has at most one triangle, since the vertices of two distinct
   triangles would induce a subgraph with bichromatic neighborhoods.

   Abboud et al.~\cite[Theorem 2]{abboud2026triangle} showed that if a
   $k$-vertex pattern $H$ admits a degenerate coloring, then Triangle Detection
   in $H$-free graphs can be solved by a randomized algorithm in
   $\tOrder(n^{3-2^{-(k-3)}})$ time when $H$ is triangle-free, and in
   $\tOrder(n^{3-2^{-(k-4)}})$ time otherwise. Thus, their saving over cubic
   time is $n^{\Order(2^{-k})}$. We improve the saving to $n^{\Omega(1/k)}$
   and make the algorithm deterministic.

   \begin{theorem}
      \label{thm:degenerate}
      Let $H$ be a $k$-vertex pattern that admits a degenerate coloring. There is a
      deterministic combinatorial algorithm for Triangle Detection in $H$-free graphs with
      running time
      \[
         \begin{cases}
            \tOrder(n^{3-\frac{1}{k-2}}) & \text{if $H$ is triangle-free and $k\geq 3$,}\\
            \tOrder(n^{3-\frac{1}{k-3}}) & \text{if $H$ contains a triangle and $k\geq 4$}.
         \end{cases}
      \]
   \end{theorem}

   A concrete family illustrates how this upper bound compares with
   \Cref{cor:triangle-lb}. Let $C_5^{(k)}$ be a balanced blowup of $C_5$: replace
   each vertex of $C_5$ by an independent set of size roughly $k/5$ to get a $k$-vertex graph,
   with the five sets as equal in size as possible, and replace each edge by a biclique. The graph $C_5^{(k)}$ is
   triangle-free, admits a degenerate coloring (since degenerate colorings of $C_5$ can be extended to its blowups), and has
   \[
      v(C_5^{(k)})=k
      \qquad\text{and}\qquad
      e(C_5^{(k)}) = k^2/5+O(k).
   \]
   For all sufficiently large $k$, we have $e(C_5^{(k)})>3v(C_5^{(k)})$. Consequently, the
   lower bound in \Cref{cor:triangle-lb} for such graphs is $n^{3-\Theta(1/k)}$,
   while \Cref{thm:degenerate} gives an upper bound of $n^{3-\Theta(1/k)}$. Thus the two
   bounds agree in their dependence on $k$, leaving a gap of only $n^{O(1/k)}$.

\paragraph{From induced $H$-freeness to bounded \VC-dimension.}
The preceding results use filtering to prove hardness under two different
structural promises. There is also an algorithmic connection between them. Our
key observation is that, for bipartite $H$, Triangle Detection in induced
$H$-free graphs can be reduced to instances whose relevant neighborhood set
systems have bounded \VC-dimension. Applying the bounded-\VC{} techniques
of~\cite{anand2026structural} then gives a truly subcubic algorithm for every
bipartite forbidden pattern.

\begin{theorem}
   \label{thm:subcubic-bipartite}
   Let $H$ be a bipartite graph on $k \geq 3$ vertices.
   There is a randomized algorithm for Triangle Detection in $n$-vertex
   induced $H$-free graphs that succeeds \whp and runs in time
   \[
      \tOrder\left(n^{3-\frac{1}{k+2\log_2 k-2}}\right).
   \]
\end{theorem}
For the special case of  the $k$-vertex path, denoted by $P_k$, we employ specialized arguments to get a faster algorithm.

   \begin{theorem}
      \label{thm:paths}
      For every integer $k\ge 4$, there is a randomized algorithm for Triangle Detection in induced $P_k$-free graphs that succeeds \whp and runs in time 
      \[
         \tOrder\left(n^{3-\frac{1}{\lfloor k/2\rfloor-1}}\right).
      \]
\end{theorem}

      \subsection{Technical Overview}
         \subsubsection{The Filter Technique}
   Our lower bounds for structured OMv and Triangle Detection are based on a common 
   filtering framework. We view Boolean $n\times n$ matrices 
   as  adjacency matrices of bipartite graphs with parts of size $n$. 
   Starting from an arbitrary
   instance $G$ (either the input graph in Triangle Detection, or the underlying graph of the input matrix in OMv), 
   we design a \emph{filter graph} $F$ on the same vertex set that satisfies the desired structural promise 
   (bounded \VC-dimension, bounded \VC-density, or $H$-freeness). 
   We then obtain \emph{filtered instances} from $G$ by randomly relabeling 
   the vertices of $F$ and retaining precisely the edges of $G$ that occur in the
   relabeled copy of $F$. In the matrix viewpoint, a filtered instance corresponds to $M \land \pi(F)$ (coordinatewise), where $\pi(F)$ is obtained by 
   randomly permuting the rows and columns of $F$. The key property that $F$ needs to satisfy is that subgraphs of $F$ inherit its structure. 

   We repeat this operation using independent relabelings to ensure that every solution of the original instance survives
   in at least one filtered instance. For OMv, the witnesses that must be 
   covered are the $1$-entries of the matrix; for Triangle Detection, they are triangles. 
   Once all witnesses are covered, the OR of the answers on the filtered instances
   recovers the answer on the original instance. More precisely, 
   from an arbitrary matrix $M \in \set{0,1}^{n \times n}$, we create $K$ filtered matrices
   $M_1,\ldots,M_K$ such that for every vector $v \in \{0,1\}^n$,
   \[
      Mv = \bigvee_{i=1}^K M_i v.
   \]
   With $K$ relabelings, an algorithm for structured inputs of running time $T_{\textup{str}}$ gives 
   a general algorithm of running time $\tOrder(K(n^2+T_{\textup{str}}))$. The quadratic terms corresponds to 
   preparing the filtered instances but since the output size for OMv is $n^2$, it is absorbed by $T_{\textup{str}}$. 
   Thus, we get a lower bound of 
   \[T_{\textup{str}} \geq  \tilde \Omega( T_{\textup{OMv}}/K ).\] 
   A filter containing more witnesses has a larger probability of retaining any fixed witness, so fewer
   relabelings are needed and the resulting lower bound is stronger.

   Let us now explain how to choose the filter $F$ for each problem.

   \paragraph{OMv in bounded \VC-dimension.}
      View $M$ as a bipartite graph. 
      Fix $D=d+1$. If $D$ columns are shattered, then $M$ contains a bipartite subgraph $P$ with $2^D$ left vertices and $D$ right vertices, in which every subset of the right vertex set is the neighborhood of exactly one left vertex.
      Notice that for every subgraph $H \subseteq P$,
      an $H$-free graph is also $P$-free and therefore its \VC-dimension is at most $D-1 = d$.
      We choose the filter $F$ to be a randomly constructed $H$-free graph for some specifically chosen pattern $H \subseteq P$.
      Which pattern should we choose? For $F \sim G(n,n,p)$, taking $p=o(n^{-v(H)/e(H)})$ makes the expected number of copies of $H$ tend to zero 
      and hence $H$-free with high probability. 
      In order to reduce the number of relabelings needed, we want $F$ to be as dense as possible. 
      Taking $p=n^{-v(H)/e(H)}$ gives $n^{2-v(H)/e(H)}$ expected edges, so we want $H \subseteq P$ to minimize the 
      ratio $v(H)/e(H)$.

      The natural choice for $H$ is a biclique since bicliques are the densest bipartite graphs. 
      For a biclique of dimension $a \times b$, we have \[v(H)/e(H) = (a+b)/ab = 1/a + 1/b.\]
      If we could set $a = D^2$ and
      $b = D$, the ratio would be $1/D + O(1/D^2) = 1/d + O(1/d^2)$.
      However, since $H$ is restricted to be a subgraph of $P$, and $P$ does not contain a $D^2 \times D$ biclique\footnote{In fact, the largest balanced biclique in $P$ has parts of size roughly $D-\log D$.}
      , 
      we choose an approximate construction. Instead of a $D^2 \times D$ biclique, we take the $\binom{D}{2}$ left vertices of degree $D-2$ 
      and the $D$ right vertices. This gives 
      \[
         \frac{v(H)}{e(H)} = \frac{\binom{D}{2} + D}{\binom{D}{2}(D-2)} = 1/(D-2) + O(1/D^2) = 1/d + O(1/d^2).
      \]
      Thus $\tOrder(n^{1/d+\Order(1/d^2)})$ relabelings suffice for the
      relative bound with this loss exponent.
      In the actual proof, we use a slightly better pattern that optimizes the lower-order terms, 
      previously used by Janzer and Pohoata~\cite[Section 3]{janzer2024zarankiewicz}.

      For $d=2$, we instead use the deterministic extremal $K_{2,2}$-free construction 
      of Brown and Erd\H{o}s--R\'enyi--S\'os~\cite{brown1966on,erdos1966problem}, with
      $\Omega(n^{3/2})$ edges. A $K_{2,2}$-free bipartite graph has \VC-dimension
      at most $2$, so $\tOrder(n^{1/2})$ relabelings suffice for \Cref{thm:omv-vc-2-lb}.

      Could more general filters reduce the loss exponent to $1/d$ for $d\geq3$?
      In \Cref{sec:general-filter-limitation}, we show that every general
      $d$-filter---a matrix of \VC-dimension $d$ whose \VC-dimension cannot
      increase when entries are zeroed out---has $o(n^{2-1/d})$ entries equal
      to $1$. A conjecture of Janzer and Pohoata~\cite{janzer2024zarankiewicz}
      would strengthen this to a polynomial loss, placing a corresponding
      barrier on the lower bounds obtainable by this filtering approach.

   \paragraph{OMv in bounded \VC-density.}
      We use the observation that $K_{D,d}$-free matrices have \VC-density at most $d$, where $D=(d-1)!+1$; see~\cite[Appendix C.1.1]{alon2016sign} and~\cite[Section 7]{fox2017semi}.
      For $d\geq2$, the projective norm graph construction of Alon, R\'onyai and Szab\'o~\cite{alon1999norm}
      gives a $K_{D,d}$-free bipartite filter with $\Theta(n^{2-1/d})$ edges,
      so $\tOrder(n^{1/d})$ relabelings suffice. For $d=1$, the bound follows
      from the output size and $T_{\mathrm{OMv}}=\Order(n^3)$.

   \paragraph{Triangle Detection in $H$-free graphs.}
      We sample an $H$-free graph $F\sim G(n,p)$ with 
      \[p=n^{-v(H)/e(H)}.\]
      In this setting, the witnesses are triangles. 
      By definition, a triangle of $G$ survives in a single filtered instance with probability 
      \[p^3 = n^{-3v(H)/e(H)}.\] 
      Hence, the number of relabelings needed to ensure, with high probability, that every triangle survives in some filtered instance is $\tilde \Order(1/p^3)$. 
      Thus, an algorithm of running time $T_{\textup{str}}$ for $H$-free graphs gives an algorithm of running time $\tilde\Order( 1/p^3 (n^2+T_{\textup{str}}) )$ for general graphs, giving the asserted lower bound in \Cref{thm:triangle-lb}.

      The extremal problem of maximizing the number of triangles in an
      $n$-vertex $H$-free graph is known as a Generalized Turán
      Problem~\cite{gerbner2026survey}. Improved constructions of triangle-dense
      $H$-free graphs yield stronger lower bounds whenever they run in $\tOrder(n^2)$ time. 
      For fixed $s\geq4$, Alon and Shikhelman~\cite{alon2016many} prove that 
      the projective norm graph underlying the construction in the
      preceding paragraph contains $\Omega(n^{3-3/s})$ triangles. Thus
      $\tOrder(n^{3/s})$ relabelings suffice for \Cref{thm:triangle-kst}
      when $t\geq(s-1)!+1$.

   \subsubsection{Shortcut Gadgets}
      We now explain how to extend the lower bounds from bounded \VC-dimension to bounded \VCB-dimension using shortcut gadgets. 
      \paragraph{Triangle Detection in bounded \VCB-dimension}
         We begin with a hard instance $G$ whose (ordinary) \VC-dimension is $\Order(d)$. Such instances exist by the following observation:
         Since $H$-free graphs have bounded \VC-dimension when $H$ is a biclique, we may apply the preceding result with, say, $H=K_{d/2,d/2}$.
         The challenge now is that the distance \VC-dimension can be
         arbitrarily larger than ordinary \VC-dimension. Our goal is to construct
         a triangle-preserving graph $T(G)$ with $\VCB(T(G))=\Order(d)$.
         To this end, we introduce shortcut gadgets that reduce the diameter without creating triangles. 
         These gadgets are highly structured and therefore keep the \VC-dimension at $\Order(d)$ while reducing the \VCB-dimension, 
         as large radius balls become the whole graph and hence do not contribute to the \VCB-dimension.
         By color-coding~\cite{alon1995color}, we may assume that $G$ is tripartite.
         One simple gadget adds, for each part in the tripartition of $G$, a hub adjacent to every vertex of that part.
         Each such vertex provides a shortcut between vertices in the same part, making the distance between every such pair at most $2$. Shortcuts between different parts require more care to avoid creating triangles.
         The main gadget, which we apply repeatedly, is the following gadget for bipartite graphs. Given a bipartite graph on vertex sets $X \sqcup Y$, add a copy $Q(X)$ of $X$, join $Q(X)$ to $X$ by a matching, and join $Q(X)$ to $Y$ by the complement $\overline{E(X,Y)}$ of the original edge set. This introduces no triangles and makes every pair $(x,y)\in X\times Y$ have distance at most $2$.
         Finally, another layer of shortcuts is used to control distances between newly added vertices (e.g., between pairs in $(X,Q(X))$).
         The final construction takes a tripartite graph $G$ and applies the bipartite gadget to each of the three bipartite subgraphs in cyclic 
         order, and then adds a constant number of hubs, each adjacent to every vertex of a specified independent set. In the analysis, we show that radius-$2$ balls centered at non-hub vertices vary only on the hubs, and radius-$3$ balls contain the whole graph. Since there are only a constant number of hubs, the \VCB-dimension is essentially determined by the radius-$1$ balls, whose intersections are controlled by ordinary \VC-dimension. 

      \paragraph{APSP in bounded \VCB-dimension}
         We use the standard reduction from Triangle Detection in a tripartite graph
         with parts $A,B,C$ to APSP: delete $E(B,C)$, compute all distances, and
         test whether some original edge $bc\in E(B,C)$ now has distance two.
         We apply the same construction as in the preceding paragraph and delete $E(B,C)$. 
         The point requiring care is that deleting these edges could increase the 
         \VCB-dimension, because distances between pairs in $B\times C$ and $Q(B)\times C$ may increase.
         Our bipartite shortcut gadget is designed to control this change:
         every nonedge between $B$ and $C$ already has distance two by the gadget, 
         so only pairs that correspond to edges of the original graph can move beyond distance two. In particular,
         the exceptional part of the set system of radius-$2$ balls is determined by 
         the original induced subgraph $G[B \sqcup C]$, which has bounded \VC-dimension. 

   \subsubsection{Triangle Detection in Induced \texorpdfstring{$H$}{H}-Free Graphs for Bipartite Patterns}
      Let $H$ be a fixed bipartite graph. We next outline a truly subcubic algorithm for Triangle Detection in induced $H$-free graphs.
      The main observation is that if $Q$ is bipartite and induced $H$-free, then its neighborhood set system has \VC-dimension bounded
      by a constant depending only on $H$.
      Indeed, a sufficiently large shattered set provides enough vertices realizing the prescribed neighborhoods of the vertices of $H$, 
      while bipartiteness guarantees that there are no edges within either part. 
      However, our worst-case input graph $G$ is arbitrary and not necessarily bipartite. 
      Nonetheless, it turns out we can at least assume that $G$ is properly $3$-colored (i.e., tripartite).
      For worst-case Triangle Detection this assumption occurs without loss of generality just by
      color-coding~\cite{alon1995color}, but this classic argument does not preserve induced $H$-freeness so cannot directly be applied in our setting. 
      Instead, we employ an alternative approach 
      of~\cite{abboud2026equivalent} to efficiently reduce Triangle Detection on the full graph $G$ to several instances of Triangle Detection on properly $3$-colored induced subgraphs of $G$, which preserves induced $H$-freeness.
      
      Now, let $A$, $B$, and $C$ denote the color classes of a fixed instance.
      We can focus on the bipartite graph between parts $A$ and $B$. 
      This bipartite graph is an induced subgraph of $G$ and is therefore induced $H$-free. 
      This implies that the adjacency matrix of this subgraph has bounded \VC-dimension. 
      We can now apply a standard reduction from Triangle Detection to OMv, where the matrix $M$ of the OMv instance is the bi-adjacency matrix for the graph between $A$ and $B$, and the vectors $v_i$ correspond to distinct vertices in $C$.
      This lets us reduce Triangle Detection in induced $H$-free graphs to OMv in matrices of bounded \VC-dimension, and obtain our new algorithms.

      For the special case of $P_k$-free graphs, we use an improved bound on the \VC-dimension and improved color coding based on the approach of~\cite{abboud2026equivalent}. Specifically, we employ vertex separators with constant chromatic number (usually referred to in the literature as ``Gy\'{a}rf\'{a}s path arguments'') to reduce the problem to properly $3$-colored induced subgraphs of the input graph $G$ more efficiently.

      \subsection{Related Work}
         Our filtering framework is related to several reductions that transfer
hardness from arbitrary instances to structured ones. We describe these
connections through a common viewpoint, and include two classical
antecedents: color coding and Valiant--Vazirani isolation.

\paragraph{Combining an input with a template.}
   Let $X$ be an arbitrary input and let $T$ be a structured \emph{template}. 
   A reduction constructs instances of the form
   \[
      X_i=\Phi_{i}(X,T),
   \]
   where $\Phi_{i}$ is a binary operation, possibly incorporating 
   intersection of edge or vertex sets, conjunction of 
   matrices or formulas, intersection with a translate of an integer set, etc. 
   The template $T$ and each $\Phi_i$ may be chosen randomly. 

   The template and operation are chosen to impose the desired structure 
   while preserving the answer to the original problem across the resulting 
   collection of instances. 
   In the simplest setting, every output satisfies the structural promise, no
   operation creates a false witness, and each original witness survives
   with a controlled probability. Repetition then allows the original answer
   to be recovered with high probability by taking the OR of the answers. 

   The examples below also illustrate variants of this approach. 

\paragraph{Color coding as filtering.}
   The color-coding method of Alon, Yuster, and
   Zwick~\cite{alon1995color} falls into our filtering framework. 
   For detecting a fixed $k$-vertex pattern $H$ as a non-induced subgraph, 
   they give a self-reduction to $k$-partite graphs by a random coloring.
   A fixed witness becomes colorful with probability $k!/k^k$ and hence survives.
   To obtain this result via filtering, take a complete $k$-partite filter $F$
   with $n/k$ vertices in each part, up to rounding.
   The witness survives whenever its vertices lie in distinct parts of a
   uniformly relabeled copy $\pi(F)$. The probability of this event is
   roughly $k!/k^k$ for fixed $k$.

\paragraph{Isolation as filtering.}
   The reduction of Valiant and Vazirani~\cite{valiant1985np} from SAT to
   Unique SAT admits a related interpretation, based on the lecture notes of Sudan~\cite{sudan2007lecture12}. A filter in this setting is a Boolean formula whose goal is to reduce the number of satisfying assignments  
   of a formula to at most one. The approximate number of satisfying assignments is ``guessed'', so we have a filter for each guess. 

   Starting from a formula $\varphi \colon \set{0,1}^n \to \set{0,1}$, 
   for each $1 \leq m \leq n$, we sample a pair $(A_m,b_m)$ independently and uniformly at random, 
   where 
   \[
      A_m \in \mathbb{F}_2^{m \times n}, \quad b_m \in \mathbb{F}_2^m.
   \]
   Define the Boolean filter $F_m \colon \set{0,1}^n \to \set{0,1}$ by
   $F_m(x)=1$ if and only if $A_m x+b_m=0$.
   The filtered instances are then 
   \[
      \varphi(x) \land F_1(x)\; , \; \varphi(x) \land F_2(x) \; , \; \ldots , \; \varphi(x) \land F_n(x).
   \]
   It is then possible to show, using pairwise independence, that at least one of the filtered instances 
   has a unique satisfying assignment with constant positive probability, assuming $\varphi$ is satisfiable. This is the core of the reduction to Unique SAT.

\paragraph{Behrend sets as a template.}
   Dudek et al.~\cite{dudek2020all} use translates of Behrend-type sets as
   templates and intersect them with the three color classes of tripartite
   3SUM instances. This proves that tripartite 3SUM remains hard on instances
   where each color class is free from three-term arithmetic progressions,
   and is subsequently used to prove hardness for various LDTs (e.g. AVERAGE).

   Jin and Xu~\cite{jin2023removing} reduce 3SUM to 3SUM in Sidon sets, also using Behrend-type sets as templates. 
   One notable distinction is that they apply them to hashed instances, giving an example of combining several restrictions 
   to obtain the final structural promise.

\paragraph{Regular hypergraphs through products with templates.}
   Fischer, K\"unnemann, Red\v{z}i\'{c}, and
   Stie{\ss}~\cite{fischer2025the} prove that Hyperclique Detection remains
   hard under strong regularity promises using a product construction, applied to 
   an arbitrary instance $G$ and a structured template $T$. This reduction is deterministic
   and outputs a single instance. The crux of their proof is defining the template 
   and the product in such a way that the output hypergraph inherits regularity properties 
   from $T$ while preserving witnesses of $G$. 

\paragraph{Extremal constructions as filters.}
   The use of $H$-free graphs as filters that enforce bounded \VC-dimension also appears in earlier work in extremal combinatorics.
   Alon, Moran, and Yehudayoff~\cite[Appendix~C.2]{alon2016sign} also use $H$-free 
   graphs as filters for Boolean matrices to prove sign-rank lower bounds. 
   Janzer and Pohoata~\cite[Section~3]{janzer2024zarankiewicz} use the same forbidden pattern 
   that we use in the proof of~\Cref{thm:omv-vc-lb} to give a lower bound for the
   Zarankiewicz Problem on graphs with bounded \VC-dimension.

\paragraph{Low depth circuits.}
    Many constructions of depth-three circuits for Boolean functions can be interpreted as instantiations of the filter framework \cite{Amano2023,GGM2023,GKPRT2026,optimal-maj-circuit}.
    These circuits are typically a single large disjunction of several CNF-formulas (conjunctions of variables and their negations). Intuitively, each CNF is a particular filtered instance that only accepts a \emph{subset} of the satisfying assignments of the function,
    and the overall disjunction corresponds to combining enough filtered instances together to cover \emph{all} satisfying assignments of the function.

   \section{Methodology}
      \label{sec:AI}
      \paragraph{Origins of the project.}
   The main questions in this paper arose from prior work on Triangle Detection in
   induced $H$-free graphs \cite{abboud2026triangle,abboud2026equivalent}. 
   We realized the connection between this problem and problems on instances with
   bounded \VC-dimension, obtained the results in \Cref{sec:induced}, and proved a lower bound of
   $n^{3-2/d-O(1/2^d)}$ for OMv in matrices of bounded \VC-dimension,
   without using LLMs.

\paragraph{LLM use.}
   The fourth author prompted GPT 5.6 Sol to optimize the argument of our original OMv lower bound mentioned above.
   The model
   proposed the forbidden pattern of Janzer and
   Pohoata~\cite{janzer2024zarankiewicz}, contributing to the final bound
   in \Cref{thm:omv-vc-lb}, and suggested combining Brown's
   deterministic construction of $K_{2,2}$-free graphs~\cite{brown1966on}
   with random relabelings for the case $d=2$. We then developed the
   general filtering framework for our lower bounds to unify these methods.
   The model also located existing arguments in the
   literature on the \VC-density of $K_{(d-1)!+1,d}$-free graphs.
   These arguments, combined with our framework, yielded the tight
   lower bound in \Cref{thm:shattering}.

   For the results on distance \VC-dimension, the first and fourth authors supplied the idea of adding
   triangle-preserving shortcuts to a graph of bounded \VC-dimension to bound its distance
   \VC-dimension.
   These authors specified the desired properties for gadgets in the reduction, and after multiple rounds of interaction with the model recovered the constructions used for the proofs in \Cref{sec:distance-vc}.
   The improvement in \Cref{thm:degenerate} to the main algorithm
   of~\cite{abboud2026triangle} was also obtained by the fourth author prompting GPT 5.6 Sol. 

   The fourth author also used Codex with GPT6 Astra
   to fill in outlines for initial 
   sketches of theorem statements and proofs throughout the paper, and following substantial revision employed the tool several times to correct writing mistakes and simplify proofs, after which all the writing was thoroughly edited again without using LLMs.

   \section{Preliminaries}
      \label{sec:preliminaries}
      \paragraph{Arithmetic. }
For  integers $k\ge 1$, let $[k]=\{1,\ldots,k\}$. 
We use $\sqcup$ for unions of pairwise disjoint sets.
By default, $\log$ denotes the base two logarithm.

We  use the following upper bound for sums of binomial coefficients:

    \begin{proposition}[Binomial Asymptotics]
        \label{binomial-sum}
        For any integers $k,D\ge 1$, we have 
            \[\sum_{i=0}^D \binom{k}{i} \le (ek/D)^D.\]
    \end{proposition}

For a proof, see \cite[Proposition 1.4]{Jukna2011}.

\paragraph{Matrices. }
For a Boolean matrix $M$, let $\wt(M)$ denote the number of entries in $M$ equal to $1$. 
For Boolean matrices $M$ and $M'$ with the same dimensions, let $M\land M'$
and $M\lor M'$ denote their coordinatewise AND and OR, respectively.
We write $M'\leq M$ if the inequality holds coordinatewise.
Equivalently, $M'$ can be obtained from $M$
by zeroing out some of its entries. 
We write $J_{s,t}$ for the
$s\times t$ matrix with all entries equal to $1$. 
We say
that a matrix is $J_{s,t}$-free if it does
not contain $J_{s,t}$ as a submatrix.

For an $n\times n$ matrix $M$ and a pair $\pi=(\pi_{\text{row}},\pi_{\text{col}})$
of permutations of $[n]$, let $\pi(M)$ denote the
matrix obtained by permuting the rows of $M$ by $\pi_{\text{row}}$ and its
columns by $\pi_{\text{col}}$, so that
\[
   \pi(M)[i,j]=M[\pi_{\text{row}}^{-1}(i),\pi_{\text{col}}^{-1}(j)].
\]
We call such a pair $\pi$ a \emph{relabeling}. If $\pi_{\text{row}}$ and
$\pi_{\text{col}}$ are independent, uniform random permutations of $[n]$, 
we call $\pi$ a \emph{uniformly random relabeling}.

\paragraph{Graphs. }
For a graph $H$, let $v(H)=|V(H)|$ and $e(H)=|E(H)|$ denote its numbers of
vertices and edges respectively. For vertex sets $X,Y\subseteq V(H)$, let
$E(X,Y)$ denote the set of edges with one endpoint in $X$ and the other in
$Y$. 

For a bipartite graph $G$ with ordered left and right parts
$L=\{\ell_1,\ldots,\ell_n\}$ and $R=\{r_1,\ldots,r_n\}$ respectively,
its \emph{bipartite adjacency matrix} is the binary matrix
$M\in\set{0,1}^{n\times n}$ with $M[i,j]=1$ if and only if
$\{\ell_i,r_j\}\in E(G)$. Conversely, every binary matrix defines a
bipartite graph in this way. Under this correspondence, $\wt(M)=e(G)$, 
and zeroing out $1$-entries corresponds to deleting edges.

For a graph $G=(V,E)$, a vertex $v\in V$, and an integer $r\geq 0$, let
\[
   N_r^G[v]=\set{u\in V\mid \operatorname{dist}_G(v,u)\leq r}
\]
denote the closed ball of radius $r$ centered at $v$.

We write $N^G[v]=N_1^G[v]$ for the closed neighborhood of $v$ and
$N^G(v)=\set{u\in V\mid uv\in E}$ for its open neighborhood.
We omit the superscript $G$ when the graph is clear from context. 
In
particular, $N(v)=N[v]\setminus\set{v}$. For $S\subseteq V$, let
\[
   N(S)=\bigcup_{v\in S}N(v)
   \qquad\text{and}\qquad
   N[S]=\bigcup_{v\in S}N[v].
\]

\begin{definition}[Erd\H{o}s--R\'enyi Random Graphs]
   For a positive integer $n$ and $p\in[0,1]$, $G(n,p)$ is the distribution
   on graphs with a fixed vertex set of size $n$ in which every unordered pair
   of distinct vertices is included as an edge independently with probability $p$.

   Similarly, $G(n,n,p)$ is the distribution on bipartite graphs with two
   fixed parts of size $n$ in which every pair with one vertex in each part is
   included as an edge independently with probability $p$.
\end{definition}

      \begin{definition}[\VC-dimension]
         \label{def:vc}
         The \VC-dimension of a set system $\mathcal{F}$ is the cardinality of the largest
          set it shatters. 
          A set $A$ is shattered by $\mathcal{F}$ if $\set{A \cap S \mid S \in \mathcal{F}}$
         contains all $2^{|A|}$ subsets of $A$.
         The \VC-dimension of a binary matrix is defined as the \VC-dimension of a set-system whose characteristic vectors are the matrix's rows.
         The \VC-dimension of a graph $G$ is the \VC-dimension of its open-neighborhood 
         set system $\set{N^G(v)\mid v\in V(G)}$.
      \end{definition}
      Unless otherwise stated, we define the \VC-dimension of a bipartite graph $G$ with a given bipartition into left and right parts as the \VC-dimension of the left vertices' neighborhoods. Equivalently, this is the \VC-dimension of the bipartite adjacency matrix of $G$. 
      For general graphs, \Cref{def:vc}  uses the standard definition in terms of their open-neighborhood set system. 

The following standard amplification follows simply by running multiple
independent copies, capping their running times, and taking their majority answer. 
\begin{lemma}[Error Reduction]
   \label{lem:standard-amplification}
   Let $\mathcal A$ be a randomized algorithm for a decision problem that
   runs in at most $T(n)$ time and answers correctly \whp
   Then for any constant $\gamma>0$, $\mathcal A$ can be amplified so that
   with probability at least $1-n^{-\gamma}$, it answers correctly in $\Order(T(n)\log n)$ time. 
\end{lemma}

\begin{remark} 
    \label{remark:OMv-amplification}
   The same amplification principle applies to OMv as well. We construct
   independent copies of the data structure and cap the running time of each
   construction. Given a query vector $v$, we return the coordinatewise majority
   of their answers.
\end{remark}

\begin{lemma}
   \label{lem:triangle-size-reduction}
   Fix a constant $c\geq1$. For integers $n\geq3$ and $N\leq cn$,
   Triangle Detection on $N$ vertices reduces in $\Order(n^2)$ time
   to $\Order(1)$ instances of Triangle Detection on at most $n$ vertices.
\end{lemma}
\begin{proof}
   Partition the $N$ vertices into $\Order(1)$ blocks of size at most
   $\lfloor n/3\rfloor$. For each union of at most three blocks, take the
   induced subgraph. 
   Every triangle of the input lies in one such union, and every triangle
   of a subgraph lies in the input. There are $\Order(1)$ unions, 
   and each subgraph can be constructed in $\Order(n^2)$ time.
\end{proof}

With constant-factor amplification, an algorithm for Triangle Detection
on $n$ vertices that succeeds \whp in time $R(n)$ therefore gives an
algorithm on $N$ vertices that succeeds \whp in time $\Order(n^2+R(n))$.

   \section{Lower Bounds for OMv in Matrices of Bounded \VC-Dimension}
      \label{sec:omv-lb}
      In this section, we prove \Cref{thm:omv-vc-lb} and \Cref{thm:omv-vc-2-lb}.
We first establish a reduction for general matrix families, which we will
also use for bounded \VC-density in \Cref{sec:density}.

\begin{lemma}[Filtering Matrices]
   \label{lem:filter-permutation}
   Let $M,F\in\set{0,1}^{n\times n}$, and 
   let $\pi_1,\ldots,\pi_K$ be independent uniformly random relabelings of $F$. 
   Then, 
   \[
      \Pr\left[\bigvee_{k=1}^K\bigl(M\land\pi_k(F)\bigr)\neq M\right]
      \leq n^2\exp\left(-\frac{K\wt(F)}{n^2}\right).
   \]
\end{lemma}
\begin{proof}
   Observe that $0$-entries of $M$ are preserved in $\bigvee_{k=1}^K\bigl(M\land\pi_k(F)\bigr)$ since $M\land\pi_k(F) \leq M$. 
   For the $1$-entries, if $M[i,j]=1$ and $\pi = (\pi_{\text{row}},\pi_{\text{col}})$ is a uniformly random relabeling 
   of $F$, then 
   \[
      (M\land\pi(F))[i,j]=\pi(F)[i,j]=F[\pi_{\text{row}}^{-1}(i),\pi_{\text{col}}^{-1}(j)].
   \]
   Since the two preimages are independent, uniformly random indices in $[n]$,
   this entry is $1$ with probability $\wt(F)/n^2$. Therefore, the probability that the
   $(i,j)$-entry is $0$ in all $K$ filtered matrices is at most
   \[
      (1-\wt(F)/n^2)^K\leq\exp(-K\cdot\wt(F)/n^2).
   \]
   A union bound over the at most $n^2$ entries of $M$ equal to $1$ proves the claimed bound. 
\end{proof}

\begin{lemma}[OMv Lower Bounds From Filters]
   \label{lem:omv-filter-lb}
   Let $\mathcal M$ be a family of square Boolean matrices closed under
   row and column permutations. Suppose that, for constants $\beta>0$ and
   $0\leq\rho<1$, and every sufficiently large integer $n$, there is a randomized
   $\tOrder(n^2)$-time algorithm that outputs an $n\times n$ Boolean matrix
   $F$ satisfying
   \[
      F'\in\mathcal M \quad\text{for every Boolean }F'\leq F
      \qquad\text{and}\qquad
      \wt(F)\geq\beta n^{2-\rho}
   \]
   with probability at least $9/10$.
   If OMv in $n\times n$ matrices from $\mathcal M$ can be solved \whp in
   time $R(n)$, then unrestricted OMv can be solved \whp in time
   $\tOrder(n^\rho(n^2+R(n)))$. 
\end{lemma}
\begin{proof}
   Set
   \[
      K=\left\lceil\frac{6}{\beta}n^\rho\ln n\right\rceil.
   \]
   Thus, $K=\tilde{\Theta}(n^\rho)$.

   Let $\mathcal A$ solve OMv in $n\times n$ matrices from $\mathcal M$
   \whp in time $R(n)$.
   By~\Cref{remark:OMv-amplification}, with just an $\Order(\log n)$ multiplicative
   overhead, we can amplify each invocation so that its failure probability is at
   most $1/(10K)$ over the entire online sequence.

   Consider an arbitrary OMv instance with matrix $M$. Construct $F$ as in
   the hypothesis, sample independent uniformly random relabelings $\pi_1,\ldots,\pi_K$ of $F$ as in
   \Cref{lem:filter-permutation}, and set
   \[
      M_j=M\land\pi_j(F)\qquad(j\in[K]).
   \]
   Observe that if the filter construction succeeds and $F$ satisfies the guarantees of \Cref{lem:omv-filter-lb}, each $M_j \in \mathcal M$. 
   That is because $M_j \leq \pi_j(F)$, therefore $\pi_j^{-1}(M_j) \leq F$, and hence $M_j \in \mathcal M$ by the first property of $F$ and that $\mathcal M$ is closed under relabelings. 

   Run an amplified invocation of $\mathcal A$ on each $M_j$, capping each
   copy at the assumed $\Order(R(n) \log n)$ runtime bound with amplification overhead. 
   This is necessary because if the 
   filter construction fails (which happens with probability at most $1/10$),
   some $M_j$ may lie outside $\mathcal M$, in which case the runtime 
   guarantee for $\mathcal A$ might not hold.

   By assumption, with probability at least $9/10$, the filter construction succeeds 
   and we have $\wt(F)\geq\beta n^{2-\rho}$ and $M_j\in\mathcal M$ for every $j$. 
   Now condition on a fixed $F$ satisfying these properties.
   By \Cref{lem:filter-permutation},
   \[
      \Pr\left[\bigvee_{j=1}^K M_j\neq M\right]
      \leq n^2\exp\left(-\frac{K\wt(F)}{n^2}\right)
      \leq n^2e^{-6\ln n}
      =n^{-4}\leq1/10.
   \]

   Pass each arriving vector $v$ to all $K$ invocations and return the
   coordinatewise OR of their answers before receiving the next vector.
   Whenever $M=\bigvee_{j=1}^K M_j$, we have
   \[
      Mv=\bigvee_{j=1}^K M_jv.
   \]
   The filter construction fails with probability at most $1/10$. Conditioned
   on its success, the relabelings fail to give $M=\bigvee_{j=1}^K M_j$ with probability
   at most $1/10$, and at least one invocation of $\mathcal A$ fails with
   probability at most $K/(10K)=1/10$.
   Thus we succeed with probability at least $7/10\geq2/3$ by a union bound over these three cases.

   The total time for constructing $F$, forming the filtered matrices,
   processing their outputs, and running the amplified algorithms is
   \[
      \tOrder\bigl(K(n^2+R(n))\bigr)
      = \tOrder\bigl(n^\rho(n^2+R(n))\bigr).
   \]
\end{proof}

Let us now describe the general filter for \VC-dimension $d \geq 3$, presented in the language of 
bipartite graphs. Fix $d\geq3$, set $D=d+1$, and let $H_d$ be the bipartite graph with a right part $R$ of 
size $D$ and one left vertex $x_S$ for each subset $S\subseteq R$ with
$|S|\geq D-3$, where $N(x_S)=S$. Thus, the family of the left vertices' neighborhoods is the
Hamming ball of radius $3$ around the all-ones vector (where we associate subsets of $R$ with their characteristic vectors). 

The numbers of left
vertices and edges of $H_d$ are
\begin{equation}
    \label{eq:s-w-def}
   s=\sum_{i=0}^3\binom{D}{i}
   \qquad\text{and}\qquad
   w=\sum_{i=0}^3(D-i)\binom{D}{i},
\end{equation}
respectively. As observed in ~\cite{janzer2024zarankiewicz} and shown below, exclusion of this pattern from a bipartite graph implies bounded \VC-dimension. 
\begin{lemma}
   \label{lem:hd-free-vc}
   If a bipartite graph is $H_d$-free, then its \VC-dimension is at
   most $D-1=d$.
\end{lemma}
\begin{proof}
   Suppose that a set $X$ of $D$ right vertices is shattered. For every
   $S\subseteq X$ with $|S|\geq D-3$, choose a left vertex $x_S$ such that
   $N(x_S)\cap X=S$. These vertices are distinct and, together with $X$,
   contain a copy of $H_d$, a contradiction.
\end{proof}

 Now, our goal is to construct a dense $H_d$-free bipartite graph $F_d$ with $n$ vertices in each part (to ultimately use as a filter). 
 The standard 
alteration construction (see~\cite[Section 2.5]{furedi2013history} or 
~\cite[Theorem 1.9.1]{zhao2023graph}) would sample a graph from $G(n,n,p)$ and then 
delete an edge from every copy of $H_d$ to make it $H_d$-free. Although this construction
yields a slightly denser graph, it is unsuitable for our reduction because we do not know how to efficiently find all edges
contained in copies of $H_d$.
We therefore use the following sparser but oblivious construction, which samples the filter without needing any post-processing. 

\begin{lemma}[Filter Construction for Dimension $d\ge 3$]
   \label{lem:random-hd-free-filter}
   Fix $d\ge3$, let $D=d+1$, and let
   \[
      \alpha=\frac{D+s}{w}.
   \]
   There is a constant $\beta=\beta(d)>0$ and a randomized construction that,
   given a sufficiently large integer $n$, runs in $\Order(n^2)$ time and
   outputs a bipartite graph $F_d$ with $n$ vertices in each part such that,
   with probability at least $9/10$, the graph $F_d$ is $H_d$-free and
   \[
      e(F_d)\geq\beta n^{2-\alpha}.
   \]
\end{lemma}
\begin{proof}
   Choose $c>0$ so that $c^w\leq1/11$, and set
   \[
      p=cn^{-\alpha}
      \qquad\text{and}\qquad
      F_d\sim G(n,n,p).
   \]
   We show that, with probability at least $9/10$, the graph $F_d$ is
   $H_d$-free and $e(F_d)\geq pn^2/2$.

   If $X$ is the number of labeled copies of $H_d$ in $F_d$, then by linearity
   \[
      \mathbb{E}[X]\leq n^{D+s}p^w
      =c^w n^{D+s-w\frac{D+s}{w}}
      =c^w\leq1/11. 
   \]
   Markov's inequality then gives $\Pr[X\geq1]\leq1/11$.

   Since $d\geq3$, the 
   definitions of $s$ and $w$ from \cref{eq:s-w-def} imply that
    $\alpha\in (0,1)$, so $pn^2=cn^{2-\alpha}$ is unbounded as a function of $n$.
   Since $\mathbb{E}[e(F_d)]=pn^2$, a Chernoff bound gives
   \[
      \Pr[e(F_d)<pn^2/2]\leq\exp(-pn^2/8)=o(1).
   \]
   Taking a union bound over the two failure events gives success probability
   at least \[10/11-o(1)\geq9/10\] 
   for all sufficiently large $n$.

    From our choice of $p=cn^{-\alpha}$,
    we thus get that $e(F_d) \ge \beta n^{2-\alpha}$ with the desired probability if we set $\beta = c/2$.
   Sampling $F_d$ in this way takes $\Order(n^2)$ time, which proves the desired result.
\end{proof}

    Given the efficient filter construction from \Cref{lem:random-hd-free-filter}, we can now substitute it into our framework to prove our lower bound for OMv with bounded \VC-dimension.

\begin{proof}[Proof of \Cref{thm:omv-vc-lb}]
   Fix $d\ge3$. For the graph $H_d$ defined before \cref{eq:s-w-def}, we have
   \[
      \alpha
      =\frac{D+s}{w}
      =\frac{D^3+11D+6}{D^2(D^2-3D+8)}
      =\frac{d^3+3d^2+14d+18}{(d+1)^2(d^2-d+6)}
      =\frac{1}{d}+\Order(1/d^2).
   \]
   Moreover, $0<\alpha<1$ since $d\geq3$. Let $F$ be the $n\times n$
   bipartite adjacency matrix of the graph $F_d$ from
   \Cref{lem:random-hd-free-filter}. The matrix $F$ can be constructed in
   $\Order(n^2)$ time, and with probability at least $9/10$, the graph $F_d$
   is $H_d$-free and
   \[
      \wt(F)=e(F_d)\geq\beta n^{2-\alpha}
   \]
   for constant $\beta$.
   On this event, every binary matrix $F'\leq F$ is the bipartite adjacency
   matrix of an $H_d$-free subgraph of $F_d$. Thus, $\VC(F') \leq d$
   by \Cref{lem:hd-free-vc}.
   The construction satisfies the requirements of
   \Cref{lem:omv-filter-lb} with $\rho=\alpha$ and $\mathcal M$ the family
   of square Boolean matrices of \VC-dimension at most $d$, which is
   closed under row and column permutations. 

   Now suppose that OMv in matrices of \VC-dimension at most $d$ is solvable \whp in $R(n)$ time. 
   By \Cref{lem:omv-filter-lb}, unrestricted OMv is solvable \whp in $\tOrder( n^\alpha(n^2 + R(n)) )$ time.  
   The $n^2$ term is absorbed by $R(n)$ because the output size of OMv is $n^2$ and therefore $R(n) \geq n^2$. 
   By the definition of $T_{\mathrm{OMv}}$, this gives $R(n)={\tilde\Omega}(T_{\mathrm{OMv}}/n^\alpha )$.
\end{proof}

We next define the filter used for $d=2$. In this case, an extremal $K_{2,2}$-free graph suffices, 
since every bipartite $K_{2,2}$-free graph has \VC-dimension at most $2$. We use the classical 
construction of Brown and Erdős--Rényi--Sós~\cite{brown1966on,erdos1966problem}. 
\begin{lemma}[Filter Construction for Dimension 2]
   \label{lem:explicit-c4-filter} 
   For every sufficiently large $n$, there is a bipartite $K_{2,2}$-free graph $F_2$ with $n$
   vertices in each part and
   \[
      e(F_2)=\Omega(n^{3/2}).
   \]
   Moreover, $F_2$ can be constructed deterministically in $\Order(n^{3/2})$ time,
   linear in the size of its adjacency-list representation.
\end{lemma}
\begin{proof}
   Let $p$ be the largest prime with $p^2+p\leq n$. By Bertrand's postulate ~\cite[Chapter 2]{aigner1999proofs}, we have 
   $\sqrt{n}/2 \leq p \leq \sqrt{n}$, so we can find such a prime $p$ by the sieve of
   Eratosthenes in $\tOrder(\sqrt n)$ time.

   Take the points $\mathcal P=\mathbb F_p^2$, and let $\mathcal L$ consist of
   the lines $y=ax+b$ for $a,b\in\mathbb F_p$ and the vertical lines $x=c$ for
   $c\in\mathbb F_p$. Let $F_2$ be their incidence graph.
   Since two distinct points lie on a unique affine line, $F_2$ is
   $K_{2,2}$-free. Moreover,
   \[
      |\mathcal P|=p^2,
      \qquad
      |\mathcal L|=p^2+p,
      \qquad\text{and}\qquad
      e(F_2)=p(p^2+p)=\Omega(n^{3/2}).
   \]
   Pad both parts to size $n$ with isolated vertices. Finally, enumerating the
   $p$ points on every line constructs all incidences in
   $\Order(p^3+p^2)=\Order(n^{3/2})$ time.
\end{proof}

\begin{lemma}
   \label{lem:c4-free-vc}
   If a bipartite graph is $K_{2,2}$-free, then its \VC-dimension is at most $2$.
\end{lemma}
\begin{proof}
   Suppose to the contrary the graph has \VC-dimension at least three. 
   Then it has a set of three right vertices $x,y,z$ that are shattered. The distinct left vertices
   realizing the restricted neighborhoods $\{x,y\}$ and $\{x,y,z\}$ have the two common
   neighbors $x$ and $y$, creating a $K_{2,2}$, a contradiction.
\end{proof}

Combining these lemmas proves the lower bound for OMv for \VC-dimension at most two.

\begin{proof}[Proof of \Cref{thm:omv-vc-2-lb}]
   Let $F_2$ be the graph from \Cref{lem:explicit-c4-filter}, and let $F$ be
   its $n\times n$ bipartite adjacency matrix. Choose a constant $\beta>0$ so that
   \[
      \wt(F)=e(F_2)\geq\beta n^{3/2}.
   \]
   Every binary matrix $F'\leq F$ is the bipartite adjacency matrix of a
   $K_{2,2}$-free subgraph of $F_2$. Thus, $\VC(F')\leq 2$
   by \Cref{lem:c4-free-vc}.
   Since $F$ can be constructed deterministically in $\Order(n^2)$ time,
   the construction satisfies the requirements of \Cref{lem:omv-filter-lb}
   with $\rho=1/2$ and $\mathcal M$ the family of square Boolean matrices
   of \VC-dimension at most $2$, proving the theorem.
\end{proof}

   \subsection{Optimizing Lower-Order Terms for \texorpdfstring{$3 \leq d \leq 8$}{d between 3 and 8}}

      For $3\leq d\leq8$, the exponent $\alpha$ in \Cref{thm:omv-vc-lb} can be
      reduced slightly by using a smaller pattern. Let $D=d+1$ and let
      $H_d^{(2)}$ have a right part $R$ of size $D$ and one left vertex $x_S$
      for every $S\subseteq R$ with $|S|\geq D-2$, where $N(x_S)=S$.  Thus,
      the left neighborhoods form the Hamming ball of radius $2$ around the
      all-$1$ vector.  Writing $s_2$ and $w_2$ for the numbers of left vertices
      and edges, respectively, we have
      \[
         s_2=\sum_{i=0}^2\binom{D}{i}
         \qquad\text{and}\qquad
         w_2=\sum_{i=0}^2(D-i)\binom{D}{i}.
      \]
      The proofs of \Cref{lem:hd-free-vc,lem:random-hd-free-filter} apply with
      $H_d^{(2)}$ in place of $H_d$. We can therefore apply
      \Cref{lem:omv-filter-lb} to matrices of \VC-dimension at most $d$
      with $\rho=\alpha_2$, where
      \[
         \alpha_2=\frac{D+s_2}{w_2}
         =\frac{(D+1)(D+2)}{D(D^2-D+2)}
         =\frac{(d+2)(d+3)}{(d+1)(d^2+d+2)}.
      \]
      Consequently, for every $3\leq d\leq8$, the lower bound in
      \Cref{thm:omv-vc-lb} holds with $\alpha_2$ in place of $\alpha$.
      For example, for $d=3$, $\alpha_2=\frac{15}{28} < \alpha=\frac{19}{32}$. 

      For $d=4$, a stronger bound follows from Brown's
      construction~\cite{brown1966on}, which yields a bipartite $K_{3,3}$-free
      graph $F_4$ with $n$ vertices in each part and $\Omega(n^{5/3})$ edges
      for every sufficiently large $n$. This graph can be constructed
      deterministically in $\Order(n^2)$ time. Every subgraph of $F_4$ is
      $K_{3,3}$-free and hence has \VC-dimension at most $4$, so the bipartite
      adjacency matrix of $F_4$ satisfies the requirements of
      \Cref{lem:omv-filter-lb} for matrices of \VC-dimension at most $4$
      with $\rho=1/3$. Consequently, any algorithm that solves OMv in
      $n\times n$ matrices of \VC-dimension at most $4$ \whp requires
      at least $\tilde{\Omega}(T_{\mathrm{OMv}}/n^{1/3})$ time.

   \subsection{A Limitation of General Filters}
      \label{sec:general-filter-limitation}
      Our lower bounds use $H$-free bipartite graphs as filters. For $d\geq3$,
      the filters used to prove \Cref{thm:omv-vc-lb} give an overhead of
      $\tOrder(n^{1/d+\Order(1/d^2)})$. Could denser filters bring the
      overhead down to $\tOrder(n^{1/d})$?

      A priori, it is not clear why one should restrict attention to filters
      arising from $H$-free bipartite graphs. More generally, consider a
      matrix $F$ satisfying the following conditions:
      \begin{enumerate}
         \item $F \in \set{0,1}^{n \times n}$.
         \item $\VC(F)=d$.
         \item $\VC(F')\leq d$ for every $F'\leq F$.
      \end{enumerate}
      Note that condition 3 is not implied by condition 2 above, since in general a matrix $F'\leq F$
      can have larger \VC-dimension than $F$. 
      For example, this can occur when
      $F$ is the all-ones matrix. The next proposition bounds the density
      of every such filter, and hence the efficiency of our reduction.
      \begin{proposition}[Limitations for Filters]
         \label{prop:general-filter-limitation} 
         For every $d\geq3$, every $F$ that satisfies conditions 1-3 above has 
         $\wt(F)=o(n^{2-1/d})$.
      \end{proposition}
      \begin{proof}
         We claim that condition 3 implies $F$ is
         $J_{2^{d+1},d+1}$-free. 
         Indeed, suppose not.
         Then we could choose $F'\leq F$
         whose restriction to the corresponding $d+1$ columns contains all
         $2^{d+1}$ binary row patterns. This would imply
         $\VC(F')\geq d+1$, a contradiction.

         A result of Janzer and Pohoata~\cite{janzer2024zarankiewicz} states
         that if $\VC(F)=d$ and $F$ is
         $J_{2^{d+1},2^{d+1}}$-free, then
         $\wt(F)=o(n^{2-1/d})$. Since $J_{2^{d+1},d+1}$-freeness implies
         $J_{2^{d+1},2^{d+1}}$-freeness, the result applies to $F$.
      \end{proof}
      Covering all $n^2$ positions requires at least $n^2/\wt(F)$ relabelings,
      since each relabeling covers at most $\wt(F)$ positions.
      The conjecture of Janzer and Pohoata~\cite[Conjecture 1]{janzer2024zarankiewicz}
      would give $\wt(F)=\Order(n^{2-1/d-c_d})$ for some $c_d>0$,
      forcing $\rho\geq1/d+c_d$ in \Cref{lem:omv-filter-lb}.

   \section{Lower Bounds for OMv in Matrices of Bounded \VC-Density}
      \label{sec:density}
      In this section, we prove \Cref{thm:shattering}. We begin with a bound on the
shatter function of biclique-free matrices.

\begin{lemma}
   \label{lem:bounded-shatter-function}
   Let $d\geq1$ be an integer and set $D=(d-1)!+1$.
   If a matrix $M\in\set{0,1}^{n\times n}$ is $J_{D,d}$-free, then
   $\pi_M(t)\leq 2Dt^d$ for every $1\leq t\leq n$.
   In particular, the family of all square $J_{D,d}$-free Boolean matrices
   has \VC-density at most $d$.
\end{lemma}

This result follows from~\cite[Appendix C.1.1]{alon2016sign} and the discussion in~\cite[Section 7]{fox2017semi}. We include a short proof below for completeness.
\begin{proof}
   For $1\leq t < d$, we have   
   \[\pi_M(t) \leq 2^t\leq 2Dt^d\] 
   by definition.
   We show that in faat 
   \[\pi_M(t) < D \binom{t}{d} + \sum_{i = 0}^{d-1}\binom{t}{i}\] for every $d \leq t\leq n$.
   
   Suppose for contradiction that 
   \[\pi_M(t) \geq D \binom{t}{d} + \sum_{i = 0}^{d-1}\binom{t}{i}\] 
   for some $d\leq t\leq n$.
   Fix a set of $t$ columns realizing $\pi_M(t)$ distinct row patterns.
   There are at most $\sum_{i = 0}^{d-1}\binom{t}{i}$ patterns of weight at most $d-1$. Hence, there are at least 
   \[\pi_M(t) - \sum_{i=0}^{d-1} \binom{t}{i} \geq D \binom{t}{d}\]
   distinct row patterns of weight at least $d$.
   For each such pattern, fix a subset of $d$ columns in which all entries are equal to $1$.
   Since there are $\binom{t}{d}$ possible subsets and at least
   $D \binom{t}{d}$ such patterns, some subset of $d$ columns has only $1$-entries in at least $D$ distinct row patterns.
   This contradicts the assumption that $M$ is $J_{D,d}$-free.

   The claimed bound follows from
   \[
      D\binom{t}{d}+\sum_{i=0}^{d-1}\binom{t}{i}
      \leq Dt^d+dt^{d-1}
      \leq (D+1)t^d
      \leq 2Dt^d.
   \]
\end{proof}

We now shift our attention to graphs. We use the \emph{projective norm graph} to construct our filter. 

\begin{lemma}[Projective Norm Graph Filter]
   \label{lem:norm}
   Let $d>1$ be an integer and set $D=(d-1)!+1$.
   For every sufficiently large $n$, there exists a $K_{D,d}$-free bipartite
   graph $Q_d$ with $n$ vertices in each part and $\Theta(n^{2-1/d})$ edges.
   Moreover, $Q_d$ can be constructed explicitly in $\tOrder(n^{2-1/d})$ time.
\end{lemma}
\begin{proof}
   Let $p$ be a prime to be chosen later.
   Let $\ProjNorm_d$ denote the projective norm graph constructed in~\cite{alon1999norm}.
   This graph has vertex set $\Field_{p^{d-1}} \times \Field^{\times}_{p}$, where $\Field_{p^{d-1}}$ is the finite field with $p^{d-1}$ elements and $\Field^{\times}_{p}$ is the multiplicative group of nonzero elements of $\Field_p$.
   Two vertices $(X,x), (Y,y) \in \Field_{p^{d-1}} \times \Field^{\times}_{p}$ are adjacent if and only if $N(X + Y) = xy$, where
   \[N(Z) = Z\cdot Z^p\cdot Z^{p^2}\cdots Z^{p^{d-2}} = Z^{(p^{d-1} - 1)/(p-1)}\]
   is the norm of $Z \in \Field_{p^{d-1}}$ over $\Field_{p}$.
   It is a standard fact that the norm function $N(\cdot)$ maps elements of $\Field_{p^{d-1}}$ to $\Field_p$ \cite[Proposition 11.2.2]{IrelandRosen-NTbook}.
   For each vertex $(X,x)$ and every $Y\in\Field_{p^{d-1}}$ with $Y\neq-X$, there is a unique 
   \[y = \frac{N(X+Y)}{x}\] 
   such that $(X,x)$ is adjacent to $(Y,y)$.
   If $Y=-X$, then $(X,x)$ is not adjacent to $(Y,y)$ for any choice of $y$, since in this case $N(X+Y) = 0$ cannot be the product of two elements of $\Field_p^\times$.

   Thus, every vertex in $\ProjNorm_d$ has degree $p^{d - 1} - 1$. Since $\ProjNorm_d$ has $p^{d - 1}(p-1)$ vertices, this implies the graph has $\frac{1}{2}(p^{d - 1}-1)p^{d - 1}(p-1)$ many edges. Moreover, by~\cite[Theorem 5]{alon1999norm} $\ProjNorm_d$ contains no copy of $K_{D,d}$.

   Let $V$ and $E$ denote the vertex and edge sets of $\ProjNorm_d$, respectively.
   
   Let $Q_d=(V_L \cup V_R,E(Q_d))$ be the bipartite double cover of $\ProjNorm_d$, defined by
   \[
      V_L=\{v_L:v \in V\},
      \qquad
      V_R=\{v_R:v \in V\},
   \]
   and
   \[
      u_Lv_R \in E(Q_d) \Leftrightarrow uv\in E.
   \]
   Hence $|V(Q_d)| = 2p^{d-1}(p - 1)$ and $e(Q_d) = (p^{d - 1}-1)p^{d - 1}(p-1)$.
   Moreover, since taking the bipartite double cover preserves biclique-freeness, $Q_d$ is $K_{D,d}$-free.

   Finally, choose $p$ to be the largest prime satisfying $|V_R| = |V_L| = p^{d - 1}(p-1) \leq n$.
   By Bertrand's postulate~\cite[Chapter 2]{aigner1999proofs}, there exists such a prime
   satisfying $p \geq n^{1/d}/2$. Thus, $Q_d$ has $n'\leq n$ vertices in each part and $e(Q_d) = \Omega(n^{2-1/d})$.
   By adding $n-n'$ isolated vertices to each part of $Q_d$, we obtain the claimed graph.

   To construct $\ProjNorm_d$, we first compute $p$ using the sieve of Eratosthenes in $\tilde{\Order}(n)$ time.
   We then compute the neighbors of each $(X,x) \in \Field_{p^{d-1}} \times \Field_{p}^{\times}$ exactly as described in the degree calculation above.
   Arithmetic computations over $\Field_{p^{d-1}}$, including the norm computation by exponentiation by squaring, take $\Order(\log n)$ time per operation.
   Thus, the total construction time is $\tOrder(n^{2-1/d})$.
\end{proof}

We now combine these lemmas to prove the lower bound.
\begin{proof}[Proof of \Cref{thm:shattering}]
   Fix an integer $d\geq1$ and set $D=(d-1)!+1$. Let $\mathcal M_d$ be the
   family of all square $J_{D,d}$-free Boolean matrices. By
   \Cref{lem:bounded-shatter-function}, this family has \VC-density at most $d$.

   For $d=1$, writing the $n$ output vectors gives a total-time lower bound
   of $\tilde{\Omega}(n^2)$. Since $T_{\mathrm{OMv}}=\Order(n^3)$, this
   implies the required $\tilde{\Omega}(T_{\mathrm{OMv}}/n)$ bound. We
   therefore assume $d\geq2$ for the remainder of the proof.

   Construct $Q_d$ using \Cref{lem:norm}, and let $F$ be its $n\times n$
   bipartite adjacency matrix. The construction of $F$ takes
   $\tOrder(n^2)$ time. Choose a constant $\beta>0$ such that
   \[
      \wt(F)=e(Q_d)\geq \beta n^{2-1/d}
   \]
   for all sufficiently large $n$. The family $\mathcal M_d$ is closed
   under row and column permutations, and every binary matrix $F'\leq F$
   belongs to $\mathcal M_d$, since zeroing out entries preserves
   $J_{D,d}$-freeness. Thus, \Cref{lem:omv-filter-lb} applies with
   $\mathcal M=\mathcal M_d$ and $\rho=1/d<1$, proving the theorem.
\end{proof}

   \section{Lower Bounds for Triangle Detection in \texorpdfstring{$H$}{H}-Free Graphs}
      \label{sec:triangle-lb}
      In this section, we prove~\Cref{thm:triangle-lb,thm:triangle-kst}.

For a graph $F$, let $t(F)$ denote its number of triangles.

\begin{lemma}[Filtering Graphs for Triangles]
   \label{lem:triangle-relabeling}
   Let $F$ and $G$ be graphs on vertex set $[n]$. Relabel $F$ by a uniformly random
   permutation $\pi:[n] \to [n]$ of its vertices, and retain an edge of $G$ exactly when it
   appears in the relabeled copy $\pi(F)$. Then every fixed triangle of $G$ is
   retained with probability
   \[
      \frac{t(F)}{\binom n3}.
   \]
   Moreover, after $K$ independent relabelings, the probability that some
   triangle of $G$ is never retained is at most
   \[
      n^3\exp\left(-\frac{K t(F)}{\binom n3}\right).
   \]
\end{lemma}
\begin{proof}
   A triangle $(i,j,k)$ in $G$ is retained if the preimages $(\pi^{-1}(i),\pi^{-1}(j),\pi^{-1}(k))$ 
   form a triangle in $F$. Since $\pi$ is a uniform random permutation, the preimages form a uniform random three-element subset of $[n]$. 
   Hence, the probability that $(\pi^{-1}(i),\pi^{-1}(j),\pi^{-1}(k))$ forms a triangle in $F$ is $t(F)/\binom n3$. Independence across relabelings and a union bound over at most $n^3$ triangles of $G$ give the asserted bound, as in the proof of \Cref{lem:filter-permutation}.
\end{proof}

\begin{lemma}[Triangle Lower Bounds From Filters]
   \label{lem:triangle-filter-lb}
   Let $H$ be a fixed graph. Suppose there are
   constants $\beta>0$ and $\rho \geq 0$ and a randomized combinatorial 
   construction that, on sufficiently large integers $n$, runs in
   $\tOrder(n^2)$ time and outputs an $n$-vertex graph $F$ such that the graph $F$ is $H$-free and 
   \[
      t(F)\geq\beta n^{3-\rho}
   \]
   with
   probability at least $9/10$.
   If Triangle Detection in $n$-vertex $H$-free graphs can be solved \whp
   in time $R(n)$, then unrestricted Triangle Detection can be solved
   \whp in time $\tOrder(n^\rho(n^2+R(n)))$. 
\end{lemma}
\begin{proof}
   Set 
   \[
      K=\left\lceil\frac{6}{\beta}n^\rho\ln n\right\rceil.
   \]
   Thus, $K=\tilde{\Theta}(n^\rho)$.

   Let $\mathcal A$ solve Triangle Detection in $n$-vertex $H$-free
   graphs \whp in time $R(n)$. Use \Cref{lem:standard-amplification} to
   reduce each invocation's failure probability to at most $1/(10K)$
   with an $\Order(\log n)$ factor in its running time.

   Consider an arbitrary $n$-vertex graph $G$. Construct $F$ as in the
   hypothesis and form
   $G_1,\ldots,G_K$ using independent random relabelings of $F$ as in
   \Cref{lem:triangle-relabeling}. Run an amplified copy of
   $\mathcal A$ on each $G_j$, capping every copy at its asserted amplified 
   $\Order(R(n)\log n)$ runtime. 
   This is necessary because if the filter construction
   fails, some $G_j$ may not satisfy the $H$-freeness promise, in which case
   the runtime guarantee for $\mathcal A$ might not hold.

   By assumption, with probability at least
   $9/10$, the graph $F$ is $H$-free and
   \[
      t(F)\geq\beta n^{3-\rho}.
   \]
   Conditioned on this event, every $G_j$ is a subgraph of a relabeled copy
   of $F$ and is therefore $H$-free. Moreover, by
   \Cref{lem:triangle-relabeling}, the probability that some triangle
   of $G$ is retained in none of the filtered graphs is at most
   \[
      n^3\exp\left(-\frac{K t(F)}{\binom n3}\right)
      \leq n^3\exp(-K\beta n^{-\rho})
      \leq n^3e^{-6\ln n}
      =n^{-3}\leq1/10.
   \]

   Report that $G$ contains a triangle if and only if at least one invocation
   reports one. Since each $G_j$ is a subgraph of $G$, this answer is correct
   whenever the filter construction succeeds, every triangle of $G$ occurs
   in some $G_j$, and all invocations succeed. The filter construction fails
   with probability at most $1/10$. Conditioned on its success, the probability that some
   triangle of $G$ is retained in none of the filtered graphs is at most
   $1/10$, and the invocations of $\mathcal A$ fail jointly with probability at most 
   $K/(10K)=1/10$. A union bound therefore gives success
   probability at least $7/10\geq2/3$.

   Constructing $F$, forming the filtered graphs, and running the amplified
   invocations takes time
   \[
      \tOrder\bigl(K(n^2+R(n))\bigr)
      = \tOrder\bigl(n^\rho(n^2+R(n))\bigr).
   \]
   All steps outside the calls to $\mathcal A$ are combinatorial. 
\end{proof}

We next describe how to construct good filters and derive our lower bounds.

\begin{lemma}[Filter Construction]
   \label{lem:triangle-random-filter}
   Fix a graph $H$ satisfying $e(H)>v(H)$. There is a constant
   $\beta=\beta(H)>0$ and a randomized combinatorial construction that,
   given a sufficiently large integer $n$, runs in $\Order(n^2)$ time and
   outputs an $n$-vertex graph $F$ such that, with probability at least
   $9/10$, the graph $F$ is $H$-free and
   \[
      t(F)\geq\beta n^{3-3v(H)/e(H)}.
   \]
\end{lemma}
\begin{proof}
   Choose $c>0$ so that $c^{e(H)}\leq1/11$, and set
   \[
      p=cn^{-v(H)/e(H)}
      \qquad\text{and}\qquad
      F\sim G(n,p).
   \]
   We show that, with probability at least $9/10$, the graph $F$ is $H$-free
   and $t(F)\geq\frac12\binom n3p^3$.

   If $X$ is the number of labeled copies of $H$ in $F$, then by linearity
   \[
      \mathbb E[X]\leq n^{v(H)}p^{e(H)}=c^{e(H)}\leq1/11.
   \]
   Markov's inequality then gives $\Pr[X\geq1]\leq1/11$.

   Since $0<v(H)/e(H)<1$, we get that
   \[
      np=cn^{1-v(H)/e(H)}
   \]
   is a growing, unbounded function of $n$.
   
   Let $Y=t(F)$ and $\mu=\mathbb E[Y]=\binom n3p^3$. The lower-tail estimate
   for subgraph counts in~\cite[Theorem~1]{janson1990poisson}, applied to
   triangles, gives
   \[
      \Pr[Y<\mu/2]
      \leq
      \exp\bigl(-\Omega(\min\{n^2p,n^3p^3\})\bigr)
      =o(1),
   \]
   where the last equality follows from our observation on the growth rate of $np$.

   Taking a union bound over the two failure events gives success probability
   at least 
   \[10/11-o(1)\geq9/10\] 
   for all sufficiently large $n$.
   Moreover, for all sufficiently large $n$,
   \[
      \frac12\binom n3p^3\geq\left(\frac{c^3}{24}\right)n^{3-3v(H)/e(H)}.
   \]

    The inequality above implies that we get the desired lower bound for $t(F)$ with probability at least $9/10$ by setting $\beta=c^3/24$.
   Sampling $F$ in this way takes $\Order(n^2)$ time combinatorially, which proves the claim.
\end{proof}

We can now establish our main lower bound for Triangle Detection. 
\begin{proof}[Proof of \Cref{thm:triangle-lb}]
   Suppose Triangle Detection in $n$-vertex $H$-free graphs can be
   solved \whp in time $R(n)$, and set $\rho=3v(H)/e(H)$.
   If $e(H)>v(H)$, \Cref{lem:triangle-random-filter,lem:triangle-filter-lb}
   give an algorithm that solves unrestricted Triangle Detection \whp
   in time $\tOrder(n^\rho(n^2+R(n)))$, which gives 
   \[
      R(n) \geq \tilde{\Omega}\left(\frac{T_{\mathrm{TD}}}{n^\rho}\right)-n^2.
   \]
   If $e(H) \leq v(H)$, $\rho\geq3$ and the asserted bound follows trivially since $T_{\mathrm{TD}} = \Order(n^3)$. 
\end{proof}

We next prove \Cref{thm:triangle-kst} by using the projective
norm graph from the proof of \Cref{lem:norm} as a filter
in \Cref{lem:triangle-filter-lb}.
\begin{proof}[Proof of \Cref{thm:triangle-kst}]
   Let $F$ be the projective norm graph $\ProjNorm_s$ from the proof of
   \Cref{lem:norm}. 
   It is $K_{s,t}$-free and can be constructed deterministically in
   $\tOrder(n^{2-1/s})$ time. By \cite[Lemma~4.4]{alon2016many},
   $t(F)=\Omega(n^{3-3/s})$.

   Suppose Triangle Detection in $n$-vertex $K_{s,t}$-free graphs can be
   solved \whp in time $R(n)$. Applying \Cref{lem:triangle-filter-lb}
   with $\rho=3/s$ gives
   $T_{\mathrm{TD}}=\tOrder(n^{3/s}(n^2+R(n)))$, which yields the
   asserted lower bound.
\end{proof}

The number of triangles in this filter is optimal up to a constant factor
by~\cite[Lemma~4.2]{alon2016many}.

   \section{APSP, Triangle Detection, and OMv in Bounded Distance \VC-Dimension}
      \label{sec:distance-vc}
      In this section we establish our lower bounds for Triangle Detection, APSP, and OMv in graphs of bounded \VCB-dimension respectively.
We organize our arguments as follows:

\begin{description}
    \item[Useful \VC-Dimension Bounds]
        In \Cref{sub:aux}, we collect useful facts about the \VC-dimension of simple graphs and their compositions.
    These results help bound the \VC-dimension of graphs obtained from gadget constructions in our proofs later on.
    
    \item[Shortcut Construction]
        In \Cref{sub:shortcut}, we design ``shortcut gadgets'' which reduce the diameter of an input graph, while (1) preserving whether the input contains triangles and (2) not increasing the \VC-dimension by much.
        This is the main technical part of our argument.

    \item[Bipartite Pattern Avoidance]
        In \Cref{sub:triangle-jr-free}, we show that Triangle Detection is hard even on input graphs that avoid a certain specially chosen bipartite subgraph.

    \item[Triangle Detection]
        In \Cref{sub:triangle-dist} we prove \Cref{thm:triangle-dist}, our lower bound for Triangle Detection on graphs of bounded \VCB-dimension.
        The proof comes from chaining the arguments in the previous two subsections: by \Cref{sub:triangle-jr-free} we can reduce from instances that avoid a certain bipartite subgraph pattern, and then using the gadgets from \Cref{sub:shortcut} we can produce hard instances with low \VCB-dimension.

    \item[APSP Detection]
        In \Cref{sub:APSP} we prove \Cref{thm:APSP-lb}, our lower bound for APSP on graphs of bounded \VCB-dimension.
        This uses similar reasoning to our Triangle Detection lower bound.

    \item[OMv]
        In \Cref{sub:omv-dist} we prove \Cref{thm:omv-vc-to-dvc}, our lower bound for OMv on graphs of bounded \VCB-dimension.
        This follows by a simple reduction from the bounded \VC-dimension case.
        
\end{description}

\paragraph{Definitions of \VC-dimension for graphs}
   Throughout this section, for any graph $G$ \emph{including bipartite graphs} we let $\VC(G)$ denote the \VC-dimension of the set system of all open neighborhoods in $G$.
   We let $\VCB(G)$ denote the distance \VC-dimension of $G$.

\subsection{Auxiliary \VC-Dimension Bounds}
   \label{sub:aux}
   We state the auxiliary \VC-dimension bounds used below. Their proofs 
   appear in \Cref{app:auxiliary-vc-proofs}. Here $\VC$ refers to ordinary
   graph \VC-dimension.

   \begin{claim}
      \label{claim:induced-subgraph-vc}
      If $H'$ is an induced subgraph of a graph $H$, then
      \[
         \VC(H')\leq \VC(H).
      \]
   \end{claim}
   \begin{claim}
      \label{claim:isolated-vertices-vc}
      Let $H'$ be obtained from a graph $H$ by adding isolated vertices. Then
      \[
         \VC(H')\leq\VC(H)+1.
      \]
   \end{claim}

   \begin{lemma}
      \label{lem:vc-union}
      Let $H_1$ and $H_2$ be graphs on the same vertex set, and let
      $d_i=\VC(H_i)$. Then \[
          \VC(H_1\cup H_2)
          \leq 5(d_1+d_2).
      \]
   \end{lemma}

   \begin{claim}
      \label{claim:vc-bipartite-complement}
      Let $H$ be bipartite with bipartition $X\sqcup Y$, and let
      $\overline{H}^{\,\mathrm{bip}}$ be its bipartite complement, with
      \[
         E\bigl(\overline{H}^{\,\mathrm{bip}}\bigr)
         =\set{xy:x\in X,\ y\in Y,\ xy\notin E(H)}.
      \]
      Then
      \[
         \VC\bigl(\overline{H}^{\,\mathrm{bip}}\bigr)\leq \VC(H)+1.
      \]
   \end{claim}

   \begin{claim}
      \label{claim:vc-matching}
      A graph consisting of a matching has \VC-dimension at most $1$.
   \end{claim}

   \begin{claim}
      \label{claim:vc-co-matching}
      A graph obtained from a balanced biclique by deleting a matching has
      \VC-dimension at most $2$.
   \end{claim}

   \begin{claim}
      \label{claim:stars}
      A star has \VC-dimension at most $1$.
   \end{claim}

\subsection{Shortcut Constructions}
   \label{sub:shortcut}
   We first describe the bipartite gadget and then combine three copies
   into the tripartite construction used by both reductions.

\subsubsection{Bipartite Shortcut Gadget}
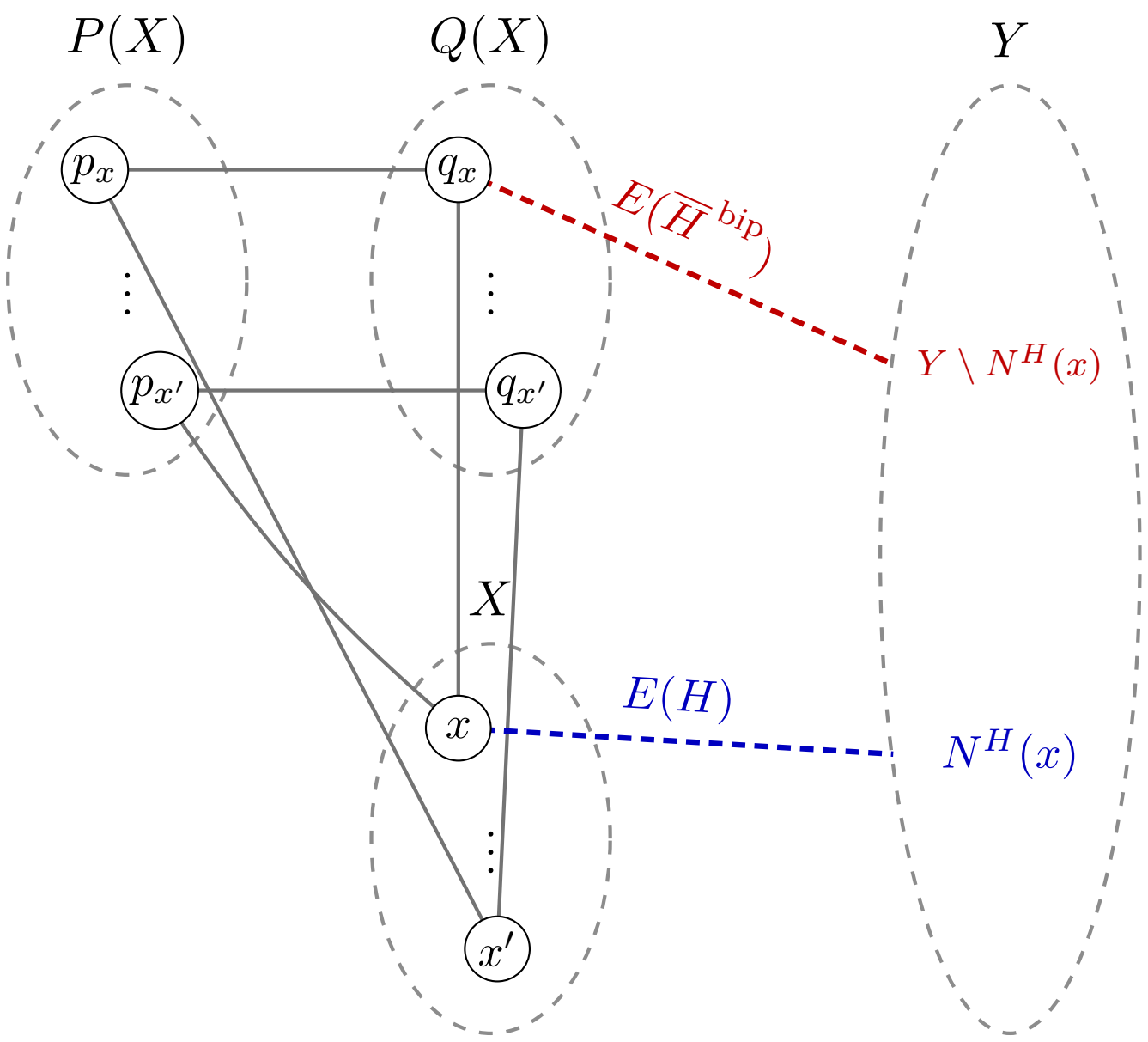
\begin{figure}[t]
      \centering
      \begin{tikzpicture}[
   vertex/.style={
      circle, draw=black, fill=white, minimum size=5mm,
      inner sep=1pt, font=\small
   },
   class/.style={draw=black!45, dashed, thick},
   auxiliary/.style={draw=black!55, line width=.8pt},
   original/.style={
      draw=blue!75!black, densely dashed, line width=1.2pt
   },
   complementary/.style={
      draw=red!75!black, densely dashed, line width=1.2pt
   }
]
   \coordinate (cpx)  at (-3.05, 3.15);
   \coordinate (cpxp) at (-2.55, 1.45);
   \coordinate (cqx)  at (-0.25, 3.15);
   \coordinate (cqxp) at ( 0.25, 1.45);
   \coordinate (cx)   at (-0.25,-1.15);
   \coordinate (cxp)  at ( 0.05,-2.85);
   \coordinate (cycomp) at (3.10, 1.65);
   \coordinate (cyn)    at (3.10,-1.35);

   \draw[class] (-2.8, 2.30) ellipse (.92cm and 1.50cm);
   \draw[class] ( 0.0, 2.30) ellipse (.92cm and 1.50cm);
   \draw[class] ( 0.0,-2.00) ellipse (.92cm and 1.50cm);
   \draw[class] ( 4.0, 0.15) ellipse (1.00cm and 3.65cm);

   \node[font=\bfseries] at (-2.8,4.15) {$P(X)$};
   \node[font=\bfseries] at ( 0.0,4.15) {$Q(X)$};
   \node[font=\bfseries] at ( 0.0,-.15) {$X$};
   \node[font=\bfseries] at ( 4.0,4.15) {$Y$};

   \draw[auxiliary] (cpx) -- (cqx);
   \draw[auxiliary] (cpxp) -- (cqxp);
   \draw[auxiliary] (cqx) -- (cx);
   \draw[auxiliary] (cqxp) -- (cxp);
   \draw[auxiliary] (cpx) -- (cxp);
   \draw[auxiliary] (cpxp) to[bend right=8] (cx);

   \draw[original] (cx) -- node[midway, above, sloped, font=\small,
      text=blue!75!black] {$E(H)$} (cyn);
   \draw[complementary] (cqx) -- node[midway, above, sloped, font=\small,
      text=red!75!black] {$E(\overline{H}^{\,\mathrm{bip}})$} (cycomp);

   \node[vertex] at (cpx)  {$p_x$};
   \node[vertex] at (cpxp) {$p_{x'}$};
   \node[vertex] at (cqx)  {$q_x$};
   \node[vertex] at (cqxp) {$q_{x'}$};
   \node[vertex] at (cx)   {$x$};
   \node[vertex] at (cxp)  {$x'$};

   \node at (-2.8,2.30) {$\vdots$};
   \node at ( 0.0,2.30) {$\vdots$};
   \node at ( 0.0,-2.00) {$\vdots$};
   \node[text=red!75!black, align=center, font=\scriptsize]
      at (4.0,1.65)
      {$Y\setminus N^H(x)$};
   \node[text=blue!75!black, font=\small]
      at (4.0,-1.35) {$N^H(x)$};
\end{tikzpicture}
      \caption{The shortcut gadget $R(H)$. The blue dashed line represents
      the original edges of $H$ between $X$ and $Y$, while the red dashed line
      represents the edges of $\overline{H}^{\,\mathrm{bip}}$ between $Q(X)$ and $Y$. 
      Gray edges come from the
      matchings incident to $Q(X)$, and the co-matching between $X$ and $P(X)$.}
      \label{fig:shortcut-gadget}
   \end{figure}

   \begin{definition}[Bipartite Shortcut Gadget]
      \label{def:gadget}
      Let $H$ be a bipartite graph with bipartition $X\sqcup Y$. Add a copy
      $Q(X)=\set{q_x\mid x\in X}$ and the matching $\set{xq_x\mid x\in X}$.
      Join $q_x$ to $y\in Y$ if and only if $xy\notin E(X,Y)$. Add another copy
      $P(X)=\set{p_x\mid x\in X}$ and a co-matching between $P(X)$ and $X$: 
      $\set{p_{x} x' \mid  x,x' \in X ,\ x \neq x'}$. 
      Finally, add the matching $\set{q_xp_x\mid x\in X}$.
      Denote the resulting graph by $R(H)$, as depicted in \Cref{fig:shortcut-gadget}.
   \end{definition}

   The role of $Q(X)$ is to bring every pair $(x,y)\in X\times Y$ within
   distance two: either $x\leadsto y$ is a direct edge so $x,y$ are already at distance one in $H$, or we have a new shortcut path $x\leadsto q_x\leadsto y$ so $x,y$ are at distance two in $R(H)$.
   The role of $P(X)$ is to do the same for pairs in $X\times Q(X)$:
   matched pairs $(x,q_x)$ are already adjacent, and for each $x\neq x'$ we have a new shortcut path
   $x\leadsto p_{x'}\leadsto q_{x'}$.
   Importantly, adding $Q(X)$ and $P(X)$ does not produce any new triangles in the graph.

   \begin{lemma}[Gadget Properties]
      \label{lem:gadget-bip}
      The graph $R(H)$ is triangle-free.
      Moreover, we have the following
      cross-class distances: for every $x\in X$ and $y\in Y$,
      \[
         \dist_{R(H)}(x,y)
         =\begin{cases}1,&xy\in E(X,Y),\\2,&xy\notin E(X,Y),\end{cases}
         \qquad
         \dist_{R(H)}(q_x,y)
         =\begin{cases}2,&xy\in E(X,Y),\\1,&xy\notin E(X,Y).\end{cases}
      \]
      Moreover, for every $x\in X$,
      \[
         \dist_{R(H)}(x,q_x)=1,\qquad
         \dist_{R(H)}(x,p_x)=2,\qquad
         \dist_{R(H)}(q_x,p_x)=1.
      \]
      For distinct $x,x'\in X$,
      \[
         \dist_{R(H)}(x,q_{x'})=2,\qquad
         \dist_{R(H)}(x,p_{x'})=1,\qquad
         \dist_{R(H)}(q_x,p_{x'})=2.
      \]
   \end{lemma}
   \begin{proof}
      Since $H$ is bipartite and $Q(X)$ is independent, any triangle containing a node $y\in Y$ must also
      contain some $x\in X$ and $q_{x'}\in Q(X)$. 
      In order for $x$ and $q_{x'}$ to be adjacent we need
      $x=x'$. 
      However, exactly one of the edges $xy$ and $q_xy$ can appear in $R(H)$, so such a triangle cannot exist.
      A triangle
      containing a node $p_x\in P(X)$ is also impossible: its neighbors are $q_x$ and the vertices in
      $X\setminus\set{x}$, but $q_x$ is not adjacent to any node in $X$ besides $x$. 
      Finally, any triangle in $R(H)$ must have a node in $Y$ or $P(X)$, because  $X\cup Q(X)$ induces a matching.
      Combining all of this together shows that $R(H)$ has no triangles.

      All pairs declared to be at distance $1$ are adjacent. The distance-$2$
      assertions are witnessed, using only edges of $R(H)$, by the paths
      \[
         x\leadsto q_x\leadsto y;\qquad q_x\leadsto x\leadsto y;\qquad
         x\leadsto q_x\leadsto p_x;\qquad x\leadsto p_{x'}\leadsto q_{x'};\qquad q_x\leadsto x\leadsto p_{x'},
      \]
      in the respective nonadjacent cases.
   \end{proof}

   \begin{lemma}[Gadget \VC-dimension]
      \label{lem:gadget-VC}
      Let $H$ be a bipartite graph with bipartition
      $X\sqcup Y$ such that $X\neq\emptyset$ and $\VC(H)\leq r$.
      Let $R'$ be the subgraph of $R(H)$ with edge set
      $E(R(H)) \setminus E(H)$, consisting only of the newly introduced edges. Then
      \[
         \VC(R')=\Order(r).
      \]
   \end{lemma}
   \begin{proof}
      We write $R'$ as the union of four graphs. Let $M_1$ and $M_2$
      be the matchings 
      \[
         E(M_1)=\set{xq_x:x\in X}
         \qquad\text{and}\qquad
         E(M_2)=\set{q_xp_x:x\in X},
      \]
      let $C$ be the co-matching 
      \[
         E(C) = \set{p_{x} x' \mid  x,x' \in X ,\ x \neq x'},
      \]
      and let $B$ be the 
      graph between $Q(X)$ and $Y$ whose edges correspond to $E(\overline{H}^{\,\mathrm{bip}})$. 

      Regard the four edge sets defined above as graphs with vertex sets $V(R(H))$.
      This may involve adding isolated vertices to the edge sets, but
      by \Cref{claim:isolated-vertices-vc}
      this can only increase each graph's \VC-dimension by at most one.
      By \Cref{claim:vc-matching,claim:vc-co-matching},
      $M_1, M_2,$ and $C$ each have constant \VC-dimension.
      Moreover, since $\VC(H)\le r$, \Cref{claim:vc-bipartite-complement} implies that
      $B$ has \VC-dimension{} at most $O(r)$. 
      Since
      \[
         R'=M_1\cup M_2\cup C\cup B,
      \]
      three applications of \Cref{lem:vc-union} 
      together with the previous discussion 
      yield
      $\VC(R')=\Order(r)$.
   \end{proof}

\subsubsection{The Tripartite Shortcut Construction}
   \label{sub:tripartite-shortcut}
   \begin{definition}[Tripartite Shortcut Construction]
      \label{def:tripartite-shortcut}
      Let $G$ be an $n$-vertex tripartite graph on nonempty parts $A,B,C$.
      Let \(\sigma(A)=B\), \(\sigma(B)=C\), and \(\sigma(C)=A\).
      For each $X\in\set{A,B,C}$, apply the gadget from~\Cref{def:gadget}
      to $G[X\sqcup\sigma(X)]$.
      Denote the resulting graph by $G_{\mathrm{base}}$, and call
      \[
         A,B,C,Q(A),Q(B),Q(C),P(A),P(B),P(C)
      \]
      its nine \emph{base classes},
      as depicted in \Cref{fig:tripartite-construction-base}.
      Write $V_{\mathrm{base}} = V(G_{\mathrm{base}})$ for the
      set of $3n$ base vertices. 

      \begin{figure}[t]
         \centering
         \begin{tikzpicture}[
   class/.style={
      draw=black!45, dashed, thick, fill=white
   },
   matching/.style={draw=green!50!black, line width=1pt},
   comatching/.style={
      draw=orange!85!black, densely dashed, line width=1.2pt
   },
   original/.style={
      draw=blue!75!black, densely dashed, line width=1.2pt
   },
   complementary/.style={
      draw=red!75!black, densely dashed, line width=1.2pt
   }
]
   \coordinate (A)  at ( 0.00, 2.30);
   \coordinate (B)  at ( 2.80,-1.60);
   \coordinate (C)  at (-2.80,-1.60);
   \coordinate (QA) at ( 0.00, 3.75);
   \coordinate (PA) at ( 0.00, 5.20);
   \coordinate (QB) at ( 4.45,-2.59);
   \coordinate (PB) at ( 6.10,-3.58);
   \coordinate (QC) at (-4.45,-2.59);
   \coordinate (PC) at (-6.10,-3.58);

   \draw[original] (A) -- (B);
   \draw[original] (B) -- (C);
   \draw[original] (C) -- (A);

   \draw[complementary] (QA) to[bend left=13] (B);
   \draw[complementary] (QB) to[bend left=13] (C);
   \draw[complementary] (QC) to[bend left=13] (A);

   \draw[matching] (A)  -- (QA);
   \draw[matching] (QA) -- (PA);
   \draw[matching] (B)  -- (QB);
   \draw[matching] (QB) -- (PB);
   \draw[matching] (C)  -- (QC);
   \draw[matching] (QC) -- (PC);

   \draw[comatching] (A) to[out=165, in=195, looseness=1.4] (PA);
   \draw[comatching] (B) to[bend left=45] (PB);
   \draw[comatching] (C) to[bend left=45] (PC);

   \draw[matching] (2.40,5.40) -- (3.10,5.40);
   \node[anchor=west, font=\scriptsize] at (3.20,5.40) {matching};
   \draw[comatching] (2.40,4.98) -- (3.10,4.98);
   \node[anchor=west, font=\scriptsize] at (3.20,4.98) {co-matching};

   \draw[class] (A)  ellipse (.82cm and .55cm);
   \draw[class] (B)  ellipse (.82cm and .55cm);
   \draw[class] (C)  ellipse (.82cm and .55cm);
   \draw[class] (QA) ellipse (.82cm and .55cm);
   \draw[class] (PA) ellipse (.82cm and .55cm);
   \draw[class] (QB) ellipse (.82cm and .55cm);
   \draw[class] (PB) ellipse (.82cm and .55cm);
   \draw[class] (QC) ellipse (.82cm and .55cm);
   \draw[class] (PC) ellipse (.82cm and .55cm);

   \node[font=\bfseries] at (A)  {$A$};
   \node[font=\bfseries] at (B)  {$B$};
   \node[font=\bfseries] at (C)  {$C$};
   \node[font=\bfseries] at (QA) {$Q(A)$};
   \node[font=\bfseries] at (PA) {$P(A)$};
   \node[font=\bfseries] at (QB) {$Q(B)$};
   \node[font=\bfseries] at (PB) {$P(B)$};
   \node[font=\bfseries] at (QC) {$Q(C)$};
   \node[font=\bfseries] at (PC) {$P(C)$};
\end{tikzpicture}
         \caption{The base graph $G_{\mathrm{base}}$, obtained by applying
         the shortcut gadget to the three pairs $(A,B)$, $(B,C)$, and
         $(C,A)$. Blue dashed lines represent the original edges of $G$,
         and red dashed lines represent the corresponding bipartite
         complements. Green lines represent matchings, and orange dashed
         lines represent co-matchings.}
         \label{fig:tripartite-construction-base}
      \end{figure}
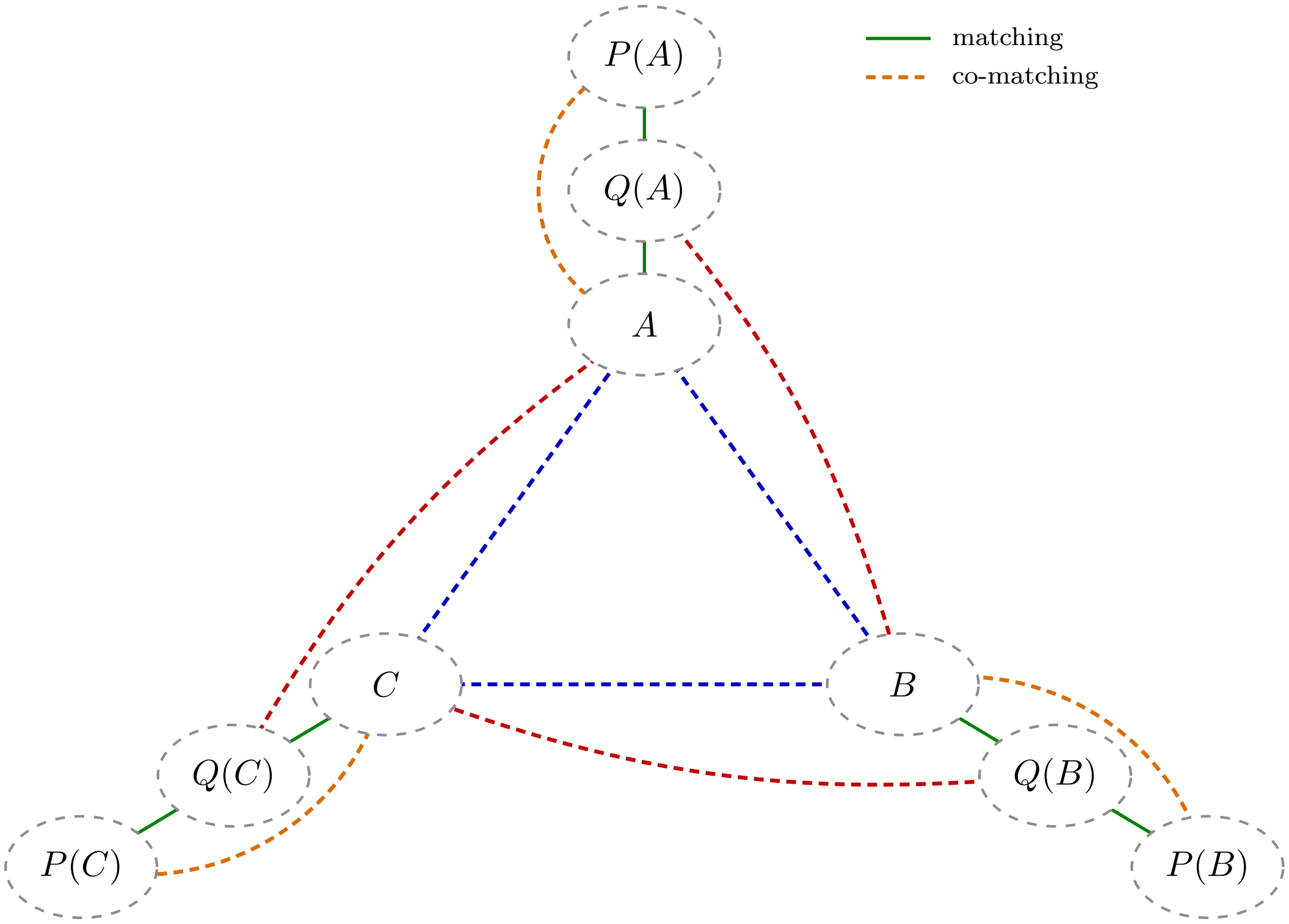

      Add the following eleven nodes we call \emph{hubs}. The hub $h_Q$ is adjacent to
      $Q(A)\cup Q(B)\cup Q(C)$, and $h_P$ is adjacent to
      $P(A)\cup P(B)\cup P(C)$. For each $X\in\set{A,B,C}$, add a hub $h_X$
      adjacent to
      \[
         X\cup\bigcup_{Y\in\set{A,B,C}\setminus\set{X}}P(Y).
      \]
      Finally, for every ordered pair of distinct $X,Y\in\set{A,B,C}$, add a
      hub $h_{Q(X),P(Y)}$ adjacent to $Q(X)\cup P(Y)$. When $Y=\sigma(X)$,
      also make this hub adjacent to $\sigma^{-1}(X)$ (i.e.,
      $h_{Q(A),P(B)}$, $h_{Q(B),P(C)}$, and $h_{Q(C),P(A)}$ are also adjacent
      to $C$, $A$, and $B$ respectively). 
      Hubs are not adjacent to each other.
      
      We call the resulting graph $T(G)$, as depicted in \Cref{fig:tripartite-construction}.
      We can construct $T(G)$ in $\Order(n^2)$ time just by following the definition above.
   \end{definition}

   To show lower bounds for structured Triangle Detection and APSP, it turns out it suffices to show hardness on tripartite graphs.
   Intuitively, to show hardness for these problems over graphs with bounded distance \VC-dimension, we want to add edges to the graph in a way that forces all nodes to be within distance two of each other (since this will simplify the structure of the closed-ball set system defining distance \VC-dimension from \cref{eq:closed-ball-system}, and make it easy to  bound this parameter  by relating it to standard \VC-dimension) while not increasing the distance \VC-dimension by too much.

    The gadget construction from \Cref{lem:gadget-VC} shows how to do this for a single bipartite graph with partition $X\sqcup Y$.
    The natural idea to generalize this to tripartite graphs with vertex partition $A\sqcup B\sqcup C$ is to apply this gadget three times, on the graphs induced on $A\sqcup B$, $B\sqcup C$, and $C\sqcup A$ respectively.
    This is what we do in the first part of the construction from \Cref{def:tripartite-shortcut}.
    This process does not work on its own, because each application of \Cref{lem:gadget-VC} introduces some new vertex sets $Q(X)$ and $P(X)$ for $X\in\set{A,B,C}$, and pairs of nodes taken from these auxiliary sets for different choices of $X$ might be far apart from one another.
    
    The \emph{hubs} from \Cref{def:tripartite-shortcut} are introduced to fix exactly this issue, and create shortcut paths between the auxiliary vertices of different gadgets through a constant number of central nodes.
   The hubs $h_Q$ and $h_P$ handle pairs of $Q$-vertices and pairs of
   $P$-vertices, respectively, while $h_X$ handles pairs within $X$ and
   between $X$ and $P(Y)$ for $Y\neq X$.
   The hubs $h_{Q(X),P(Y)}$ handle the remaining pairs between $Q$-classes
   and $P$-classes; their extra adjacencies cover the remaining pairs
   between $Q$-classes and original classes.
   Each hub has an independent neighborhood, so these shortcuts create
   no triangles, and together with the gadgets, force all base vertices within distance two of each other.

      \begin{figure}[t]
         \centering
         \begin{tikzpicture}[
   class/.style={
      draw=black!45, dashed, thick, fill=white
   },
   auxiliary/.style={draw=black!55, line width=.8pt},
   complete/.style={draw=purple!75!black, line width=1pt},
   original/.style={
      draw=blue!75!black, densely dashed, line width=1.2pt
   },
   complementary/.style={
      draw=red!75!black, densely dashed, line width=1.2pt
   },
   hub/.style={
      circle, draw=black!65, fill=white, line width=.8pt,
      minimum size=3.5mm, inner sep=0pt
   },
   hub edge/.style={complete}
]
   \coordinate (A)  at ( 0.00, 2.30);
   \coordinate (B)  at ( 2.80,-1.60);
   \coordinate (C)  at (-2.80,-1.60);
   \coordinate (QA) at ( 0.00, 3.40);
   \coordinate (PA) at ( 0.00, 4.50);
   \coordinate (QB) at ( 4.05,-2.35);
   \coordinate (PB) at ( 5.30,-3.10);
   \coordinate (QC) at (-4.05,-2.35);
   \coordinate (PC) at (-5.30,-3.10);

   \coordinate (hP) at ( 6.35,-4.45);
   \coordinate (hC) at (-1.75,-2.75);
   \coordinate (hQCPA) at (-3.40,2.75);
   \coordinate (hQBPA) at ( 2.20,1.80);

   \draw[hub edge] (hP) to[bend right=38] (PA);
   \draw[hub edge] (hP) -- (PB);
   \draw[hub edge] (hP) to[bend left=5] (PC);

   \draw[hub edge] (hC) -- (C);
   \draw[hub edge] (hC) to[out=90, in=215, looseness=1.1] (PA);
   \draw[hub edge] (hC) -- (PB);

   \draw[hub edge] (PA) -- (hQCPA);
   \draw[hub edge] (hQCPA) -- (QC);
   \draw[hub edge] (hQCPA) -- (B);

   \draw[hub edge] (PA) to[bend left=10] (hQBPA);
   \draw[hub edge] (hQBPA) -- (QB);

   \draw[original] (A) -- (B);
   \draw[original] (B) -- (C);
   \draw[original] (C) -- (A);

   \draw[complementary] (QA) to[bend left=18] (B);
   \draw[complementary] (QB) to[bend left=13] (C);
   \draw[complementary] (QC) to[bend left=22] (A);

   \draw[complete] (2.40,4.56) -- (3.10,4.56);
   \node[anchor=west, font=\scriptsize] at (3.20,4.56) {complete};

   \draw[class] (A)  ellipse (.82cm and .55cm);
   \draw[class] (B)  ellipse (.82cm and .55cm);
   \draw[class] (C)  ellipse (.82cm and .55cm);
   \draw[class] (QA) ellipse (.82cm and .55cm);
   \draw[class] (PA) ellipse (.82cm and .55cm);
   \draw[class] (QB) ellipse (.82cm and .55cm);
   \draw[class] (PB) ellipse (.82cm and .55cm);
   \draw[class] (QC) ellipse (.82cm and .55cm);
   \draw[class] (PC) ellipse (.82cm and .55cm);

   \node[font=\bfseries] at (A)  {$A$};
   \node[font=\bfseries] at (B)  {$B$};
   \node[font=\bfseries] at (C)  {$C$};
   \node[font=\bfseries] at (QA) {$Q(A)$};
   \node[font=\bfseries] at (PA) {$P(A)$};
   \node[font=\bfseries] at (QB) {$Q(B)$};
   \node[font=\bfseries] at (PB) {$P(B)$};
   \node[font=\bfseries] at (QC) {$Q(C)$};
   \node[font=\bfseries] at (PC) {$P(C)$};

   \node[hub] at (hP) {};
   \node[anchor=north, font=\scriptsize] at (6.35,-4.68) {$h_P$};
   \node[hub] at (hC) {};
   \node[anchor=north, font=\scriptsize] at (-1.75,-2.98) {$h_C$};
   \node[hub] at (hQCPA) {};
   \node[anchor=south, font=\scriptsize] at (-3.40,2.98)
      {$h_{Q(C),P(A)}$};
   \node[hub] at (hQBPA) {};
   \node[anchor=west, font=\scriptsize] at (2.43,1.70)
      {$h_{Q(B),P(A)}$};
\end{tikzpicture}
         \caption{The tripartite shortcut construction, obtained from the 
            base graph by adding $11$ hubs. To avoid clutter, 
            only a 
            representative subset and its incident edges are shown. We also 
            omit the matchings and co-matchings of the base graph, as they are illustrated in
            \Cref{fig:tripartite-construction-base}.}
         \label{fig:tripartite-construction}
      \end{figure}
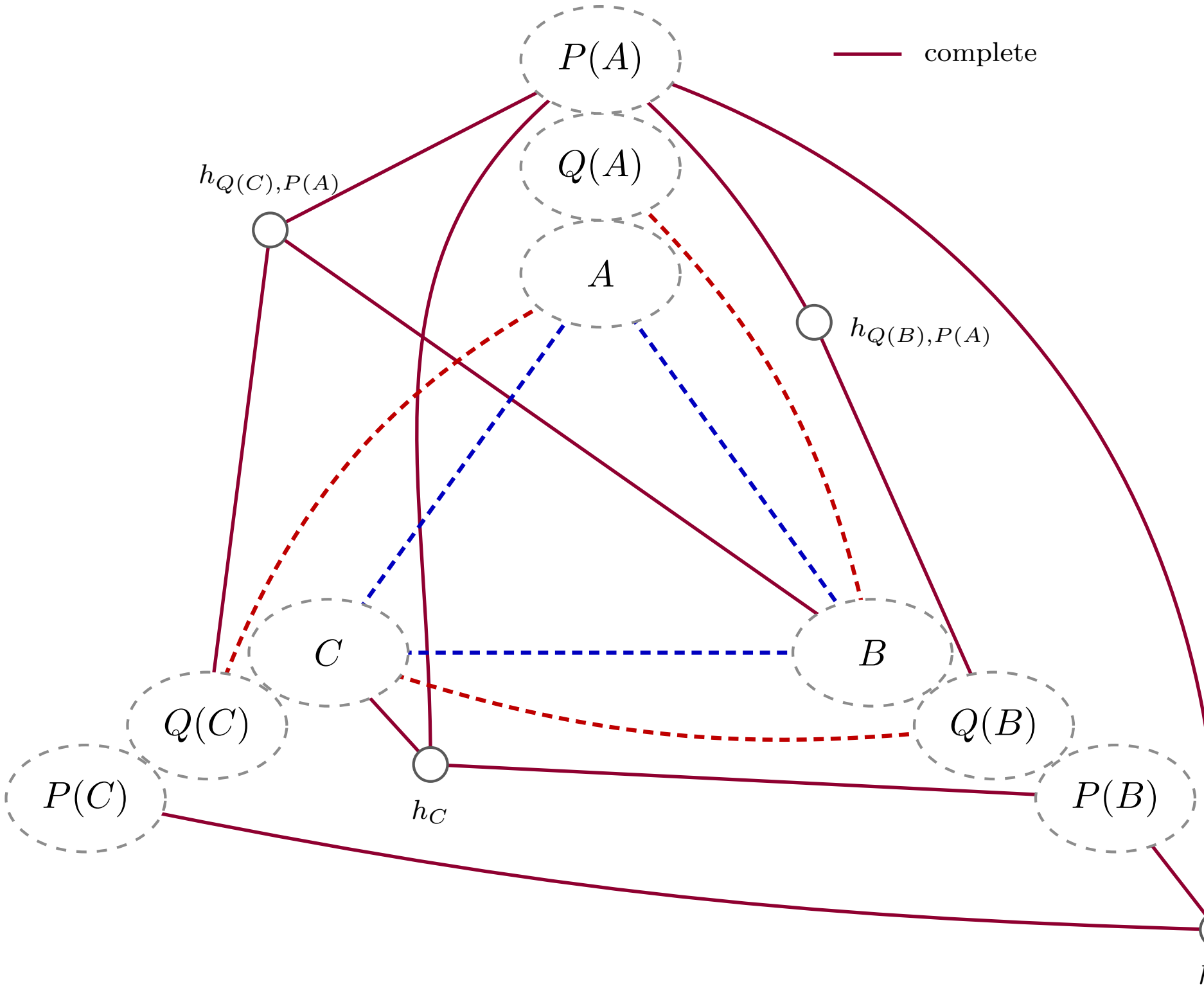
   Throughout the following results, we use the definitions and notation
   from \Cref{def:tripartite-shortcut}.

   \begin{claim}[Preserving Triangles]
      \label{claim:tripartite-triangles}
      The triangles of $T(G)$ are exactly the triangles of $G$.
   \end{claim}
   \begin{proof}
      We first prove that every triangle in $G_{\mathrm{base}}$ is a 
      triangle in $G$. 
      We prove this directly by showing that neighborhoods of vertices in the $Q$ and $P$ classes are independent in $G_{\mathrm{base}}$. 
      Consequently, every triangle in $G_{\mathrm{base}}$ lies in $A\cup B\cup C$ and is a triangle of $G$. 
      Indeed, by the definition of the shortcut gadget, for every $X\in\set{A,B,C}$ and $x \in X$,
      \[
         N^{G_{\mathrm{base}}}(q_x)
         =\set{x,p_x}\cup\bigl(\sigma(X)\setminus N^G(x)\bigr),
         \qquad
         N^{G_{\mathrm{base}}}(p_x)
         =\set{q_x}\cup\bigl(X\setminus\set{x}\bigr).
      \]
      Both sets are independent in $G_{\mathrm{base}}$. 

      It remains to show that adding the hubs creates no triangle. Consider 
      first the hubs $h_Q$ and $h_P$. The sets 
      $Q(A)\cup Q(B)\cup Q(C)$ and $P(A)\cup P(B)\cup P(C)$ are independent,
      so the neighborhoods of $h_Q$ and $h_P$ are independent. Now consider $h_X$ 
      for each $X\in\set{A,B,C}$. 
      There are no edges between $X$ and $P(Y)$ when
      $Y\neq X$, and there are no edges within or between the $P$-classes.
      Hence the neighborhood of $h_X$ is independent.

      Finally, consider a hub $h_{Q(X),P(Y)}$ for $X\neq Y$. If
      $Y\neq\sigma(X)$ (e.g., $h_{Q(B),P(A)}$ in \Cref{fig:tripartite-construction}), then
      \[
         N^{T(G)}(h_{Q(X),P(Y)})=Q(X)\cup P(Y).
      \]
      Both classes are independent, and there are no edges between $Q(X)$ and
      $P(Y)$, so this neighborhood is independent. If $Y=\sigma(X)$ (e.g., $h_{Q(C),P(A)}$ in \Cref{fig:tripartite-construction}), let
      $R=\sigma^{-1}(X)$. In this case,
      \[
         N^{T(G)}(h_{Q(X),P(Y)})=Q(X)\cup P(Y)\cup R.
      \]
      Again there are no edges between $Q(X)$ and $P(Y)$. Moreover, $Q(X)$
      has neighbors among $A\cup B\cup C$ only in $X\cup\sigma(X)$, and
      $P(Y)$ has neighbors there only in $Y$. Hence $R\cup Q(X)$ and $R \cup P(Y)$ are independent,
      so this neighborhood is also independent. Thus no hub
      of the form $h_{Q(X),P(Y)}$ belongs to a triangle. Together with the
      preceding cases and the fact that no edges are added between hubs, this
      shows that no triangle of $T(G)$ contains a hub. 
      
      This discussion, together with the fact that $T(G)$ never adds or removes edges between vertices in the original graph $G$, proves that the sets of triangles in $T(G)$ and $G$ are the same.
   \end{proof}

   \begin{lemma}[\VC-dimension with Shortcuts]
      \label{lem:tripartite-vc}
      Let $G$ be a tripartite graph as in \Cref{def:tripartite-shortcut}.
      If $\VC(G)\leq r$, then 
      \[
         1 \leq \VC(T(G))=\Order(r).
      \]
      Moreover, for every $X\in\set{A,B,C}$ we have
      \[
         1 \leq \VC\bigl(T(G)-E_G(X,\sigma(X))\bigr)=\Order(r).
      \]
   \end{lemma}
   \begin{proof}
      We decompose $T(G)$ into a constant number of subgraphs of \VC-dimension
      $\Order(r)$, and then apply \Cref{lem:vc-union} to bound the \VC-dimension
      of their union. Whenever we take a union, we add isolated vertices to
      the constituent graphs as needed so that they have the same vertex set.
      By \Cref{claim:isolated-vertices-vc}, this increases the \VC-dimension
      of each graph by at most one.

      First, we decompose $G_{\mathrm{base}}$ into $G$ and the new vertices
      and edges introduced by the three bipartite shortcut gadgets.
      For every $X\in\set{A,B,C}$, put $Y=\sigma(X)$ and let
      $R_{(X,Y)}$ be the subgraph of $R(G[X\sqcup Y])$ consisting only
      of the newly introduced edges.
      By \Cref{claim:induced-subgraph-vc}, $\VC(G[X\cup Y]) \leq r$, and therefore
      by \Cref{lem:gadget-VC}, $\VC(R_{(X,Y)}) = \Order(r)$.
      Thus, repeated applications of \Cref{lem:vc-union} to $R_{(A,B)},R_{(B,C)},R_{(C,A)},$ and $G$ give
      \[
         \VC(G_{\mathrm{base}})=\Order(r).
      \]

      We now decompose $T(G)$ into $G_{\mathrm{base}}$ and the hub vertices
      and their incident edges. Each hub, together with its incident edges,
      forms a star. Since there are $11$ hubs, \Cref{claim:stars} and repeated
      applications of \Cref{lem:vc-union} yield
      \[
         \VC(T(G))=\Order(r),
      \]
      proving the first assertion.

      For the second assertion, fix $X\in\set{A,B,C}$ and put
      $Y=\sigma(X)$. The graph $G-E_G(X,Y)$ is the union of 
      two bipartite induced subgraphs of $G$,
      each of which has \VC-dimension at most $r$ by \Cref{claim:induced-subgraph-vc}. Thus, \Cref{lem:vc-union}
      gives $\VC(G-E_G(X,Y))=\Order(r)$.
      
        At this point, we can just use similar reasoning to the previous two paragraphs, and decompose $T(G) - E_G(X,Y)$ into the graph $G-E_G(X,Y)$ added to the various bipartite shortcut gadgets and stars centered at hub vertices.
        The same reasoning shows that each of these added components has \VC-dimension at most $O(r)$, and so once more by \Cref{lem:vc-union} their combination $T(G) - E_G(X,Y)$ will also have $\VC$-dimension at most $O(r)$.  

      Finally, both $T(G)$ and $T(G)-E_G(X,Y)$ induce a nonempty
      matching on $A\cup Q(A)$, of \VC-dimension $1$.
      Thus \Cref{claim:induced-subgraph-vc} gives both lower bounds.
   \end{proof}

   \begin{claim}[Tripartite Shortcut Distances]
      \label{claim:tripartite-distances}
      In $T(G)$, every two base vertices are at distance at most two from one another,
      and every hub is at distance at most three from every vertex.
      Moreover, for every $X\in\set{A,B,C}$, each pair of base vertices within the subgraph
      $T(G)-E_G(X,\sigma(X))$, outside the
      two class pairs $(X,\sigma(X))$ and $(Q(X),\sigma(X))$, is still
      at distance at most two.
      Similarly, within the subgraph $T(G)-E_G(X,\sigma(X))$ hubs are still at distance at most three from all vertices.
   \end{claim}
   \begin{proof}
      We first verify that every two base vertices are at distance at most two.
      
      For each $X\in\set{A,B,C}$, \Cref{lem:gadget-bip} proves this for pairs of nodes corresponding to any of
      \[
         (X,\sigma(X)),\quad (X,Q(X)),\quad
         (Q(X),\sigma(X)),\quad (X,P(X)),\quad
         (Q(X),P(X)).
      \]
      The hubs $h_Q$ and $h_P$ handle the pairs $(Q(X),Q(Y))$ and $(P(X),P(Y))$ 
      for all $X,Y \in \set{A,B,C}$. 
      The hub $h_X$ handles pairs $(X,X)$ and $(X,P(Y))$ for $Y\neq X$. 
      The hub $h_{Q(X),P(Y)}$ handles
      $(Q(X),P(Y))$ when $X\neq Y$. Finally,
      $h_{Q(X),P(\sigma(X))}$ handles
      $(Q(X),\sigma^{-1}(X))$ by its additional adjacencies. These cases
      exhaust all pairs of base classes.

      Since every hub has a base neighbor and every two base vertices are
      at distance at most two, every base vertex is within distance three
      of every hub.

      Each hub is adjacent to all vertices of $Q(X)$ or $P(X)$ for at
      least two choices of $X\in\set{A,B,C}$. Thus any two hubs have a
      common such choice of $X$. If they are adjacent to the same class,
      they have a common neighbor. Otherwise, the matching between
      $Q(X)$ and $P(X)$ gives a path of length three between them.
      Hence every pair of hubs is at distance at most three as well.

      This proves the first part of the claim about distances in $T(G)$.
      It remains to prove the second part of the claim, about distances in $T(G)$ after some set of edges $E_G(X,\sigma(X))$ is deleted. 

      Fix $X\in\set{A,B,C}$, and put
      $Y=\sigma(X)$. 
      We first consider pairs of base vertices.
      
      Deleting $E_G(X,Y)$ leaves the
       two other bipartite gadgets and all edges incident to hubs unchanged. In the affected
      gadget, all distance bounds from \Cref{lem:gadget-bip}, except
      those for $(X,Y)$ and $(Q(X),Y)$, use only matching and co-matching
      edges, which remain present. Thus every pair of base vertices
      outside $(X,Y)$ and $(Q(X),Y)$ is still at distance at most two.
        It remains to consider distances from hubs to base vertices.

      The fact that pairs of hubs are at distance at most three still holds, because the proof of this fact only used hub
      edges and possibly one matching edge, which are unaffected by the deletion. 
      To bound distances between hubs and base vertices, fix a base vertex $v$. 
      As shown in the previous paragraph, the distances from $v$ to all $P$-vertices remain
      at most two. 
      If $h$ is a hub node distinct from $h_Q$, then $h$ has a neighbor in a
      $P$-class,  so $\dist(v,h)\leq3$ even after deletion.
      For $h_Q$, we have the stronger bound $\dist(v,h_Q)\leq2$, because
      every vertex in a $Q$-class is adjacent to $h_Q$, and every other
      base vertex has a matching neighbor in a $Q$-class. All these
      edges survive the deletion.

      This completes the proof of the desired result.
   \end{proof}

    Using the distance properties of $T(G)$ established thus far, we can relate its distance \VC-dimension to the vanilla \VC-dimension of the original graph $G$.
    Before presenting this result, we show one more technical lemma that will help us with calculations related to \VC-dimension.

    \begin{lemma}[Binomial Sum Bound]
        \label{lem:vc-silly-bound}
        Let $k, D\ge 1$ be integers, and $M \ge 1$ a constant. 
        Set
        \[S = \sum_{i=0}^D \binom{k}{i}.\]
        Suppose that $2^k\le MS^2$.
        Then $k = O(D)$.
    \end{lemma}
    \begin{proof}
        By \Cref{binomial-sum}, we have 
            \[S\le (ek/D)^D.\]
        Thus the inequality from the lemma simplifies to 
            \[2^k \le MS^2 \le M(ek/D)^{2D}.\]
        Set $t=k/D$.
        By taking the logarithm of both sides of the above inequality and then dividing by $D$, we get that 
            \[t\le 2\log(2et) + \log(M)/D.\]
        Since $M$ is a constant, the above inequality is saying $t = \Order(\log t)$, which forces $t = \Order(1)$.
        Thus we have $k = tD = \Order(D)$ as desired.
    \end{proof}
    
    We now proceed with the main result of this section.

   \begin{lemma}[Distance \VC-dimension to Standard \VC-dimension]
      \label{lem:tripartite-balls}
      If $\VC(G)\leq r$, then
      \[
         \VCB(T(G))=\Order(r).
      \]
   \end{lemma}
   \begin{proof}
      Set $D=\VC(T(G))$. By \Cref{lem:tripartite-vc},
      $D\geq1$ and $D=\Order(r)$. Fix a $k$-vertex set $X$ shattered
      by the closed balls of $T(G)$ (as defined in \cref{eq:closed-ball-system}). We count their distinct intersections with
      $X$, separating balls of radius zero, radius one, and radius at least two.

      Write $V=V(T(G))$. The three quantities to bound are
      \[
         \bigl|\set{N_0[v]\cap X:v\in V}\bigr|,\qquad
         \bigl|\set{N_1[v]\cap X:v\in V}\bigr|,\qquad
         \bigl|\set{N_b[v]\cap X:v\in V,\ b\geq2}\bigr|.
      \]
      Radius-zero balls $N_0[v] = \set{v}$ have size one, and thus
      contribute at most $k+1$ distinct intersections since $X$ has size $k$. 
      Since $N_1[v] =  N_0[v]\cup N(v)$, radius-one balls contribute 
      \begin{align*} 
            &\bigl| \set{N_1[v] \cap X : v \in V} \bigr| \leq 
            \bigl| \set{N_0[v] \cap X : v \in V} \bigr| \cdot \bigl| \set{N(v) \cap X : v \in V} \bigr| \leq \\ 
            &(k+1) \cdot \bigl| \set{N(v) \cap X : v \in V} \bigr| \leq 
            (k+1)\sum_{i=0}^D\binom{k}{i}
      \end{align*} 
      intersections, 
      where the last inequality follows from the Sauer--Shelah lemma
        \cref{eq:sauer-shelah-intro}  applied to
      the open neighborhoods of $T(G)$.

      By \Cref{claim:tripartite-distances}, a ball of radius at least two centered at a base vertex contains all 
      base vertices and can vary only in its intersection with the at most $11$ 
      hubs, giving at most $2^{11}$ intersections. Formally, 
      \begin{align*}
         &\bigl|\set{N_b[v]\cap X:v\in V_{\mathrm{base}},\ b\geq2}\bigr| = \\
         &\bigl|\set{(X\cap V_{\mathrm{base}})\sqcup
            (X \cap N_b[v] \setminus V_{\mathrm{base}} ):
            v\in V_{\mathrm{base}},\ b\geq2}\bigr| \leq \\
         &2^{|X\setminus V_{\mathrm{base}}|}\leq 2^{11}.
      \end{align*}

      It remains to count radius-two balls centered at hubs, which
      contribute at most $11$ additional intersections. By
      \Cref{claim:tripartite-distances}, every ball of radius at least three
      equals $V$, whose intersection with $X$ has already been counted
      among balls centered at base vertices.

      Since $X$ is shattered, these bounds give
      \[
         2^k \leq k+1 + 
         (k+1)\sum_{i=0}^D\binom{k}{i}+
         2^{11}+11\;.
      \]
      Set 
        \[S = \sum_{i=0}^D\binom{k}{i}\]
    so that the above inequality becomes
        \[2^k \le k+1  + (k+1)S + 2^{11} + 11.\]
    Since $D\ge 1$, we have $k+1\le S$, and we can simplify the above inequality to 
        \[2^k \le MS^2\]
    for some constant $M\ge 1$.
    Now by applying \Cref{lem:vc-silly-bound}, we deduce that $k = \Order(D) = \Order(r)$.
    
      Thus every shattered set has size $\Order(r)$,
      which proves that $\VCB(T(G))=\Order(r)$ as desired.
   \end{proof}

\subsection{Hard Tripartite Instances of Bounded \VC-Dimension}
   \label{sub:triangle-jr-free}
   In this subsection, we reduce unrestricted Triangle Detection to Triangle
   Detection in tripartite graphs that exclude a fixed biclique as a subgraph
   and contain at most one triangle.
   These graphs have bounded \VC-dimension. We begin by defining the forbidden biclique.

   \begin{definition}[Forbidden Biclique $J_r$]
      \label{def:jr}
      For every integer $r\geq1$, define
      \[
         J_r=K_{\lceil r/2\rceil,\,2^{\lfloor r/2\rfloor+1}}.
      \]
   \end{definition}

   \begin{lemma}
      \label{lem:jr-free-vc}
      For every integer $r\geq1$, if a graph $G$ is $J_r$-free, then $\VC(G)\leq r$.
   \end{lemma}
   \begin{proof}
      Suppose for the sake of contradiction that $G$ has a shattered set $X$ of size $r+1$. Fix
      $R\subset X$ of size $\lceil r/2\rceil$. For every $Y$ satisfying
      $R\subseteq Y\subseteq X$, choose $v_Y$ such that
      $N(v_Y)\cap X=Y$. There are
      \[
         2^{|X|-|R|}=2^{\lfloor r/2\rfloor+1}
      \]
      such vertices. They are distinct, lie outside $R$, and are all adjacent
      to every vertex of $R$. Hence they form a copy of $J_r$ together with
      $R$, a contradiction.
   \end{proof}

   \begin{lemma}
      \label{lem:triangle-jr-free}
      Let $r\geq3$ be an integer. If Triangle Detection in $n$-vertex tripartite $J_r$-free graphs
      containing at most one triangle can be solved \whp in time $R(n)$,
      then unrestricted Triangle Detection can be solved \whp in time
      $\tOrder(n^{\rho_r}(n^2+R(n)))$, where
      \[
         \rho_r=\frac{3v(J_r)}{e(J_r)}=\frac{3}{\lceil r/2\rceil}
         +\frac{3}{2^{\lfloor r/2\rfloor+1}}=\Order(1/r).
      \]
   \end{lemma}
   \begin{proof}
      Since $r\geq3$, the two parts of $J_r$ have sizes at least two and
      four, respectively, so $e(J_r)>v(J_r)$. Applying
      \Cref{lem:triangle-random-filter,lem:triangle-filter-lb} with $H=J_r$
      reduces unrestricted Triangle Detection to Triangle Detection in
      $J_r$-free graphs, producing $\tOrder(n^{\rho_r})$ instances on $n$
      vertices with an overhead of $\tOrder(n^{2+\rho_r})$ time.

      It remains to reduce the latter problem to its more restricted setting, where in addition to $J_r$-freeness, 
      the graph is tripartite and contains at most one triangle. 

      Let $\mathcal A$ solve Triangle Detection on tripartite 
      $J_r$-free graphs containing at most one triangle \whp in time $R(n)$.
      Consider an arbitrary $n$-vertex $J_r$-free graph $G$. Apply the
      standard color-coding reduction~\cite{alon1995color} by assigning each
      vertex independently and uniformly to one of three color classes and
      deleting monochromatic edges. This preserves $J_r$-freeness, and every
      fixed triangle survives with constant probability. We amplify the success probability 
      to $9/10$ by independent repetitions, resulting in a constant number of subgraphs.  

      On each resulting subgraph, apply a reduction to Unique Triangle
      Detection. This is done using ideas similar to those of the isolation
      lemma~\cite{valiant1985np}. We use the formulation
      in~\cite[Lemma~7 of the full version]{abboud2026equivalent} for our
      setting. The reduction produces $\tOrder(1)$ subgraphs of $G$ such that
      with probability at least $9/10$, one of them contains exactly one
      triangle if the original subgraph does.

      Let $K=\tOrder(1)$ be an upper bound on the number of subgraphs.
      We run an amplified copy of $\mathcal A$ on each subgraph
      using the amplification from~\Cref{lem:standard-amplification} to reduce each invocation's
      failure probability to at most $1/(10K)$. We cap the running times 
      at the asserted $\Order(R(n)\log n)$ runtime with amplification, since some subgraphs 
      may contain more than one triangle. Report YES if at least one invocation reports YES. 

      If $G$ is triangle-free, every produced subgraph is triangle-free.
      Otherwise, color coding and isolation give a subgraph containing exactly 
      one triangle with probability at least $8/10$. A union bound over the
      calls therefore gives success probability at least
      $7/10\geq2/3$. The running time for $G$ is $\tOrder(n^2+R(n))$. 
      Summing over all $\tOrder(n^{\rho_r})$ instances gives the asserted running time. 
      All steps outside the calls to $\mathcal A$ are combinatorial.
   \end{proof}

\subsection{Triangle Detection in Bounded \VCB-Dimension}
   \label{sub:triangle-dist}
   We now combine the hardness of tripartite $J_r$-free instances with
   the properties of $T(G)$ to prove \Cref{thm:triangle-dist},
   restated below.

   \phantomsection\label{restatement:triangle-dist}
   \triangleDistanceVCLowerBound*

   \begin{proof}
      By \Cref{lem:tripartite-balls}, there is an absolute constant
      $K\geq1$ such that if $\VC(G)\leq r$, then $\VCB(T(G))\leq Kr$.
      For every integer $d\geq 3K$, define
      \[r=r(d)= \left\lfloor\frac d{K}\right\rfloor \geq 3.\]

      Applying \Cref{lem:triangle-jr-free} reduces unrestricted Triangle Detection to 
      Triangle Detection in tripartite $J_r$-free graphs, 
      producing $\tilde\Order(n^{\tau})$ instances with overhead $\tilde\Order(n^{2+\tau})$, 
      where $\tau = \Order(1/r) = \Order(1/d)$. It remains to give a reduction from the latter
      problem to Triangle Detection on graphs of bounded \VCB-dimension. 

      Let $\mathcal A$ solve Triangle Detection in graphs with \VCB-dimension at most $d$ \whp,
      and let $R(N)$ be its worst-case running time on such graphs with at most $N$ vertices.

      We use $\mathcal A$ to solve Triangle
      Detection in tripartite $J_r$-free graphs as follows. 
      Given such an $n$-vertex tripartite graph $G$,
      \Cref{lem:jr-free-vc} gives $\VC(G)\leq r$.
      Construct $T(G)$ as in \Cref{def:tripartite-shortcut}.
      By \Cref{claim:tripartite-triangles}, $T(G)$
      contains a triangle if and only if $G$ does.
      By \Cref{lem:tripartite-balls}, we have 
      \[
         \VCB(T(G))\leq Kr\leq d\;.
      \]
      Thus $T(G)$ satisfies the promise of $\mathcal A$ and has $N=3n+11$
      vertices.
      Run $\mathcal A$ on $T(G)$ and return its answer.
      Constructing $T(G)$ takes $\Order(n^2)$ time, so the resulting
      algorithm solves Triangle Detection \whp in time $\tilde\Order(n^2+R(N))$.
      By \Cref{lem:triangle-size-reduction}, we may first split the unrestricted
      Triangle Detection input into $\Order(1)$ instances on at most
      $\lfloor(n-11)/3\rfloor$ vertices and apply the reduction above to each.
      Every resulting call to $\mathcal A$ is then on a graph with at most $n$
      vertices, so the $R(N)$ term becomes $\Order(R(n))$.
      Summing over all $\tOrder(n^{\tau})$ instances produced by \Cref{lem:triangle-jr-free} gives a total time of 
      \[
         \tilde\Order\bigl(n^{\tau}(n^2 + R(n))\bigr).
      \]
      In particular, 
      \[
         R(n) = \tilde\Omega(T_{\mathrm{TD}}/n^{\tau}) - n^2.
      \]
   \end{proof}

    Note that in the proof above, we did not need the assumption from  \Cref{lem:triangle-jr-free} that Triangle Detection is hard even when the input graph has at most one triangle.
    This unique triangle property will be used in the next subsection, where we prove our lower bound for APSP.

\subsection{APSP in Bounded \VCB-Dimension}
   \label{sub:APSP}
   Here, we prove the following theorem. 
   \phantomsection\label{restatement:APSP-lb}
   \apspDistanceVCLowerBound*

   We use the construction from \Cref{def:tripartite-shortcut} and delete the
   edges between parts $B$ and $C$ to prove \Cref{thm:APSP-lb}. We first analyze
   distances in the modified graph and bound its \VCB-dimension.

   Let $G$ be an $n$-vertex tripartite graph with nonempty parts $A,B,C$
   and at most one triangle.
   This ``nonempty'' assumption is without loss of generality, since if a part is empty then $G$ has no triangle.
   Construct $T(G)$ as in
   \Cref{def:tripartite-shortcut}, and define the graphs
   \begin{equation}
   \label{eq:Ghat}
      \widehat G=T(G)-E_G(B,C),
      \qquad F=G[B\sqcup C].
   \end{equation}
   For each part $S\in\set{A,B,C}$ and subset $W\subseteq S$, we define
   \[Q(S)[W]=\set{q_w:w\in W}\subseteq Q(S).\]
   We use this notation in the following claims.

   \begin{claim}[Triangle Certificates in $\widehat G$]
      \label{claim:apsp-triangle-certificates}
      For every original edge $bc\in E_G(B,C)$,
      \begin{equation*}
         \dist_{\widehat G}(b,c)=2
         \quad\Longleftrightarrow\quad
         \text{$a,b,c$ form a triangle in $G$ for some $a\in A$.}
      \end{equation*}
   \end{claim}
   \begin{proof}
      Fix an original edge $bc\in E_G(B,C)$. By
      \Cref{claim:tripartite-triangles}, the common neighbors of $b$ and $c$
      in $T(G)$ are exactly the vertices $a\in A$ such that $a,b,c$
      form a triangle in $G$. Deleting $E_G(B,C)$ removes $bc$ but
      preserves every such path $b\leadsto a\leadsto c$. Thus, $\dist_{\widehat G}(b,c) = 2$ if and only
      if $bc$ belongs to a triangle in $G$. 
   \end{proof}

   \begin{claim}[Distances in $\widehat G$]
      \label{claim:apsp-distances}
      If $G$ is triangle-free, then for every $b\in B$ and $c\in C$,
      \begin{equation}
      \begin{aligned}
         N^{\widehat G}_2[b]\cap V_{\mathrm{base}}
            &=V_{\mathrm{base}}\setminus N^F(b),\\
         N^{\widehat G}_2[q_b]\cap V_{\mathrm{base}}
            &=V_{\mathrm{base}}\setminus N^F(b),\\
         N^{\widehat G}_2[c]\cap V_{\mathrm{base}}
            &=V_{\mathrm{base}}\setminus\bigl(N^F(c)\cup Q(B)[N^F(c)]\bigr).
      \end{aligned}
      \label{eq:apsp-radius-two}
      \end{equation}
      If $G$ has a unique triangle $abc$, the same formulas hold
      except for the balls centered at $b$ and $c$, which additionally
      contain $c$ and $b$ respectively. 
   \end{claim}
   \begin{proof}
      Suppose first that $G$ is triangle-free. 
      Fix $b\in B$.
      
      By \Cref{claim:tripartite-distances} applied to $X=B$, every pair of base vertices
      not coming from the pairs of parts $(B,C)$ and $(Q(B),C)$ is still at distance at most two in $\widehat{G}$.
      Hence $N^{\widehat G}_2[b]$ contains $V_{\mathrm{base}}\setminus C$.
      For every $c\in C\setminus N^F(b)$, the path $b\leadsto q_b\leadsto c$ shows that
      $c\in N^{\widehat G}_2[b]$.
      For every $c\in N^F(b)$, the edge $bc$ was deleted in $\widehat{G}$. 
      Since $G$ is
      triangle-free, \Cref{claim:apsp-triangle-certificates} implies that $b$
      and $c$ are not at distance two. Hence $c\notin N^{\widehat G}_2[b]$.
      This proves the first equality in \cref{eq:apsp-radius-two}.

      For the second equality, the observation from the first sentence of the previous paragraph shows that each ball $N^{\widehat G}_2[q_b]$ contains
      $V_{\mathrm{base}}\setminus C$. Moreover, $q_b$ is adjacent to every
      vertex of $C\setminus N^F(b)$ by definition.
      It remains to show that for every $c\in N^F(b)$, there is no path of
      length at most two from $q_b$ to $c$.
      Fix $c\in N^F(b)$. 
      By definition, $q_b$ and $c$ are nonadjacent.
      Since $C$ is independent, any common neighbor of these two nodes must lie outside $C$.
      The neighbors of $q_b$ outside $C$ are exactly
      \[
         \set{b,p_b,h_Q,h_{Q(B),P(A)},h_{Q(B),P(C)}}.
      \]
      None of these vertices has a neighbor in $N^F(b)$ in $\widehat G$,
      so $q_b$ and $c$ have no common neighbor.
      This proves the second equality in \cref{eq:apsp-radius-two}.
      
      The third equality in \cref{eq:apsp-radius-two} follows by symmetry of distances applied to
      the preceding arguments.
      That is, in the previous arguments we described exactly what nodes in $C$ are not within distance two in $\widehat{G}$ of each node in $B$ and $Q(B)$, and reversing the argument lets us characterize for each $c\in C$ what nodes in $B$ and $Q(B)$ are not within distance two of $c$ in $\widehat{G}$.

        This proves the first part of the claim.
        Now assume $G$ has a unique triangle $abc$.
        
      \Cref{claim:apsp-triangle-certificates} shows that among all edges in $E(F)$, only $bc$ has endpoints at distance two in $\widehat G$.
      Thus the first equality 
      from \cref{eq:apsp-radius-two}
      has only one exception: the ball centered at
      $b$ additionally contains node $c$. 
      The second equality from \cref{eq:apsp-radius-two} still holds because the argument establishing it did not rely on triangle-freeness.
      Finally, by symmetry of distances again, the third equality in \cref{eq:apsp-radius-two} has only one exception:
      the ball centered at $c$ additionally contains node $b$.
   \end{proof}

   \begin{claim}[Diameter in $\widehat{G}$]
      \label{claim:apsp-diameter}
      The graph $\widehat G$ has diameter at most three.
   \end{claim}
   \begin{proof}
      By \Cref{claim:tripartite-distances}, every radius-two ball centered
      in a base class other than $B$, $Q(B)$, or $C$ contains all of
      $V_{\mathrm{base}}$. Together with \Cref{claim:apsp-distances}, this
      implies that every pair of base vertices is at distance at most two
      unless it is of the form $(b,c)$ or $(q_b,c)$
      for some $bc\in E(F)$. For every such edge, the paths
      \[b\leadsto h_{Q(C),P(A)}\leadsto q_c\leadsto c \quad \text{and} \quad q_b\leadsto p_b\leadsto h_C\leadsto c\] 
      survive the deletion in $\widehat{G}$.
      Thus every two base vertices are at distance at most three.
      
      By \Cref{claim:tripartite-distances} the distance from hubs to all other vertices remains at most three even after the deletion producing $\widehat{G}$, so $\widehat G$ has diameter at most three as claimed. 
   \end{proof}

   \begin{lemma}[Distance \VC-dimension in $\widehat{G}$]
      \label{lem:apsp-balls}
      If $G$ contains at most one triangle and $\VC(G)\leq r$, then $\VCB(\widehat G)=\Order(r)$.
   \end{lemma}
   \begin{proof}
      Let  $D=\max(r,\VC(\widehat G))$. 
      By
      \Cref{lem:tripartite-vc} applied to the set $X=B$, we have $D=\Order(r)$. 
      Since $F$ is an induced subgraph of $G$,
      \Cref{claim:induced-subgraph-vc} gives $\VC(F)\leq \VC(G)\le r$.
      Set $k=\VCB(\widehat G)$, and fix a $k$-vertex set 
      $X\subseteq V(\widehat G)$ shattered by the closed balls of $\widehat{G}$. 
      We upper bound the number of distinct intersections they induce with $X$. 
      Write 
      \[s_t(k)=\sum_{i=0}^t\binom{k}{i}.\] 
      By similar reasoning to the proof of
      \Cref{lem:tripartite-balls}, radius-zero and radius-one balls from $\widehat{G}$
      induce at most $k+1$ and $(k+1)s_D(k)$ distinct intersections with $X$ respectively.

      Let $S$ be the set of vertices consisting of the eleven hubs and, if $G$ has a unique triangle
      $abc$, the vertices $b$ and $c$. 
      By construction, $|S|\leq13$.
      Put $X_{\mathrm{base}}=X\cap V_{\mathrm{base}}$.
      We first count the distinct intersections of radius-two balls
      centered outside $S$ with $X_{\mathrm{base}}$. We then extend this
      count to $X$ and account for centers in $S$.

      For each center $b\in B\setminus S$, the first equality from \Cref{claim:apsp-distances} proves that
      \[
         N^{\widehat G}_2[b]\cap X_{\mathrm{base}}
         =X_{\mathrm{base}}\setminus N^F(b).
      \]

      Similarly, for every $b\in B$ the second equality from \Cref{claim:apsp-distances} proves that
      \[
         N^{\widehat G}_2[q_b]\cap X_{\mathrm{base}}
         =X_{\mathrm{base}}\setminus N^F(b).
      \]

      Since $\VC(F)\leq r$, the subfamily $\set{N^F(b):b\in B}$ has $\VC$-dimension at most $r$.
      Then the Sauer--Shelah lemma \cref{eq:sauer-shelah-intro}
      shows that
      \[
         \bigl|\set{N^F(b)\cap X_{\mathrm{base}}:b\in B}\bigr|
         \leq s_r(|X_{\mathrm{base}}|) \leq s_r(k).
      \]
        Each distinct possible value for $N^F(b)\cap X_{\mathrm{base}}$ uniquely determines the value of $X_{\mathrm{base}}\setminus N^F(b)$,
        so from the earlier equations we see that radius-two balls centered at $b\in (B\cup Q(B))\setminus S$ contribute at most $s_r(k)$ distinct intersections with $X_{\mathrm{base}}$.
      
      For each center $c\in C\setminus S$, the third equality from \Cref{claim:apsp-distances}
      proves that
      \[
         N^{\widehat G}_2[c]\cap X_{\mathrm{base}}
         =X_{\mathrm{base}}\setminus
            \bigl(N^F(c)\cup Q(B)[N^F(c)]\bigr).
      \]
      A vertex $b\in X\cap B$ is excluded from this intersection exactly
      when $b\in N^F(c)$. A copy $q_b\in X \cap Q(B)$ is excluded under the same
      condition. 
      Thus the intersection from the right-hand side of the above equation is uniquely determined by the value of $N^F(c)\cap W$, for the set
      \[
         W=(X\cap B)\cup\set{b\in B:q_b\in X}.
      \]
      Since \[|W|\leq |X\cap B|+|X\cap Q(B)|\leq |X| = k\] 
      and $\VC(F)\leq r$,
      the Sauer--Shelah lemma \cref{eq:sauer-shelah-intro} again gives
      \[
         \bigl|\set{N^F(c)\cap W:c\in C}\bigr|
         \leq s_r(|W|)\leq s_r(k).
      \]
      Consequently, the radius-two balls centered at $c\in C\setminus S$ also contribute at most
      $s_r(k)$ distinct intersections with $X_{\mathrm{base}}$.

      By \Cref{claim:tripartite-distances}, radius-two
      balls centered outside $B\cup Q(B)\cup C$ contain all of
      $V_{\mathrm{base}}$ and therefore contribute at most one new intersection with
      $X_{\mathrm{base}}$. 
      Combining all these bounds gives
      \[
         \bigl|\set{N^{\widehat G}_2[v]\cap X_{\mathrm{base}}
            :v\in V(\widehat G)\setminus S}\bigr|
         \leq 2s_r(k)+1.
      \]

      We now count intersections with the full set $X$. 
      Since
      $X\setminus X_{\mathrm{base}}$ contains at most eleven hubs, each
      intersection counted by the above inequality extends to at most $2^{11}$ distinct
      intersections with $X$. 
      Also, the radius-two balls centered in $S$
      contribute at most $|S|\leq13$ additional intersections with $X$.
      By \Cref{claim:apsp-diameter}, $\widehat{G}$ has diameter at most three, so all balls of radius at least three
      contain $X$ and thus contribute at most one additional intersection. 
      Since $X$ is shattered, these bounds imply that
      \[
         2^k\leq
         (k+1)+(k+1)s_D(k)+2^{11}\bigl(2s_r(k)+1\bigr)+14.
      \]
      Since $D\geq r$, we have $s_r(k)\leq s_D(k)$, so the above inequality simplifies to
        \[2^k \le (k+1) + (k+2^{11} + 2)s_D(k) + 14.\]
    Since $D\ge 1$, we know that $s_D(k) \ge k+1$.
    Thus we can simplify the above inequality to
        \[2^k \le M\left(s_D(k)\right)^2\]
    for some constant $M \ge 1$.
    By applying \Cref{lem:vc-silly-bound} with $S = s_D(k)$, we deduce that \[k = \Order(D) = \Order(r).\]

          Thus every shattered set has size $\Order(r)$,
      which proves that $\VCB(\widehat{G})=\Order(r)$ as desired.
   \end{proof}

   We are now ready to prove our lower bound for APSP.

   \begin{proof}[Proof of \Cref{thm:APSP-lb}]
      By \Cref{lem:apsp-balls}, there is an absolute constant
      $K\geq1$ such that 
      \[\VCB(\widehat G)\leq Kr\] whenever $G$ is
      tripartite, contains at most one triangle, and
      satisfies $\VC(G)\leq r$.

      Fix an integer $d\geq 3K$, and define
      \[
         r=r(d)=\left\lfloor\frac d{K}\right\rfloor\geq3.
      \]

      Applying \Cref{lem:triangle-jr-free} reduces unrestricted Triangle
      Detection to Triangle Detection in tripartite $J_r$-free graphs
      containing at most one triangle, producing $\tilde\Order(n^{\tau})$ instances with 
      an overhead of $\tilde\Order(n^{2+\tau})$, where 
      $\tau = \Order(1/r) = \Order(1/d)$. We reduce the latter problem to APSP on graphs of bounded \VCB-dimension.

      Let $\mathcal A$ be an algorithm that solves APSP in graphs with \VCB-dimension
      at most $d$ \whp, and let $R(N)$ be its worst-case running time on such graphs
      with at most $N$ vertices.

      Let $G$ be an $n$-vertex tripartite $J_r$-free graph containing at most 
      one triangle, with parts $A,B,C$. 
      Construct $\widehat G$ from $G$ as defined in \cref{eq:Ghat}. 
      By \Cref{lem:jr-free-vc}, $\VC(G)\leq r$.
      Therefore by \Cref{lem:apsp-balls}
      \[
         \VCB(\widehat G)\leq Kr\leq d.
      \]

      Run $\mathcal A$ on $\widehat G$, which has $N = 3n+11$ vertices, and scan
      the original edges $bc\in E_G(B,C)$. Return YES if and only if
      $\dist_{\widehat G}(b,c)=2$
      for some $bc \in E_G(B,C)$. 
      By \Cref{claim:apsp-triangle-certificates}, 
      whenever $\mathcal A$ returns the correct distances, this test correctly determines whether $G$ contains a triangle. 
      Thus the resulting algorithm succeeds \whp 
      
      The construction and final scan take $\Order(n^2)$ time, so the
      reduction runs in time $\tilde \Order( n^2 + R(N) )$.
      By \Cref{lem:triangle-size-reduction}, we may first split the unrestricted Triangle Detection input into
      $\Order(1)$ instances on at most $\lfloor(n-11)/3\rfloor$ vertices, so each resulting APSP call is on at most
      $n$ vertices and the $R(N)$ term in the reduction becomes $\Order(R(n))$.
      Since the output size is $n^2$, we have
      $R(n) \geq n^2$, so the quadratic overhead is absorbed into $R(n)$.
      Thus, summing over all instances of \Cref{lem:triangle-jr-free} gives a running time of 
      $\tOrder(n^\tau R(n))$ for unrestricted Triangle Detection. This gives 
      \[
         R(n)\geq\tilde\Omega(T_{\mathrm{TD}}/n^\tau).
      \]
   \end{proof}

\subsection{OMv in Bounded \VCB-Dimension}
   \label{sub:omv-dist}
   In this subsection, we prove \Cref{thm:omv-vc-to-dvc}. 

    Let $\mathbf 1$ be the $n$-dimensional all-ones vector.
   For an $n\times n$ binary matrix $M$, define its \emph{padded matrix} to be the $(n+1)\times (n+1)$ matrix of the form
   \begin{equation}
   \label{eq:padding}
      M'=
      \begin{pmatrix}
         M & \mathbf 1\\
         \mathbf 1^{\mathsf T} & 1
      \end{pmatrix}.
   \end{equation}
   Thus, $M'$ is obtained by appending an all-ones row and an all-ones column to $M$.
   This simple transformation turns out to be enough to relate  distance \VC-dimension to  standard \VC-dimension. 

   \begin{lemma}[Distance \VC-dimension for Padding]
      \label{lem:dist-vc-padding}
      The padded matrix $M'$ satisfies
      \[
         \VC(M)\leq\VCB(M')\leq\VC(M)+2.
      \]
   \end{lemma}
   \begin{proof}
      Let $d=\VC(M)$.
      Let $R_i=\supp(M[i,\cdot])\subseteq[n]$. Over the Boolean semiring,
      $(M')^r$ is the all-ones matrix for every $r\geq2$.
      This is because a walk of length $r$
      from any index to any other index can go through $n+1$ and use the loop at
      $n+1$ as needed. 
      Thus
       \begin{equation}
          \mathcal B(M')=
          \underbrace{\set[\big]{\set{j}\bigm| j\in[n+1]}}_{\text{rows of }(M')^0}
          \cup\underbrace{\set[\big]{R_i\cup\set{n+1}\bigm| i\in[n]}}_{\text{rows of }(M')^1}
          \cup\underbrace{\set[\big]{[n+1]}}_{\substack{\text{rows of }\\(M')^r\text{, } r\geq 2}}.
          \label{eq:dist-vc-padding-family}
       \end{equation}

      Since $(R_i\cup\set{n+1})\cap[n]=R_i$, every subset of $[n]$ shattered by the rows of
      $M$ remains shattered by $\mathcal B(M')$. Thus,
      $\VC(M)\leq\VCB(M')$.

      We claim that this family has \VC-dimension at most $d+2$. 
      
      Assume for contradiction that there exists a shattered set $S \subseteq [n+1]$ of size $d+3$. 
      We start by showing that $(n+1)\notin S$. 
      
      Indeed, if $(n+1)\in S$, pick two distinct elements $a,b \in (S \setminus \set{n+1})$.
      Such elements must exist since $|S| \geq 3$. 
      Observe that by \cref{eq:dist-vc-padding-family}, no support $B$ of a row of $(M')^0$ satisfies 
      $B \cap S = \set{a,b}$, since every such support is a singleton. Likewise, no support of a row of $(M')^1$ or $(M')^{r}$, $r \geq 2$,
      satisfies $B \cap S = \set{a,b}$, since every such support contains the element $(n+1)$. This contradicts our assumption that $\mathcal{B}(M')$ 
      shatters $S$.

      Thus we may assume $(n+1)\not\in S$.
      We next show that the rows of $M$ shatter a set of size $d+1$, which will yield the desired contradiction.
      
      For arbitrary distinct $a,b \in S$, consider the set $X = S \setminus \set{a,b} \subseteq [n]$. 
      Let $Y \subseteq X$. 
      
      Suppose that $Y \neq \emptyset$. 
      We show that there exists a row in $M$ such that its support 
      $R_i$ satisfies $R_i \cap X = Y$. 
      Indeed, since $\mathcal{B}(M')$ shatters $S$, there exists $B \in \mathcal{B}(M')$ such that $B \cap S = Y \cup \set{a}$. 
      By \cref{eq:dist-vc-padding-family}, $B$ is not the support of a row of $(M')^0$ (as $|B| \geq 2$), 
      nor the support of a row of $(M')^{r}$ for $r \geq 2$ (as $b \notin B$). Thus, $B$ must take the form 
      $R_i \cup \set{n+1}$. Since $(n+1) \notin S$, we have $R_i \cap S = Y \cup \set{a}$. Since $a \notin X$ and 
      $Y \subseteq X$, we have $R_i \cap X = Y$ as desired.

      So suppose instead that $Y = \emptyset$. There exists $B \in \mathcal{B}(M')$ 
      such that $B \cap S = \set{a,b}$ and therefore $B \cap X = \emptyset$. 
      Since $B\cap S=\set{a,b}$, we have $|B|\geq2$, so $B$ is not the
      support of a row of $(M')^0$. Moreover, since $|S|\geq3$, there exists
      $c\in S\setminus\set{a,b}$. Then $c\notin B$, so $B\neq[n+1]$ and
      therefore $B$ is not the support of a row of $(M')^r$ for any $r\geq2$.
      \cref{eq:dist-vc-padding-family} therefore implies that $B = R_i \cup \set{n+1}$ for some $i \in [n]$. 
      Since $B \cap X = \emptyset$ and $B = R_i \cup \set{n+1}$, it follows that $R_i \cap X = \emptyset = Y$.
      
      Thus, the set $X$ of size
        \[|X| = |S| - 2 = (d+3) - 2 = d+1\]
    is shattered by the supports of the rows of $M$, and therefore $\VC(M) \geq d+1$, which  contradicts the definition of $d$. 
    Thus our assumption was false, and $\VCB(M')\le d+2$ as claimed.
   \end{proof}

We can now reduce OMv on bounded \VC-dimension to bounded distance \VC-dimension.

   \begin{proof}[Proof of \Cref{thm:omv-vc-to-dvc}]
      Let $M\in\set{0,1}^{n\times n}$ have \VC-dimension at most $d$, and form
      its padded matrix $M'$ as in \cref{eq:padding}. 
      By \Cref{lem:dist-vc-padding},
      $\VCB(M')\leq d+2$. Constructing $M'$ takes $\Order(n^2)$ time, after which we preprocess an $(n+1)\times (n+1)$ matrix using an  algorithm for OMv.

      Given a query
      $v\in\set{0,1}^n$, make a query $v'=(v,0)$ to the algorithm for $M'$. Notice 
      that the first $n$ coordinates of $M'v'$ are exactly $Mv$. 
      Discard the last coordinate. 
      Padding the query vectors and trimming
      their answers costs $\Order(n^2)$ total time. 
      The remaining runtime comes from running OMv queries on $(n+1)$-dimensional vectors.
      
      Thus with $O(n^2)$ additive overhead, we have reduced OMv on $n\times n$ matrices of $\VC$-dimension at most $d$ to OMv on $(n+1)\times (n+1)$ matrices of distance $\VC$-dimension at most $d+2$, as desired.
   \end{proof}

   \section{Triangle Detection in Induced \texorpdfstring{$H$}{H}-Free Graphs for Bipartite Patterns} 
      \label{sec:induced}
      In this section, we prove \Cref{thm:subcubic-bipartite} via the following
bound in terms of the bipartition sizes.

\begin{theorem}[Forbidding a Bipartite Pattern]
   \label{thm:subcubic-bipartite-parts}
   Let $H$ be a bipartite graph on $k\geq3$ vertices with bipartition
   $L_H\sqcup R_H$, where $1\leq |L_H|=\ell\leq r=|R_H|$.
   There is a randomized algorithm for Triangle Detection in $n$-vertex
   induced $H$-free graphs that succeeds \whp and runs in time
   \[
      \tOrder\left(n^{3-\frac{1}{2(\ell+\lceil\log_2 r\rceil-1)}}\right).
   \]
\end{theorem}

We  employ the following rectangular result of Anand et al.~{\cite[Theorem 1.3]{anand2026structural}}.
   \begin{proposition}[$\VC$-dimension Multiplication]
      \label{thm:omv-ub-rec}
      Let $M \in \set{0,1}^{m \times n}$ be a matrix of \VC-dimension at most $d$.
      We can preprocess $M$ in 
      $\tOrder(mn)$ time to support queries that take a vector
      $v \in \set{0,1}^n$ and output $Mv$. After preprocessing, \whp every
      query is answered in $\tOrder(nm^{1-1/d}+m)$ time.\footnote{The probability is over the
      construction of the data structure, not separately over the queries:
      conditioned on a successful construction, the guarantee holds
      simultaneously for all queries. The original formulation says
      ``with high probability''; here we use the convention in
      \Cref{sec:preliminaries}.}
   \end{proposition}

We also use the following result from~\cite[Theorem 6]{abboud2026equivalent}. 
\begin{theorem}[$H$-free Color-Coding]
   \label{lem:color-coding}
   There is a deterministic algorithm that, given an $n$-vertex graph $G$, runs in
   $\Order(n^{2.5})$ time and either finds a triangle in $G$ or outputs
   $\Order(n^{3/2})$ instances of Triangle Detection. Each instance
   is a properly $3$-colored induced subgraph of $G$ with $\Order(\sqrt{n})$ vertices in each color class, and
   every triangle in $G$ is present in exactly one instance.
\end{theorem}
Crucially, this reduction preserves induced $H$-freeness.

For a bipartite graph $H$, it is well known that if a bipartite graph $G$ is induced $H$-free, then
the \VC-dimension of $G$ is bounded. For example, an implicit bound follows
from~\cite[Theorem 1.1]{anand2026structural}. An explicit bound can also be derived from the proof
of~\cite[Lemma 3.4]{bousquet2015identifying}.
We include a short proof giving an explicit bound on the \VC-dimension.
\begin{lemma}[Forbidden Subgraph $\Rightarrow$ Low \VC-dimension]
   \label{lem:VC-bound}
   Let $H$ be a graph satisfying the conditions of \Cref{thm:subcubic-bipartite-parts}.
   Let $G$ be an induced $H$-free bipartite graph with bipartition $L_G\sqcup R_G$. 
   Then the set systems 
   \[\set{N(u) \mid u \in L_G}\quad\text{and}\quad\set{N(v) \mid v \in R_G}\] 
   have \VC-dimension
   at most $\ell + \lceil \log r \rceil - 1$.
\end{lemma}
\begin{proof}
   Suppose for the sake of contradiction that the set system $\set{N(u) \mid u \in L_G}$
   shatters a set $S \subseteq R_G$ of size $d = \ell + \lceil \log r \rceil$.
   We will embed the smaller part $L_H$ of $H$ into $S$, and the larger part $R_H$ of $H$ into $L_G$.
   The difficulty is that several vertices in $R_H$ may have the same neighborhood in $L_H$,
   whereas shattering merely guarantees a single realization of each subset of $S$.
   The additive slack of $\lceil \log r \rceil$ provides sufficiently many vertices
   that share the same neighborhood in a subset of $S$ to overcome this issue.

   Choose an arbitrary subset $A \subseteq S$ of size $\ell$ and identify $A$ with $L_H$.
   Let $A' \in \set{ N(v) \mid v \in R_H}$ be an arbitrary neighborhood in $H$.
   We count the number of vertices in $L_G$ whose neighborhood in $A$ is exactly $A'$.
   
   Since $S$ is shattered, for every $X \subseteq (S \setminus A)$, there exists a vertex $u \in L_G$ 
   such that 
   \[N(u) \cap S = A' \sqcup X.\] 
   In particular, there are at least 
$2^{|S \setminus A|} \geq 2^{\lceil \log r \rceil} \geq r$
   distinct vertices in $L_G$ whose neighborhood in $A$ is exactly $A'$. 
   This allows us to embed an induced copy of $H$ into $G$ as follows: For each $v \in R_H$, embed $v$ 
   into a distinct vertex in $L_G$ whose neighborhood in $A$ is exactly $N(v)$. Because there are at least $r$ such vertices,
   we never run out of vertices inside $L_G$. 
   Because $G$ is bipartite, there are no edges within the images of $L_H$ and $R_H$. 
   This implies that the embedding gives an induced copy of $H$ and contradicts the assumption that $G$ is induced $H$-free.
   Thus $\set{N(u) \mid u \in L_G}$ has \VC-dimension at most $\ell+\lceil\log r\rceil - 1$ as claimed.
   
   Symmetric reasoning shows that the same bound applies to $\set{N(v) \mid v \in R_G}$. 
\end{proof}

The next lemma gives an algorithm for Triangle Detection in tripartite graphs in which one of the
bipartite subgraphs between two parts has bounded \VC-dimension.
\begin{lemma}
   \label{lem:tripartite-vcd}
   Take an integer $d\geq 1$, and let $G$ be a tripartite graph with parts $A,B,C$, each of size at most $n$.
   If the bipartite adjacency matrix between $A$ and $B$ has \VC-dimension at most $d$,
   then Triangle Detection on $G$ can be solved by a randomized algorithm in
   $\tOrder(n^{3-1/d})$ time \whp
\end{lemma}
\begin{proof}
   To find a triangle, it suffices to determine whether there is a vertex $v \in C$ with adjacent neighbors
   $x \in A$ and $y \in B$. 
   Set $n_A=|A|$, $n_B=|B|$, and $n_C=|C|$, and let $M$ be the
   $n_A\times n_B$ bi-adjacency matrix between $A$ and $B$.
   By assumption, the \VC-dimension of $M$ is at most $d$.

   For each $v \in C$, we define $q_v \in \{0, 1\}^{n_B}$ as the indicator vector of its neighborhood
   $N(v) \cap B$.
   We compute the matrix-vector product $w=Mq_v$ over the Boolean semiring. Thus, for every $x \in A$, 
   $w_x=1$ if and only if $x$ and $v$ have a common neighbor in $B$. We check whether this holds for
   some $x\in N(v)\cap A$; if so, we have found a triangle $(x,y,v)$ for some $y\in B$.

   By~\Cref{thm:omv-ub-rec}, we can preprocess $M$ in $\tOrder(n_A n_B)$ time so that \whp
   we can answer each of the $n_C$ queries, one for each $v\in C$, in
   $\tOrder(n_B n_A^{1-1/d}+n_A)$ time. On this event, the total running time is
   \[
      \tOrder\bigl(n_A n_B+n_C(n_Bn_A^{1-1/d}+n_A)\bigr)
      =\tOrder(n^{3-1/d})
   \]
   since $n_A,n_B,n_C\le n$.
\end{proof}

We now prove the bipartition-sensitive bound and derive \Cref{thm:subcubic-bipartite}.
\begin{proof}[Proof of \Cref{thm:subcubic-bipartite-parts,thm:subcubic-bipartite}]
   Let $G$ be an $n$-vertex graph that is induced $H$-free.
   We apply ~\Cref{lem:color-coding}, which either finds a triangle in $G$, 
   or outputs $\Order(n^{3/2})$ instances
   of Triangle Detection, each of which is a properly $3$-colored induced subgraph of
   $G$ with $\Order(\sqrt{n})$ vertices in each color class.
   This reduction takes $\Order(n^{2.5})$ time. If a triangle is found, we output it and terminate. 
   
   Otherwise, consider one properly $3$-colored instance with color classes
   $A$, $B$, and $C$, and set 
   \[N = \max\set{|A|,|B|,|C|} \le \Order(\sqrt{n}).\]
   Since the instance is an induced subgraph of $G$, it remains induced $H$-free.
   Since the coloring is proper, $A$ and $B$ are independent sets.
   Thus, the subgraph induced by $A \cup B$ is bipartite and induced $H$-free.
   By~\Cref{lem:VC-bound}, the \VC-dimension of the bipartite adjacency matrix between $A$ and $B$ is
   at most $d = \ell + \lceil \log r \rceil - 1$.
   By~\Cref{lem:tripartite-vcd}, we can determine whether this instance
   contains a triangle in $\tOrder(N^{3-1/d})$ time \whp
   By~\Cref{lem:standard-amplification}, for each instance, we can amplify the success probability of this routine to 
   at least $1-n^{-2}$, with only logarithmic overhead in the runtime. 
   A union bound over the $\Order(n^{3/2})$ instances shows that, with probability at least
   \[1-\Order(n^{-1/2}) \geq 2/3\] 
   the algorithm runs in $\tOrder(N^{3-1/d})$ time for every
   instance and outputs the correct answer.
   Summing over all $\Order(n^{3/2})$ instances, the total running time for evaluating them is
   \[
       \Order(n^{3/2}) \cdot \tOrder(N^{3-1/d}) = \tOrder\left( n^{3/2} \cdot (\sqrt{n})^{3-1/d} \right) = \tOrder\left( n^{3 - \frac{1}{2d}} \right).
   \]
   Substituting $d = \ell + \lceil \log r \rceil - 1$, the overall running time, including the reduction step, is
   \[
       \tOrder\left( n^{3 - \frac{1}{2(\ell + \lceil \log r \rceil - 1)}} + n^{2.5} \right).
   \]
   Since $\ell+r=k\geq3$ and $1\leq\ell\leq r$, we have $d\geq1$.
   Thus the first term dominates, so the bound becomes
   \[
       \tOrder\left(n^{3-\frac{1}{2(\ell+\lceil\log r\rceil-1)}}\right).
   \]
   This proves \Cref{thm:subcubic-bipartite-parts}.
   Finally, since $\ell+r=k$ and $\ell\leq r$, we have
   \[d=\ell+\lceil\log_2 r\rceil-1\leq k/2+\log k-1.\]
   Hence the running time is
   \[
       \tOrder\left(n^{3-\frac{1}{k+2\log k-2}}\right),
   \]
   proving \Cref{thm:subcubic-bipartite}.
\end{proof}

\subsection{Triangle Detection on Induced \texorpdfstring{$P_k$}{Pk}-Free Graphs}
   In this subsection, we prove \Cref{thm:paths}. 
   Although the previous algorithm 
   of \Cref{thm:subcubic-bipartite}
   applies to paths because they are bipartite, it incurs
   two losses. First, the general \VC-dimension bound in \Cref{lem:VC-bound} contains a
   logarithmic additive term that is unnecessary for $P_k$-free graphs because the relevant
   neighborhoods are all distinct. The next lemma gives an improved bound.

   \begin{lemma}[Forbidden Path $\Rightarrow$ Low \VC-dimension]
      \label{lem:VC-paths}
      Take an integer $k\geq 4$, and let $G$ be an induced $P_k$-free bipartite graph with bipartition $U \sqcup V$.
      Then, the set systems $\set{N(u) \mid u \in U}$ and $\set{N(v) \mid v \in V}$ have \VC-dimension at most $\lfloor
      k/2 \rfloor - 1$.
   \end{lemma}
   \begin{proof}
      The path $P_k$ is a bipartite graph with parts $L$ and $R$ of sizes $\ell = \lfloor k/2 \rfloor$
      and $r = \lceil k/2 \rceil$, respectively.
      A key property of $P_k$ is that the neighborhoods of the $r$ vertices of $R$
       are all distinct subsets of $L$. 
      Suppose for the sake of contradiction that the set system $\set{N(u) \mid u \in U}$
      shatters a subset $S \subseteq V$ of size $\ell = \lfloor k/2 \rfloor$.
      We identify $L$ with $S$. 
      The $r$ vertices of $R$ require $r$ distinct subsets of $S$ as their neighborhoods.
      Since $S$ is shattered, there exist $r$ distinct vertices in $U$ that realize exactly these
      required neighborhoods in $S$. We map each vertex $v \in R$ to a vertex of $U$ realizing $N(v)$. 
      Because $G$ is bipartite, the $\ell$ vertices in $S$ and the $r$ chosen vertices in $U$
      form an exact induced copy of $P_k$, which is a contradiction. 
      A symmetric argument applies to $\set{N(v) \mid v \in V}$. Thus, the \VC-dimension must be
      strictly less than $\lfloor k/2 \rfloor$. 
   \end{proof}

   The second loss of the general bipartite pattern algorithm is in the color-coding step, as the color-coding of~\cite{abboud2026equivalent}
   reduces the problem to $\Order(n^{3/2})$ instances of size $\Order(\sqrt{n})$. We next show that induced
   $P_k$-free graphs admit a more efficient color-coding scheme that produces
   $\Order(\log^3 n)$ properly $3$-colored instances. The key idea is to use a vertex separator that can be easily colored with 
   few colors, and then recursively find a coloring with $\Order(\log n)$ colors. 

   The following lemma specializes the argument of~\cite[Lemma 2]{bacso2019subexponential}, which is a variation
   of the classic path-growing argument by Gy{\'a}rf{\'a}s~\cite[Theorem 2.4]{gyarfas1987problems}. We include its proof here.

   A connected component of an $n$-vertex graph is called \emph{big} if it has more than $n/2$ vertices.
   \begin{lemma}[Small Neighborhood Separator]
      \label{lem:path-separator}
      Fix $k \geq 2$ and let $G$ be a connected induced $P_k$-free graph with $n$ vertices and
      $m$ edges. In $\Order(m+n)$ time, we can find a set $X$ with $|X| \leq k-2$ such that every connected component of
      $G-N(X)$ has at most $n/2$ vertices.
   \end{lemma}

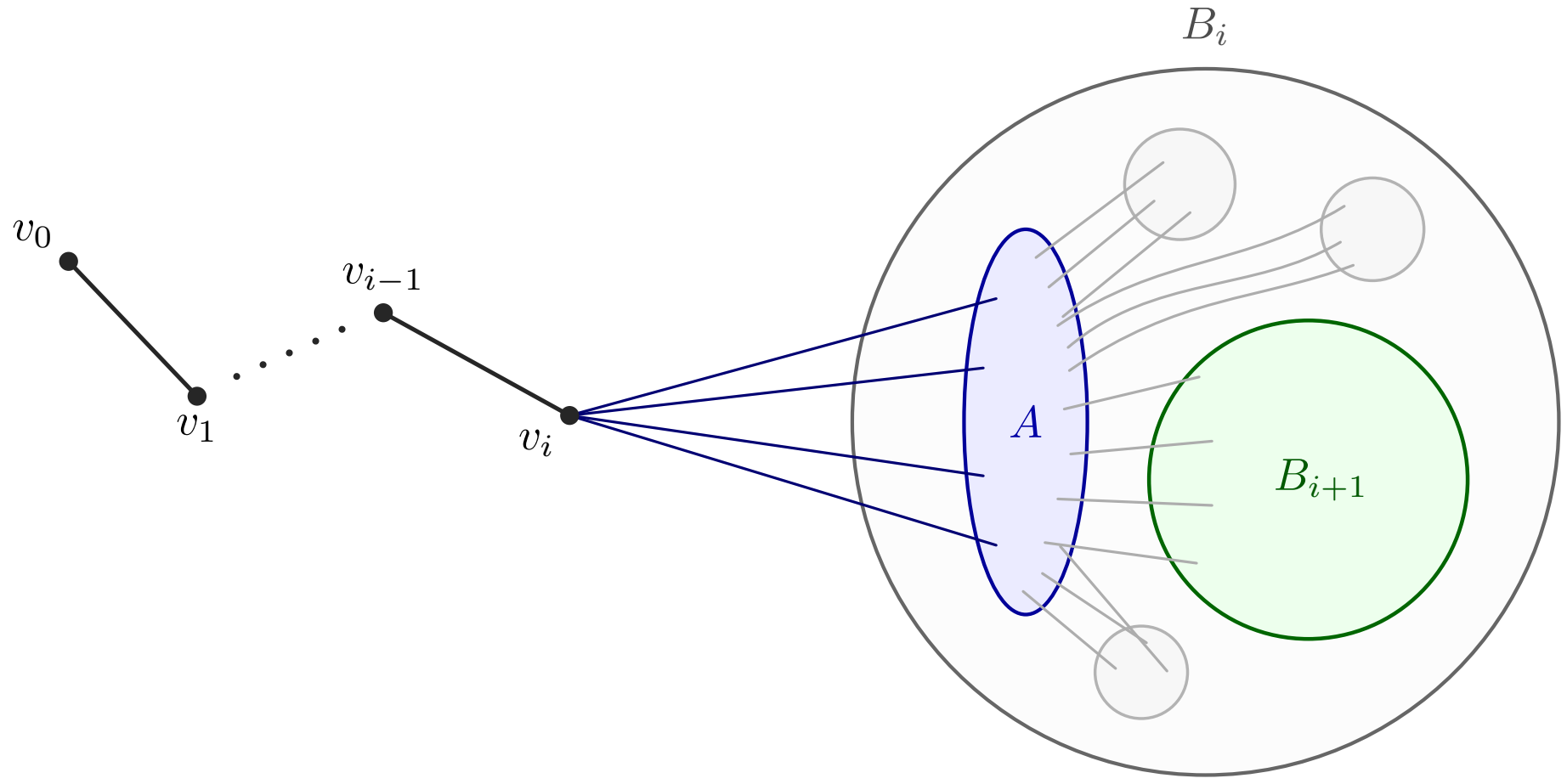
\begin{figure}[htbp]
      \centering
      \begin{tikzpicture}[
   every node/.style={font=\small},
   path edge/.style={draw=black!85, line width=.95pt, line cap=round},
   neighborhood edge/.style={draw=blue!45!black, line width=.6pt, line cap=round},
   component edge/.style={draw=black!32, line width=.6pt, line cap=round},
   component/.style={draw=black!30, fill=black!3, line width=.65pt}
]
   \draw[draw=black!60, fill=black!1, line width=.8pt]
      (4.45,0) circle (2.75cm);
   \draw[draw=blue!60!black, fill=blue!8, line width=.8pt]
      (3.05,0) ellipse (.48cm and 1.50cm);
   \draw[draw=green!40!black, fill=green!7, line width=.85pt]
      (5.25,-.45) circle (1.24cm);
   \draw[component] (4.25,1.85) circle (.43cm);
   \draw[component] (5.75,1.50) circle (.40cm);
   \draw[component] (3.95,-1.95) circle (.36cm);

   \draw[component edge] (3.23,1.05) -- (4.05,1.72);
   \draw[component edge] (3.34,.82) -- (4.33,1.63);
   \draw[component edge] (3.13,1.28) -- (4.12,2.02);

   \draw[component edge] (3.38,.58)
      .. controls (4.05,1.15) and (4.82,1.00) .. (5.50,1.40);
   \draw[component edge] (3.39,.40)
      .. controls (4.25,.99) and (4.85,.94) .. (5.60,1.22);
   \draw[component edge] (3.30,.75)
      .. controls (4.10,1.30) and (4.75,1.20) .. (5.53,1.68);

   \draw[component edge] (3.35,.10) -- (4.40,.35);
   \draw[component edge] (3.40,-.25) -- (4.50,-.15);
   \draw[component edge] (3.30,-.60) -- (4.50,-.65);
   \draw[component edge] (3.20,-.94) -- (4.38,-1.10);

   \draw[component edge] (3.03,-1.32) -- (3.75,-1.92);
   \draw[component edge] (3.18,-1.18) -- (3.99,-1.72);
   \draw[component edge] (3.32,-.97) -- (4.15,-1.94);

   \coordinate (vzero) at (-4.40,1.25);
   \coordinate (vone) at (-3.40,.20);
   \coordinate (vprevious) at (-1.95,.85);
   \coordinate (vcurrent) at (-.50,.05);
   \draw[path edge] (vzero) -- (vone);
   \node[font=\Large, text=black!85, rotate=24.1, inner sep=0pt]
      at (-2.675,.525) {$\mathinner{\cdotp\cdotp\cdotp\cdotp\cdotp}$};
   \draw[path edge] (vprevious) -- (vcurrent);

   \draw[neighborhood edge] (vcurrent) -- (2.82,.96);
   \draw[neighborhood edge] (vcurrent) -- (2.72,.42);
   \draw[neighborhood edge] (vcurrent) -- (2.72,-.42);
   \draw[neighborhood edge] (vcurrent) -- (2.82,-.96);

   \foreach \v in {vzero,vone,vprevious,vcurrent}
      \fill[black!85] (\v) circle (2.1pt);
   \node[above left, inner sep=3pt] at (vzero) {$v_0$};
   \node[below, inner sep=4pt] at (vone) {$v_1$};
   \node[above, inner sep=4pt] at (vprevious) {$v_{i-1}$};
   \node[below left, inner sep=3pt] at (vcurrent) {$v_i$};

   \node[above, inner sep=5pt, text=black!70] at (4.45,2.75) {$B_i$};
   \node[text=blue!65!black] at (3.05,0) {$A$};
   \node[text=green!35!black] at (5.35,-.45) {$B_{i+1}$};
\end{tikzpicture}
      \caption{An extension step in the path-growing argument. Removing $A$ creates several 
      connected components in $B_i$, including the big component $B_{i+1}$. We extend the path 
   by choosing $v_{i+1} \in A$ that has a neighbor in $B_{i+1}$.}
      \label{fig:path-growing}
   \end{figure}

   \begin{proof}
      Our goal is to grow an induced path $v_0,v_1,\ldots$ until the open neighborhood of the set
      $X_i := \set{v_0,\ldots,v_i}$ is a good separator, meaning that $G-N(X_i)$ has no big component.
      Since $G$ is induced $P_k$-free, this separator will be found after selecting at most $k-2$ vertices.

      For every $i\geq0$, let $B_i$ be the unique big component of
      $G-N(X_{i-1})$, whenever such a component exists, with $B_0=V(G)$.
      Observe that the big components are nested, meaning 
      \[B_0\supseteq B_1\supseteq B_2\supseteq\cdots\] 
      because a big component after further deletions 
      must lie in the previous big component.
      We maintain the invariant that $v_i$ has a neighbor in $B_i$ for every $i\geq0$, 
      to guarantee that we can keep extending the path into the next big component. 

      Choose an arbitrary vertex $v_0$. Since $G$ is connected, $v_0$ has a neighbor in $B_0$, 
      so the invariant holds initially.

      At step $i\geq0$, if $N(X_i)$ is a good separator, return $X=X_i$. Otherwise, a big component $B_{i+1}$ exists. 
      The set $A=N(v_i)\cap B_i$ is nonempty by the invariant, and deleting $N(X_i)$ removes exactly $A$ from $B_i$ 
      since $B_i$ is a subgraph of $G-N(X_{i-1})$.
      By nesting, $B_{i+1}$ is the big component of $G[B_i] - A$. 
      Since $B_i$ is connected, there is an edge between $A$ and $B_{i+1}$.
      This guarantees the existence of a vertex $v_{i+1} \in A$ which preserves the invariant. 

      Clearly, $v_{i+1}$ is adjacent to $v_i$ since $A \subseteq N(v_i)$, so the vertices form a path.
      To see why the path is induced, 
      observe that $v_{i+1}\in B_i$, which is disjoint from $N(X_{i-1})$ by definition, so it has no neighbors among
      $v_0,\ldots,v_{i-1}$. Thus the extended path is induced.
      
      \Cref{fig:path-growing} illustrates one extension step in this argument.

      The invariant supplies a neighbor of $v_i$ in $B_i$, which can be appended to form an induced path
      on $|X_i|+1$ vertices. Since $G$ is induced $P_k$-free, $|X_i|+1<k$.
      As the path can be extended whenever the separator condition fails, the process must stop
      with $|X|\leq k-2$.

      Each iteration takes $\Order(m+n)$ time to mark the deleted neighborhoods, compute the remaining
      components, and find a suitable boundary edge. There are at most $k-2$ iterations, so the total
      running time is $\Order(k(m+n))=\Order(m+n)$ for constant $k$. 
   \end{proof}

   Next, we show how to color $N(X)$ with $|X|$ colors unless we first find a triangle in $G$.

   \begin{lemma}[Coloring Neighborhoods]
      \label{lem:path-neighborhood-coloring}
      Let $G$ be a graph with $n$ vertices and $m$ edges, and let $X\subseteq V(G)$ be nonempty.
      In $\Order(|X|(m+n))$ time, we can either find a triangle in $G$ or compute a proper coloring of
      $G[N(X)]$ with at most $|X|$ colors.
   \end{lemma}
   \begin{proof}
      For each $x\in X$, mark $N(x)$ and scan its vertices' adjacency lists to test whether
      $N(x)$ contains an edge. If we find such an edge, it forms a triangle with $x$, and we return it.
      Otherwise, we have certified that $N(x)$ is independent for all $x\in X$.

      Color each vertex of $N(X)$ by one of its neighbors in $X$. Each color class lies in an
      independent set $N(x)$, so this is a proper coloring using at most $|X|$ colors.
      Each neighborhood test takes $\Order(m+n)$ time, and the colors can be assigned by scanning
      the neighborhoods. The total running time is $\Order(|X|(m+n))$.
   \end{proof}

   We now apply the separators and colorings recursively to color the whole graph. 
   \begin{lemma}[Color-Coding $P_k$-free Graphs]
      \label{lem:path-coloring}
      For every fixed $k\geq2$, there is a $\tOrder(m+n)$-time algorithm that, given an induced
      $P_k$-free graph $G$, either finds a triangle in $G$ or computes a proper coloring of $G$ with
      $\Order(\log n)$ colors.
   \end{lemma}
   \begin{proof}
      We may assume that $G$ is connected, since the same colors can be reused across different connected components.
      Apply \Cref{lem:path-separator} to obtain a set $X\subseteq V(G)$ with $|X|\leq k-2$ such that
      every component of $G-N(X)$ has at most $n/2$ vertices.
      Apply \Cref{lem:path-neighborhood-coloring} to $G$ and $X$.
      If a triangle is found, return it. Otherwise, properly color $G[N(X)]$ with at most
      $|X|$ colors, delete $N(X)$, and recurse on the remaining connected components.
      Use a fresh palette at each recursion depth, shared by all components at that depth.

      Every recursive component has at most half as many vertices as its parent, so the recursion has at most
      $\lfloor\log_2 n\rfloor+1$ nonempty levels. Each level uses $\Order(k)$ colors, giving a proper coloring
      with $\Order(k(1+\log n))=\Order(\log n)$ colors for constant $k$.
      At each level, the input components are vertex-disjoint, so
      \Cref{lem:path-separator,lem:path-neighborhood-coloring} together take $\Order(k(m+n))$ time.
      The overall running time is
      $\Order(k(m+n)(1+\log n))=\tOrder(m+n)$ for fixed $k$, as desired.
   \end{proof}

    We now combine the above results to detect triangles in induced $P_k$-free graphs quickly.

   \begin{proof}[Proof of \Cref{thm:paths}]
      Let $G$ be an $n$-vertex graph that is induced $P_k$-free.
      We apply \Cref{lem:path-coloring}, which either finds a triangle in $G$ or computes a proper coloring
      of $G$ using $\Order(\log n)$ colors. If a triangle is found, we output it and terminate.
      Otherwise, let $V_1,V_2,\dots,V_C$ be the color classes, where
      $C=\Order(\log n)$.
      Since each color class is an independent set, any triangle in $G$ must have exactly one
      vertex in each of three distinct color classes.
      We iterate over all $\binom{C}{3} = \Order(\log^3 n)$ triples of distinct color classes.

      Consider a fixed triple $(V_i, V_j, V_\ell)$.
      The bipartite graph induced by $V_i \cup V_j$ is an induced subgraph of $G$,
      and therefore is induced $P_k$-free. 
      By \Cref{lem:VC-paths}, the \VC-dimension of the bipartite adjacency matrix between $V_i$ and $V_j$
      is at most $d = \lfloor k/2 \rfloor - 1$.

      By \Cref{lem:tripartite-vcd}, we can solve Triangle Detection in this
      tripartite subgraph in $\tOrder(n^{3-1/d})$ time \whp
      For each triple, we amplify the success probability of this algorithm
      to at least $1-n^{-2}$ using \cref{lem:standard-amplification}. A union bound over the $\Order(\log^3 n)$ triples shows that,
      with probability at least 
      \[1-\Order(\log^3 n/n^2) \geq 2/3\] 
      the algorithm
      runs in $\tOrder(n^{3-1/d})$ time and outputs the correct answer for every triple.
      Over all triples, the total running time is
      \[
         \tOrder( n^{3-1/d}  \log^3 n) = \tOrder\left( n^{3-\frac{1}{\lfloor k/2 \rfloor -1}}\right)
      \]
      as desired.
   \end{proof}

   \section{Triangle Detection in \texorpdfstring{$H$}{H}-Free Graphs for Degenerately Colored Patterns}
      \label{sec:degenerate}
      In this section, we optimize and derandomize the algorithm of~\cite{abboud2026triangle} for Triangle Detection in 
$H$-free graphs for ``degenerately colorable patterns.'' 
This proves \Cref{thm:degenerate}. 

\subsubsection*{Problem Setup}
   We first recall the definition of degenerate coloring. 

   \degeneratecoloring*
   Note that any graph $H$ that admits a degenerate coloring necessarily has at most one triangle.
   This is because if $H$ has two triangles, we can pick $J$ to be the subgraph induced on the vertices of those two triangles, and now  under any proper 3-coloring no vertex of $J$ can have a monochromatic neighborhood within $J$, because the coloring will assign different colors to the nodes of each triangle.

   Our algorithm will take a given $H$-free graph $G$ and then color-code it to get $3$-colored subgraphs of $G$. 
   Each colored subgraph is \emph{colored $H$-free} for \emph{every} $3$-coloring of $H$, and in particular, free from 
   degenerately colored copies of $H$. The formal definition of colored copies is given below. 
   \begin{definition}[Colored Copies]
      \label{def:colored-copy}
      Given a $3$-colored graph $G$ and a $3$-colored pattern $H$ with colorings
      $c_G$ and $c_H$ respectively, a \emph{colored copy} of $H$ in $G$ is an
      injective homomorphism 
      \[\phi\colon V(H)\to V(G)\] 
      that preserves 
      colors. Namely, $\phi(u)\phi(v) \in E(G)$ for every $uv \in E(H)$, and 
      $c_G(\phi(v))=c_H(v)$ for every $v\in V(H)$. We say that $G$ is \emph{colored $H$-free} if 
      it does not contain a colored copy of $H$. 
   \end{definition}

   To get a deterministic algorithm, we use a deterministic color coding variant instead of the usual randomized version. 
   \begin{lemma}[Deterministic Color Coding]
      \label{lem:det-color-coding}
      Given an $n$-vertex graph $G$, we can deterministically construct, in
      $\tOrder(n^2)$ time, $\tOrder(1)$ $3$-colored subgraphs
      $G_1,\ldots,G_t$ of $G$ such that $G$ contains a triangle if and only if at
      least one $G_i$ contains a triangle. Moreover, every triangle of $G$ appears
      in at least one of the subgraphs.
   \end{lemma}
   \begin{proof}
      The deterministic color-coding construction of
      ~\cite[Section 4]{alon1995color} gives a family $\mathcal C$ of
      $\tOrder(1)$ efficiently computable coloring functions $c:V(G)\to\set{0,1,2}$ such that for every
      triple of vertices, some coloring in $\mathcal C$ assigns three distinct
      values to the triple. For each $c\in\mathcal C$, discard all monochromatic
      edges of $G$ and let $G_c$ be the resulting subgraph, equipped with the
      coloring $c$. Thus, $G_c$ is a $3$-colored subgraph of $G$ and every triangle of 
      $G$ is preserved in some $G_c$. Constructing all $\tOrder(1)$ instances takes $\tOrder(n^2)$ time.
   \end{proof}

   We henceforth assume that the input graph $G$ has a fixed $3$-coloring with
   color classes $A,B,C$, each of size at most $n$. We represent the graph by a
   data structure supporting color-specific adjacency and degree queries in
   $\tOrder(1)$ time and deletion of a vertex $v$ in $\tOrder(\deg(v))$ time. 

\subsubsection*{The Embedding Phase}
   The main idea of the algorithm is to repeatedly choose vertices of $H$
   with monochromatic neighborhoods and embed them in $G$. A degenerate
   coloring guarantees that such vertices exist. 
   Since $G$ is $H$-free, this process cannot produce a complete embedding
   of $H$, so the process terminates when a base case is reached, corresponding to some induced sub-pattern $H' \subseteq H$. 
   We next explain how to implement this procedure and use it to find triangles in $G$.

\begin{figure}[htbp]
      \centering
      \begin{tikzpicture}[
   class/.style={line width=.9pt},
   edge/.style={draw=black!65, line width=.7pt, line cap=round},
   every node/.style={font=\small}
]
   \draw[class, draw=red!75!black, fill=red!5]
      (0,2.40) ellipse (1.15cm and .65cm);
   \draw[class, draw=blue!75!black, fill=blue!5]
      (2.70,-.20) ellipse (1.20cm and .70cm);
   \draw[class, draw=green!45!black, fill=green!5]
      (-2.70,-.20) ellipse (1.20cm and .70cm);

   \coordinate (x) at (0,2.10);

   \foreach \u in {2.22,3.18}
      \draw[edge] (x) -- (\u,-.15);

   \foreach \u in {-3.65,-3.38,-3.11,-2.84,-2.57,-2.30,-2.03,-1.76}
      \draw[edge] (x) -- (\u,-.15);

   \fill[red!80!black] (x) circle (2.6pt);
   \node[anchor=west, text=red!75!black] at (.14,2.14) {$x$};

   \node[text=red!75!black] at (0,3.32) {$A=X$};
   \node[text=blue!75!black] at (2.70,-1.17) {$B=Y$};
   \node[text=green!45!black] at (-2.70,-1.17) {$C=Z$};
\end{tikzpicture}
      \caption{Low-degree cleanup with $X=A$, $Y=B$, and $Z=C$.
         The red vertex $x$ has fewer than $\Delta$ neighbors in the blue class
         $B$, while its degree into the green class $C$ is at most $|Z|$. 
         To check if $x$ participates in a triangle we check at most $\Delta |Z|$ pairs.}
      \label{fig:low-degree-cleanup}
   \end{figure}
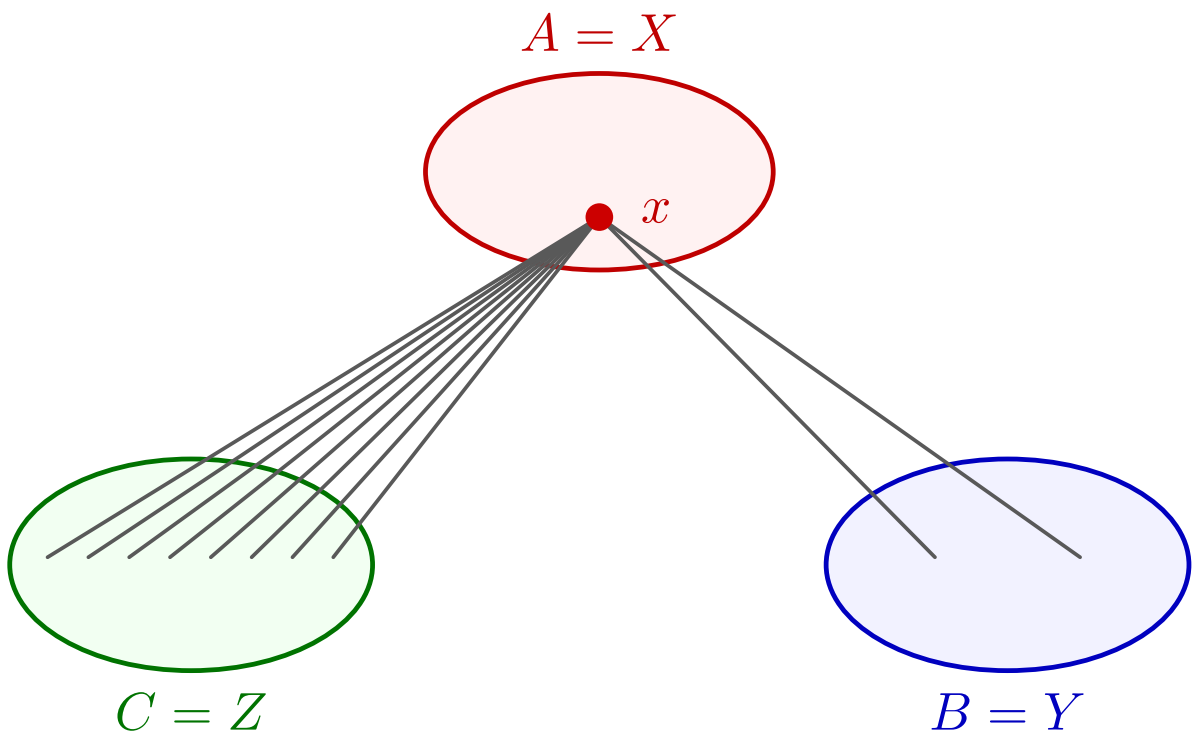

   Before each embedding step, we ensure that every vertex has at least
   $\Delta$ neighbors in each of the other two color classes, for a degree
   threshold $\Delta \ll n$. We enforce this condition through repeated
   applications of a low-degree cleanup procedure. For our choice of
   $\Delta$, the total cleanup time is truly subcubic.
   \Cref{fig:low-degree-cleanup} illustrates the case $X=A$ and $Y=B$.

   \begin{claim}[Low-degree Cleanup]
      \label{claim:deg-cleanup}
      Given distinct color classes $X,Y\in\set{A,B,C}$, there is an algorithm
      that either finds a triangle in $G$, or deletes all vertices $x\in X$ with
      fewer than $\Delta$ neighbors in $Y$.
      In the latter case, $X$ is replaced by a subset $X'\subseteq X$
      such that every $x\in X'$ has at least $\Delta$ neighbors in $Y$,
      while the other two color classes remain unchanged.
      The running time is
      \[
         \tOrder(\Delta |X||Z|),
      \]
      where $Z$ is the remaining color class not equal to $X$ or $Y$.
   \end{claim}
   \begin{proof}
      Enumerate the vertices in $X$ with fewer than $\Delta$ neighbors in $Y$.
      For each such vertex $x$, examine all pairs in
      $(N(x)\cap Y)\times(N(x)\cap Z)$.
      If a pair $(y,z)$ forms an edge, return the triangle $(x,y,z)$.
      Otherwise, delete $x$ from the graph.
      This takes $\tOrder(\Delta |Z|)$ time per vertex in $X$,
      for a total of $\tOrder(\Delta |X||Z|)$ time. 
      This process is depicted in \Cref{fig:low-degree-cleanup}. 
   \end{proof}

\begin{figure}[htbp]
      \centering
      \resizebox{.84\linewidth}{!}{\begin{tikzpicture}[
   every node/.style={font=\small},
   class/.style={line width=.8pt},
   path edge/.style={draw=violet!85!black, line width=1.4pt, line cap=round},
   pattern edge/.style={draw=black!75, line width=.65pt}
]
   \draw[class, draw=red!75!black, fill=red!5]
      (0,1.90) ellipse (.80cm and .50cm);
   \draw[class, draw=green!45!black, fill=green!5]
      (-1.50,0) ellipse (.80cm and .50cm);
   \draw[class, draw=blue!75!black, fill=blue!5]
      (1.35,0) ellipse (.90cm and .60cm);
   \node[text=red!75!black] at (0,2.70) {$A$};
   \node[text=green!45!black] at (-1.50,-.82) {$C$};
   \node[text=blue!75!black] at (1.35,-.92) {$B$};

   \draw[->, draw=black!65, line width=1.8pt] (2.60,1.05) -- (3.45,1.05);

   \begin{scope}[xshift=6.10cm]
   \coordinate (a) at (-.10,1.90);
   \coordinate (c) at (-1.30,0);
   \coordinate (bfirst) at (1.70,-.20);
   \coordinate (blast) at (.95,.20);
   \coordinate (s) at (1.50,1.90);

   \path[draw=blue!45!black, draw opacity=.30,
      fill=blue!20, fill opacity=.12, line width=.55pt]
      (s) -- (.75,0)
      arc[start angle=180,end angle=360,x radius=.60cm,y radius=.60cm]
      -- cycle;

   \draw[class, draw=red!75!black, fill=red!5]
      (0,1.90) ellipse (.80cm and .50cm);
   \draw[class, draw=green!45!black, fill=green!5]
      (-1.50,0) ellipse (.80cm and .50cm);
   \draw[class, draw=blue!75!black, fill=blue!5]
      (1.35,0) circle (.60cm);

   \draw[path edge] (bfirst) -- (c) -- (a) -- (blast);
   \draw[path edge] (blast) -- (s) -- (bfirst);

   \fill[red!80!black] (a) circle (2.8pt);
   \fill[green!45!black] (c) circle (2.8pt);
   \fill[blue!80!black] (bfirst) circle (2.8pt);
   \fill[blue!80!black] (blast) circle (2.8pt);
   \fill[red!80!black] (s) circle (3pt);
   \node[above, inner sep=4pt, text=red!75!black] at (s) {$\hat v$};

   \node[text=red!75!black] at (0,2.70) {$A\setminus\{\hat v\}$};
   \node[text=green!45!black] at (-1.50,-.82) {$C$};
   \node[text=blue!75!black] at (1.35,-.92) {$N_B(\hat v)$};
   \end{scope}

   \foreach \badgepos/\iscycle in {0/1,6.10/0} {
      \begin{scope}[xshift=\badgepos cm,yshift=-1.75cm]
         \coordinate (smallh) at (-.45,.45);
         \coordinate (smallbfirst) at (-.02,.12);
         \coordinate (smallc) at (-.19,-.37);
         \coordinate (smalla) at (-.71,-.37);
         \coordinate (smallblast) at (-.88,.12);
         \draw[pattern edge] (smallbfirst) -- (smallc) -- (smalla) -- (smallblast);
         \ifnum\iscycle=1\relax
            \draw[pattern edge] (smallblast) -- (smallh) -- (smallbfirst);
         \fi
         \fill[blue!80!black] (smallbfirst) circle (1.8pt);
         \fill[green!45!black] (smallc) circle (1.8pt);
         \fill[red!80!black] (smalla) circle (1.8pt);
         \fill[blue!80!black] (smallblast) circle (1.8pt);
         \node[anchor=west, inner sep=0pt] at (.13,-.03) {-free};
         \ifnum\iscycle=1\relax
            \fill[red!80!black] (smallh) circle (1.8pt);
         \fi
      \end{scope}
   }
\end{tikzpicture}}
      \caption{From a colored $H$-free graph to a colored
         $(H\setminus\{v\})$-free restricted graph,  
         with $H=C_5$ and $H\setminus\{v\}=P_4$.
         Purple edges show how a hypothetical colored $P_4$ would extend to a
         forbidden colored $C_5$ by mapping $v$ to the sampled vertex $\hat v$,
         which is complete to $N_B(\hat v)$.}
      \label{fig:embedding-correctness}
   \end{figure}
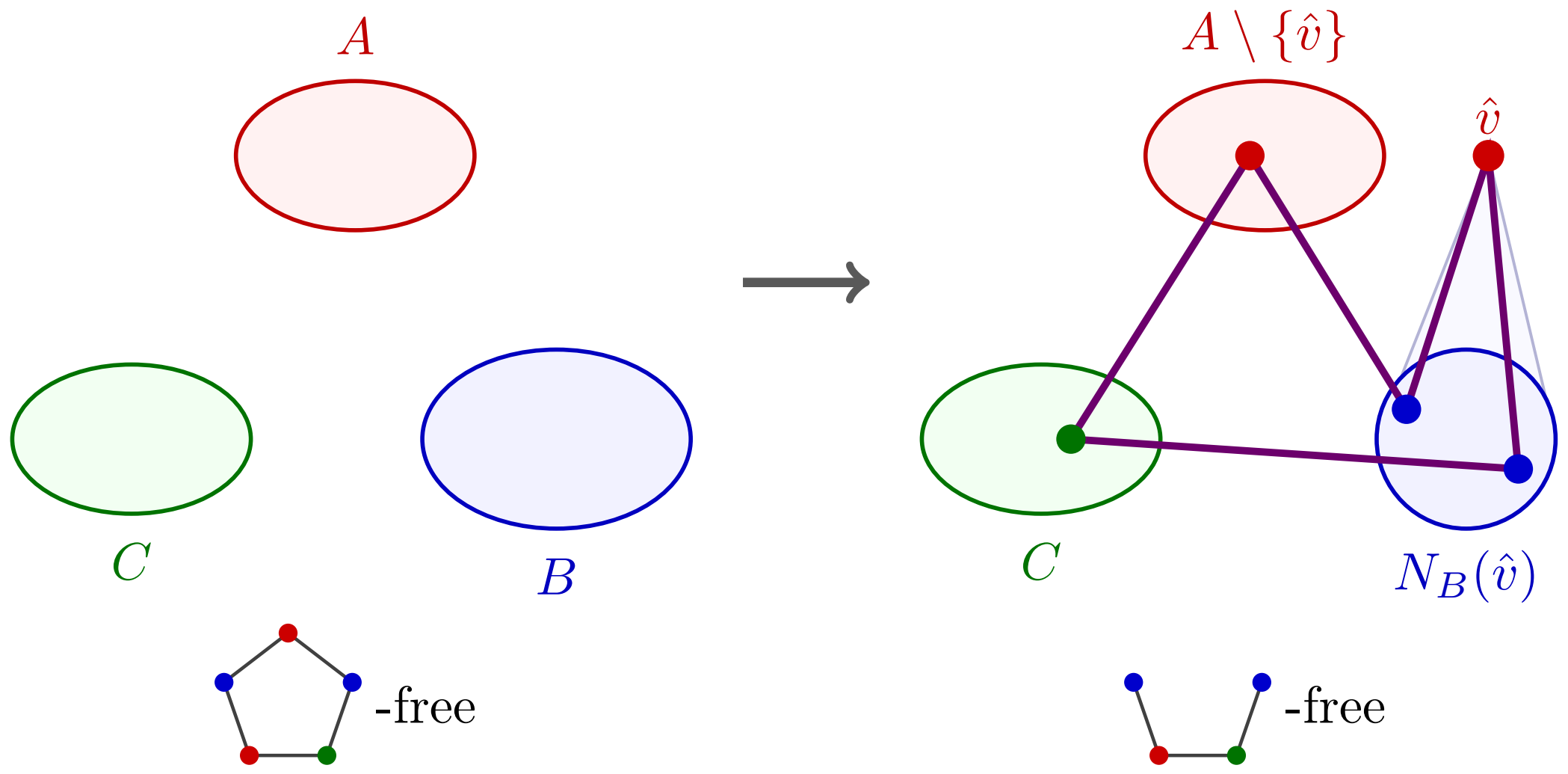

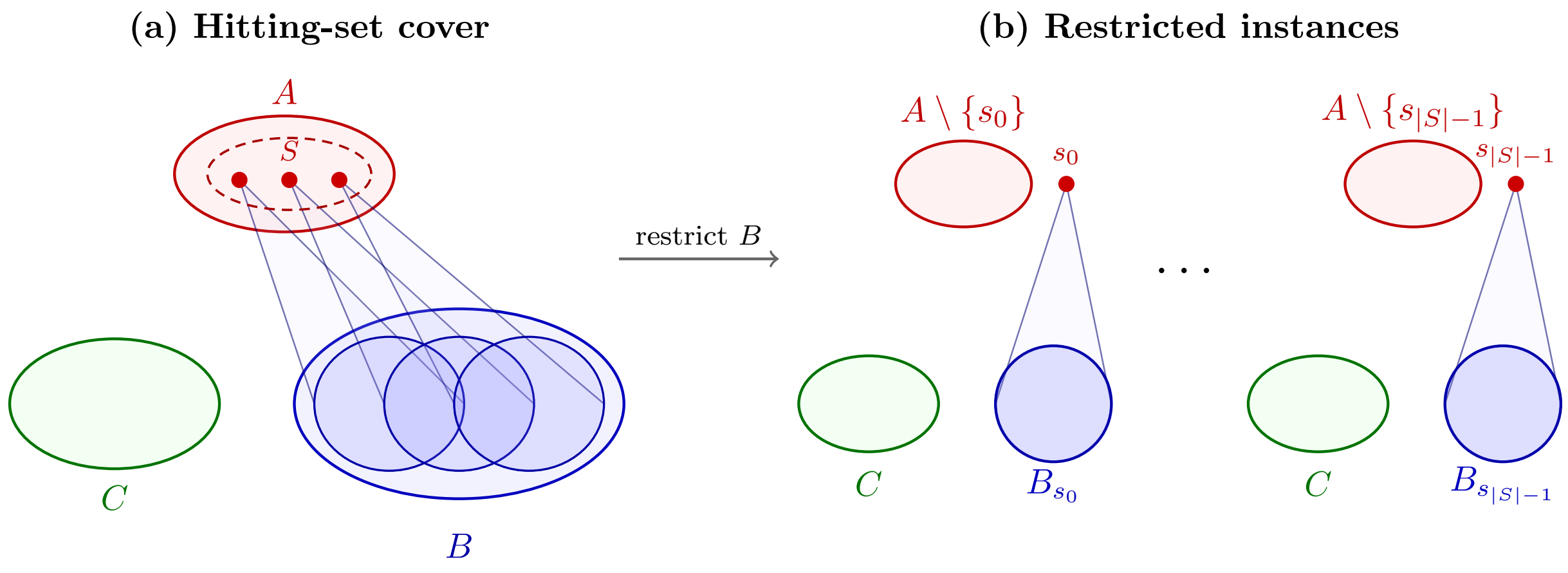
\begin{figure}[htbp]
      \centering
      \resizebox{.98\linewidth}{!}{\begin{tikzpicture}[
   class/.style={line width=.8pt},
   cone/.style={draw=blue!45!black, line width=.45pt,
      fill=blue!20, fill opacity=.10, draw opacity=.55},
   every node/.style={font=\small}
]
   \node[font=\small\bfseries] at (.25,3.75) {(a) Hitting-set cover};
   \node[font=\small\bfseries] at (9.05,3.75) {(b) Restricted instances};

   \draw[class, draw=red!75!black, fill=red!5]
      (0,2.30) ellipse (1.10cm and .58cm);
   \draw[class, draw=blue!75!black, fill=blue!5]
      (1.75,0) ellipse (1.65cm and .95cm);
   \draw[class, draw=green!45!black, fill=green!5]
      (-1.70,0) ellipse (1.05cm and .65cm);

   \foreach \sx/\bx in {-.45/1.05,.05/1.75,.55/2.45} {
      \path[cone] (\sx,2.24) -- ({\bx-.75},0)
         arc[start angle=180,end angle=360,x radius=.75cm,y radius=.67cm]
         -- cycle;
   }
   \foreach \bx in {1.05,1.75,2.45}
      \draw[draw=blue!65!black, line width=.6pt,
         fill=blue!45, fill opacity=.16]
         (\bx,0) ellipse (.75cm and .67cm);

   \draw[draw=red!65!black, densely dashed, line width=.65pt]
      (.05,2.30) ellipse (.82cm and .36cm);
   \foreach \sx in {-.45,.05,.55}
      \fill[red!80!black] (\sx,2.24) circle (2.3pt);
   \node[text=red!75!black, font=\scriptsize] at (.05,2.53) {$S$};

   \node[text=red!75!black] at (0,3.12) {$A$};
   \node[text=green!45!black] at (-1.70,-.94) {$C$};
   \node[text=blue!65!black] at (1.75,-1.42) {$B$};

   \draw[->, draw=black!60, line width=.8pt]
      (3.35,1.45) -- node[above, font=\scriptsize] {restrict $B$} (4.95,1.45);

   \foreach \shift/\centerindex in {6.8/0,11.3/{|S|-1}} {
      \begin{scope}[xshift=\shift cm]
         \draw[class, draw=red!75!black, fill=red!5]
            (0,2.20) ellipse (.68cm and .43cm);
         \draw[class, draw=green!45!black, fill=green!5]
            (-.95,0) ellipse (.70cm and .48cm);
         \draw[class, draw=blue!75!black, fill=blue!5]
            (.90,0) circle (.58cm);

         \path[cone] (1.03,2.20) -- (.32,0)
            arc[start angle=180,end angle=360,radius=.58cm]
            -- cycle;
         \draw[draw=blue!65!black, line width=.6pt,
            fill=blue!45, fill opacity=.16]
            (.90,0) circle (.58cm);
         \fill[red!80!black] (1.03,2.20) circle (2.3pt);
         \node[text=red!75!black, font=\scriptsize] at (1.03,2.48)
            {$s_{\centerindex}$};

         \node[text=red!75!black] at (0,2.91) {$A\setminus\{s_{\centerindex}\}$};
         \node[text=green!45!black] at (-.95,-.80) {$C$};
         \node[text=blue!75!black] at (.90,-.82) {$B_{s_{\centerindex}}$};
      \end{scope}
   }
   \node[font=\Large] at (9.05,1.32) {$\cdots$};
\end{tikzpicture}}
      \caption{The embedding phase of~\cite{abboud2026triangle}, assuming that every vertex
         of $B$ has at least $\Delta$ neighbors in $A$.
         Left: a hitting set $S=\{s_0,\ldots,s_{|S|-1}\}\subseteq A$ whose neighborhoods cover $B$.
         Right: the first and last restricted instances, with classes
         $(A\setminus\{s_j\},B_{s_j},C)$, where $B_{s_j}=N(s_j)\cap B$.
         These blue classes may overlap. Each center $s_j$ is complete to 
         $B_{s_j}$. Other edges are omitted.}
      \label{fig:embedding-phase}
   \end{figure}

   Under this degree condition, the algorithm of~\cite{abboud2026triangle}
   embeds a vertex $v$ of $H$ whose color corresponds to $A$ as follows.
   Suppose the neighbors of $v$ have the color of $B$.
   We sample $\tilde\Order(|A|/\Delta)$ vertices uniformly at random from $A$
   and, for each sampled vertex $\hat{v}$, first check for a triangle containing
   $\hat{v}$. If none is found, form the restricted graph with color classes
   $(A\setminus\{\hat{v}\},N_B(\hat{v}),C)$.
   The following properties hold:
   \begin{itemize}
      \item If $G$ has a triangle, then \whp there is at least one sampled vertex $\hat v$ whose neighborhood in $B$ contains a triangle vertex of $G$.
        That triangle is either found by the preliminary check or survives in
        the restricted graph. This follows from the high degree condition and the sample size.
      \item Every colored copy of $H\setminus\set{v}$ in the restricted graph 
   extends to a colored copy of $H$ by mapping $v$ to $\hat{v}$.
   \end{itemize}
   Thus, the procedure reduces the problem to colored
   $H\setminus\set{v}$-free graphs $G'$, each obtained from $G$ by deleting
   $\hat v$ and restricting $B$ to $N_B(\hat v)$.
   \Cref{fig:embedding-correctness} illustrates this reduction for $H=C_5$.
   \Cref{fig:embedding-phase} illustrates this embedding phase.

   Here, we replace the random sampling step with a deterministic hitting set algorithm. 
   We use a formulation that follows directly from~\cite[Lemma 5.1]{petruschka2026connectivity}.

   \begin{lemma}[Deterministic Hitting Set]
      \label{lem:deg-cover}
      Let $G$ be a bipartite graph with bipartition $X\sqcup Y$.
      Let $n=|X|+|Y|$, and suppose that every
      $y\in Y$ has at least $\Delta \geq 1$ neighbors in $X$. One can find, in
      $\Order(n \Delta )$ time, a set $S\subseteq X$ of size
      $\tOrder((|X|/\Delta))$ such that
      $Y\subseteq N(S)$.
   \end{lemma}
   In particular, the neighborhoods of vertices in $S$, restricted to $Y$, cover $Y$.

   \paragraph{Optimizing the algorithm.}
      We now explain the main inefficiency in the algorithm
      of~\cite{abboud2026triangle}, and how our new algorithm avoids it.
      In the setup described above, where we embed $v$ into $A$ and restrict
      $B$ to neighborhoods of sampled nodes, the graphs produced by each embedding phase
      have overlapping color classes $B'$. These overlaps arise because
      a hitting set guarantees only a cover.
      The overlaps are unnecessary: the reduction to colored
      $H\setminus\set{v}$-free graphs requires only that each triangle's
      vertex in $B$ belongs to at least one restricted class, and that
      every restricted class $B'$ has a vertex in $A$ adjacent to all
      vertices of $B'$. Our algorithm refines the cover into a partition.
      This is the main change that yields the speedup.

   \paragraph{Base cases.}
      If no triangle is found during repeated cleanup and embedding
      steps, we reach a base case.
      These base cases are handled by the following claim from~\cite[Claim 6.2]{abboud2026triangle}. 
      \begin{claim}[Base Case]
         \label{claim:deg-base}
         Let $F$ be a degenerately colored pattern with coloring $c_F$. Triangle Detection in colored $F$-free
         graphs can be solved in $\Order(n^2)$ time if $F$ is triangle-free and $|V(F)|=3$, or if
         $F$ contains a triangle and $|V(F)|=4$.
      \end{claim}

We now combine all the ideas above into a formal proof.
\begin{proof}[Proof of \Cref{thm:degenerate}]
   Let $G$ be the uncolored $n$-vertex $H$-free input graph, and fix a
   degenerate $3$-coloring of $H$. By~\Cref{lem:det-color-coding}, we
   deterministically construct $q=\tOrder(1)$ $3$-colored subgraphs
   $G_1,\ldots,G_q$ of $G$. For each $i$, let $A_i,B_i,C_i$ denote the color
   classes of $G_i$. 

   We apply the following recursive algorithm separately to every $G_i$. Its
   root recursive instance is the graph induced by $(A_i,B_i,C_i)$, with forbidden colored 
   pattern $H$.
   Set
   \begin{equation}
      \label{eq:deg-parameters}
      r=\begin{cases}
          k-3 & \text{if $H$ is triangle-free,}\\
          k-4 & \text{if $H$ contains a triangle,}
      \end{cases}
   \end{equation}
   and let $\delta=n^{-1/(r+1)}$.

   For a recursive instance $I=(A_I,B_I,C_I)$ whose forbidden colored pattern
   is $F$, execute the following four steps.
   \begin{tcolorbox}[
      breakable,
      colback=blue!4,
      colframe=blue!35!black,
      boxrule=.5pt,
      arc=1mm,
      left=8pt, right=8pt, top=5pt, bottom=5pt,
      before upper={\interlinepenalty=10000},
      before skip=8pt, after skip=8pt
   ]
   \begin{enumerate}[label=(\roman*),leftmargin=1.9em]
      \item \textbf{Easy cases.}
      If $F$ is triangle-free and $|V(F)|=3$, or if $F$ contains a triangle and
      $|V(F)|=4$, invoke~\Cref{claim:deg-base} and return its output.

      If one of $A_I,B_I,C_I$ is empty, report that the current instance is
      triangle-free.
      If $F$ has an isolated vertex $h$, choose an arbitrary vertex $s$ in $I$
      having the same color as $h$. Check for a triangle containing $s$ and, if
      there is none, create a child instance $I'$ by deleting $s$ and replacing
      the forbidden pattern with $F-h$. Recurse on $I'$ and return its output.
      Thus, below we may assume that $F$ has no isolated vertices.

      \item \textbf{Low-degree cleanup.}
      Choose $h\in V(F)$ with a monochromatic neighborhood. 
      Suppose without loss of generality that $h$ has color $A$ and its neighbors have
      color $B$. Apply~\Cref{claim:deg-cleanup} with $X=B_I$, $Y=A_I$, and
      $\Delta=\lceil\delta|A_I|\rceil$.
      If it finds a triangle,
      return it. Otherwise, let $B'$ be the surviving subset of
      $B_I$. 

      \item \textbf{Partition by centers.}
      Every vertex of $B'$ has at least $\Delta$ neighbors in $A_I$.
      Invoke~\Cref{lem:deg-cover} to find a set $S\subseteq A_I$ of
      $\tOrder(|A_I|/\Delta)=\tOrder(1/\delta)$ centers whose neighborhoods
      cover $B'$.
      Assign each vertex of $B'$ to an arbitrary adjacent center, obtaining a partition
      \begin{equation}
         \label{eq:deg-center-partition}
         B'=\bigsqcup_{s\in S}B_s,
         \qquad B_s\subseteq N(s).
      \end{equation}

      \item \textbf{Recursive calls.}
      For every $s\in S$, first check for a triangle consisting of $s$,
      a vertex of $B_s$, and a vertex of $C_I$.
      If there is none, create the child instance
      $I_s=(A_I\setminus\{s\},B_s,C_I)$ with forbidden pattern $F'=F-h$ and
      recurse on $I_s$. Return a triangle if any check or recursive call finds
      one; otherwise, report that the current instance is triangle-free.
   \end{enumerate}
   \end{tcolorbox}

   \paragraph{Correctness.}
   Each root instance $G_i$ is colored $H$-free. We prove by induction on
   $|V(F)|$ that the recursive algorithm correctly detects triangles in
   every colored $F$-free instance $I=(A_I,B_I,C_I)$ arising in the recursion.
   The base cases follow from~\Cref{claim:deg-base}, and an instance with
   an empty color class is triangle-free.

   Suppose first that $F$ has no isolated vertices. For each child $I_s$,
   any colored copy of $F-h$ would extend to a colored copy of $F$ in
   $I$ by mapping $h$ to $s$. Indeed, the images of all neighbors of
   $h$ lie in $B_s\subseteq N(s)$, and $s$ is absent from $I_s$.
   The extension therefore preserves colors and edges and is injective.
   Since $I$ is colored $F$-free, every child is colored $(F-h)$-free.

   If cleanup finds no triangle, every triangle of $I$ survives in
   $(A_I,B',C_I)$: the cleanup procedure checks for triangles through
   each vertex before deleting it, as shown in the proof of
   \Cref{claim:deg-cleanup}. By the partition of $B'$, every surviving
   triangle is either found in step (iv) or retained in a child instance.
   Correctness therefore follows from the induction hypothesis.

   Now suppose that $F$ has an isolated vertex $h$, and let $I'=I-s$
   be the child instance created in step (i). Any colored copy of $F-h$
   in $I'$ would extend to a colored copy of $F$ in $I$ by mapping
   $h$ to $s$, since their colors agree and $h$ has no incident edges.
   Thus, $I'$ is colored $(F-h)$-free, and correctness follows
   from the induction hypothesis.

   Conversely, every triangle reported by a direct check or a recursive
   call is a triangle of $I$, since every child is a subgraph of its
   parent. This proves correctness for each $G_i$. Finally,
   \Cref{lem:det-color-coding} guarantees that $G$ contains a triangle
   if and only if some $G_i$ does, proving correctness for the
   original input.

   \paragraph{Runtime.} Fix one colored input $G_i$. Each edge of the recursion
   tree corresponds to deleting one vertex from the forbidden pattern. If $H$ contains
   a triangle, that triangle survives every deletion, since none of its
   vertices has a monochromatic neighborhood. Thus, every root-to-leaf path
   has at most $r$ edges.

   Let $\mathcal I_\ell$ be the set of recursive instances at level $\ell$,
   with the root at level $0$. If $h$ is not isolated, an instance produces
   $\tOrder(1/\delta)$ children; otherwise, it produces at most one child.
   Since $H$, and hence $r$, is fixed, for every $0\leq\ell\leq r$ we have
   \begin{equation}
      \label{eq:deg-instance-count}
      |\mathcal I_\ell|=\tOrder(\delta^{-\ell}).
   \end{equation}

   Define the volume of an instance $I$ by $V(I)=|A_I||B_I||C_I|$.
   If $h$ is not isolated, name the color classes so that $h$ has color $A$
   and its neighbors have color $B$; the volume is invariant under this
   relabeling. By~\Cref{claim:deg-cleanup}, cleanup at $I$ takes
   $\tOrder(\delta V(I)+n^2)$ time, since
   $\Delta = \lceil\delta|A_I|\rceil\leq\delta|A_I|+1$.

   Charge the construction and initialization of each instance to that
   instance, at a cost of $\tOrder(n^2)$. Excluding recursive calls and the
   construction of child instances, the remaining work in steps (i),
   (iii), and (iv) also takes $\tOrder(n^2)$ time per instance. In particular,
   constructing the hitting set $S$ and the partition $(B_s)_{s\in S}$ takes
   $\tOrder(n^2)$ time, and the center checks in step (iv) take total time
   $\tOrder(\sum_{s\in S}|B_s||C_I|)=\tOrder(|B'||C_I|)$.
   Thus, the total work charged to $I$ is $\tOrder(\delta V(I)+n^2)$.

   By the volume bound in~\Cref{claim:deg-volume} below and the instance
   count in~\Cref{eq:deg-instance-count}, the total work at level $\ell$ is
   \begin{equation}
      \label{eq:deg-level-cost}
      \tOrder\left(\delta\sum_{I\in\mathcal I_\ell}V(I)
         +n^2|\mathcal I_\ell|\right)
      =\tOrder\left(\delta n^3+\frac{n^2}{\delta^\ell}\right).
   \end{equation}
   Summing~\Cref{eq:deg-level-cost} over the at most $r+1$ levels and all
   $q=\tOrder(1)$ colored instances gives a total running time of
   \begin{equation}
      \label{eq:deg-total-cost}
      \tOrder\left(\delta n^3+\frac{n^2}{\delta^r}\right).
   \end{equation}
   This bound also includes the $\tOrder(n^2)$ time needed to construct the
   colored instances.
   Our choice $\delta=n^{-1/(r+1)}$ balances the two terms:
   \[
      \delta n^3=\frac{n^2}{\delta^r}=n^{3-1/(r+1)}.
   \]
   Therefore, the total running time is
   $\tOrder(n^{3-1/(r+1)})$, which is the claimed bound in both cases.
\end{proof}

\begin{claim}[Volume Bound]
   \label{claim:deg-volume}
   Fix a root instance $G_i$ of the recursive algorithm above, and let
   $\mathcal I_\ell$ denote the recursive instances at level $\ell$.
   Writing $V(I)=|A_I||B_I||C_I|$, for every $\ell\geq0$ we have
   \[
      \sum_{I\in\mathcal I_\ell}V(I)\leq |A_i||B_i||C_i|\leq n^3.
   \]
\end{claim}
\begin{proof}
   Consider an instance $I$, and assume that $h$ is not isolated. Name the
   color classes so that $h$ has color $A$ and its neighbors have color $B$.
   Each child has classes $(A_I\setminus\{s\},B_s,C_I)$ for some $s\in S$.
   Since the sets $B_s$ partition $B'\subseteq B_I$, we have
   \[
      \sum_{\text{$I'$ child of $I$}}V(I')
      \leq |A_I||C_I|\sum_{s\in S}|B_s|
      =|A_I||B'||C_I|
      \leq V(I).
   \]
   The same inequality holds when $h$ is isolated, so the claim follows by
   induction over the recursion levels.
\end{proof}

   \section{Conclusion}
      \label{sec:conclusions}
      Assuming the OMv Conjecture (\Cref{conj:omv}), we proved an
$n^{3-1/d-\Order(1/d^2)}$ total-time lower bound for OMv in matrices $M$ with low \VC-dimension 
$\VC(M)\leq d$, for every $d\geq3$. The best known upper bound is
$\tOrder(n^{3-1/d})$. We also showed that,
assuming the conjecture of Janzer and 
Pohoata~\cite{janzer2024zarankiewicz}, our filter-based self-reduction cannot yield
lower-bound exponents above $3-1/d-f(d)$ for some $f(d)>0$.
This limitation of the reduction does not provide a faster OMv algorithm.
Closing the gap between the upper and lower bounds, through an improved
algorithm or a reduction beyond this framework, is the main open problem left by this work.

We conclude with a few more interesting questions suggested by our work:
\begin{itemize}
   \item Can specialized techniques yield lower bounds for Triangle Detection in \emph{induced} $H$-free graphs? The filtering approach inherently relies on deleting edges. Eliminating only induced copies of $H$ is substantially harder, since it may require adding edges without creating triangles.
   \item In the setting of induced $H$-free graphs, it remains open whether, for every $k \geq 1$, Triangle Detection admits a truly subcubic combinatorial algorithm on graphs with no induced $C_{2k+1}$. Another related question is whether Triangle Detection in induced $P_k$-free graphs can be solved in near-linear time for all $k$. 
   \item What meaningful notions of structure for weighted graphs or matrices lead to faster combinatorial algorithms? Can the filtering technique provide lower bounds under those notions?
\end{itemize}

   \section*{Acknowledgements}
      We thank Jan van den Brand and Emile Anand for sharing some arguments related to OMv in matrices of bounded \VC-dimension.
      The first, third, and fourth authors thank David Woodruff for informing them that an internal model by Google suggested derandomizing the main algorithm of~\cite{abboud2026triangle} 
      using a deterministic hitting set. Although it was already suggested in~\cite[Section 11]{abboud2026triangle} that such a 
      derandomization should be possible, no formal proof was presented.

   \bibliographystyle{alpha}
   \bibliography{references}
   \clearpage
   \appendix
   \section{Proofs of Auxiliary \VC-Dimension Bounds}
      \label{app:auxiliary-vc-proofs}
      This appendix contains the proofs of the auxiliary \VC-dimension bounds
stated in \Cref{sub:aux}.

   \begin{proof}[Proof of \Cref{claim:induced-subgraph-vc}]
      Let $U=V(H')$. For every $v\in U$, we have
      $N^{H'}(v)=N^H(v)\cap U$. Hence every set shattered by the open-neighborhood
      set system of $H'$ is also shattered by that of $H$.
   \end{proof}

   \begin{proof}[Proof of \Cref{claim:isolated-vertices-vc}]
      Let $A\neq\emptyset$ be shattered by $H'$. No added isolated vertex
      belongs to $A$. Fix $a\in A$. For every
      $B\subseteq A\setminus\set{a}$, a vertex whose neighborhood in $H'$
      intersects $A$ in $B\cup\set{a}$ must belong to $H$. Its neighborhood
      in $H$ intersects $A\setminus\set{a}$ in exactly $B$. Thus
      $A\setminus\set{a}$ is shattered by $H$, and
      $|A|\leq\VC(H)+1$.
   \end{proof}

   \begin{proof}[Proof of \Cref{lem:vc-union}]
      Set $D=d_1+d_2$. If some $d_i=0$, then $H_i$ is edgeless and the claim is
      immediate, so assume that $d_1,d_2\geq 1$. Suppose that a set $X$ of size $k$
      is shattered by $H_1\cup H_2$. Since
      \[
      N^{H_1\cup H_2}(v)\cap X
      =\bigl(N^{H_1}(v)\cap X\bigr)\cup\bigl(N^{H_2}(v)\cap X\bigr),
      \]
      the Sauer--Shelah lemma~\cite{sauer1972density,shelah1972combinatorial} gives
      \[
      2^k
      \leq \left(\sum_{j=0}^{d_1}\binom{k}{j}\right)\left(\sum_{j=0}^{d_2}\binom{k}{j}\right).
      \]
      If $k\geq 5D$, then by \Cref{binomial-sum} we have
      \[\sum_{j=0}^{d_i}\binom{k}{j}\leq (ek/d_i)^{d_i}\] 
      for $i\in\set{1,2}$,
      so the above inequality simplifies to 
      \[2^k \le (ek)^D/\left(d_1^{d_1}d_2^{d_2}\right).\]
      Now applying convexity of $x\mapsto x\log x$,
    we deduce that 
      \[
      2^k
      \leq \frac{(ek)^D}{d_1^{d_1}d_2^{d_2}}
      \leq \left(\frac{2ek}{D}\right)^D.
      \]
      Writing $t=k/D\geq 5$ gives $2^t\leq 2et$, which is false for any $t\ge 5$. Hence $k<5D$ as desired.
   \end{proof}

   \begin{proof}[Proof of \Cref{claim:vc-bipartite-complement}]
      Let $A$ be shattered by $\overline{H}^{\,\mathrm{bip}}$.
      If $A=\emptyset$, there is nothing to prove. Since every
      nonempty neighborhood lies in one side of the bipartition, $A$ lies
      entirely in one side; assume by symmetry that $A\subseteq Y$. Fix
      $a\in A$. For
      every $B\subseteq A\setminus\set{a}$, choose a vertex whose neighborhood
      in $\overline{H}^{\,\mathrm{bip}}$ intersects $A$ in
      \[
         \set{a}\cup\bigl((A\setminus\set{a})\setminus B\bigr).
      \]
      This set is nonempty, so the chosen vertex belongs to $X$. Its
      neighborhood in $H$ intersects $A\setminus\set{a}$ in exactly $B$.
      Thus $A\setminus\set{a}$ is shattered by $H$, and
      $|A|\leq\VC(H)+1$.
   \end{proof}

   \begin{proof}[Proof of \Cref{claim:vc-co-matching}]
      A shattered set cannot meet both sides of the bipartition.
      On a set $A$ contained in one side, every neighborhood
      induces $\emptyset$, $A$, or $A\setminus\set{a}$ for some $a\in A$.
      If $|A|\geq3$, no singleton is realized, so $A$ is not shattered.
   \end{proof}

   \begin{proof}[Proof of \Cref{claim:vc-matching}]
      Every neighborhood has size at most one, so no two-element set is
      shattered.
   \end{proof}

   \begin{proof}[Proof of \Cref{claim:stars}]
      Let $H$ be a star with center $u_S$ and leaf set $S$. Every
      neighborhood is either $S$ or $\set{u_S}$, so no two-element set is
      shattered. If $s\in S$, then $\set{s}$ is shattered
      by $N^H(u_S)$ and $N^H(s)$, proving equality when $S\neq\emptyset$.
   \end{proof}

\end{document}